%% file: A-file.tex
\documentclass[english]{elsarticle}

\usepackage[T1]{fontenc}
\usepackage[latin9]{inputenc}
\usepackage{geometry}
\usepackage{xcolor}
\usepackage{bm}
\usepackage{amsmath}
\usepackage{graphicx}

\makeatletter

\providecommand{\tabularnewline}{\\}

\journal{Journal of Computational Physics}

\biboptions{numbers,sort&compress}

\usepackage{tcolorbox}
\tcbuselibrary{theorems}

\usepackage[ruled, lined, longend]{algorithm2e}

\tcbset{colback=black!5!white,colframe=red!75!black,left=1mm,top=1mm,bottom=1mm,
right=1mm,boxsep=0mm,width=3cm,nobeforeafter}

\makeatother

\usepackage{babel}
\begin{document}

\begin{frontmatter}{}

\title{An efficient, conservative, time-implicit solver for the fully kinetic
arbitrary-species 1D-2V Vlasov-Ampère system}

\author[theory]{S. E. Anderson\corref{cor1}}

\ead{andeste@lanl.gov}

\author[theory]{W. T. Taitano}

\author[theory]{L. Chacón}

\author[design]{A. N. Simakov}

\cortext[cor1]{Corresponding author}

\address[theory]{Theoretical Division, Los Alamos National Laboratory, Los Alamos,
NM, U.S.A. 87545 }

\address[design]{Theoretical Design Division, Los Alamos National Laboratory, Los
Alamos, NM, U.S.A. 87545 }
\begin{abstract}
We consider the solution of the fully kinetic (including electrons)
Vlasov-Ampère system in a one-dimensional physical space and two-dimensional
velocity space (1D-2V) for an arbitrary number of species with a time-implicit
Eulerian algorithm. The problem of velocity-space meshing for disparate
thermal and bulk velocities is dealt with by an adaptive coordinate
transformation of the Vlasov equation for each species, which is then
discretized, including the resulting inertial terms. Mass, momentum,
and energy are conserved, and Gauss's law is enforced to within the
nonlinear convergence tolerance of the iterative solver through a
set of nonlinear constraint functions while permitting significant
flexibility in choosing discretizations in time, configuration, and
velocity space. We mitigate the temporal stiffness introduced by,
e.g., the plasma frequency through the use of high-order/low-order
(HOLO) acceleration of the iterative implicit solver. We present several
numerical results for canonical problems of varying degrees of complexity,
including the multiscale ion-acoustic shock wave problem, which demonstrate
the efficacy, accuracy, and efficiency of the scheme.
\end{abstract}

\end{frontmatter}{}

\section{Introduction\label{sec:Introduction}}

In recent years, it has become apparent that kinetic effects (i.e.,
particle long mean-free-path) can play a significant role in the evolution
of high-energy-density (HED) plasma systems, such as inertial confinement
fusion (ICF) capsule implosions \citep{Rinderknecht2018,Taitano2018b,Keenan2018,Atzeni2016,Sangster2017}.
To study these systems, radiation-hydrodynamic models are typically
used; however, to resolve the long mean-free-path effects, it is necessary
to employ a kinetic approach. Vlasov-Fokker-Planck codes, such as
iFP \citep{Taitano2018} and FPion \citep{Larroche2003}, have been
developed with the goal of resolving ion kinetic effects in weakly
collisional regimes with arbitrary Knudsen numbers. However, they
continue to treat the electrons as a quasineutral, ambipolar fluid,
including only an electron temperature equation. The fluid electron
assumption neglects important kinetic plasma effects such as nonlocal
electron heat transport, which may be necessary to correctly describe
HED plasma phenomena such as shocks and ablation fronts in ICF implosions.
This study proposes an efficient and accurate algorithmic solution
for simulating the fully kinetic \emph{1D-2V} ion-electron system.

There have been attempts to account for nonlocal electron effects
within fluid models. A common approach is limiting the electron heat
flux to some fraction of the free-streaming flux \citep{Luciani1983};
another strategy is to spatially convolve the electron heat flux \citep{Schurtz2000}.
However, in order to describe electron kinetic effects accurately,
it is necessary to solve a fully kinetic model.  To this end, the
Vlasov-Fokker-Planck(VFP)-Maxwell system of equations may be taken
as a first-principles representation of a weakly-coupled plasma. In
a one-dimensional spatial system, and assuming an electrostatic field
response, this may \textendash{} without loss of generality \textendash{}
be reduced to a 2D velocity space described by longitudinal (parallel)
and perpendicular velocities. This leads to the 1D-2V VFP-Ampère (or
VFP-Poisson) system. The 1D Ampère equation describes the evolution
of the longitudinal electric field based on the moments of all species'
velocity distribution functions, while the 1D-2V VFP equation describes
the evolution of these distribution functions. To simplify the current
development and presentation, we explore in this work only the collisionless
aspects of the algorithm, i.e., the 1D-2V Vlasov-Ampère system. This
in no way compromises our goal of developing a fully kinetic simulation
tool for plasmas of arbitrary collisionality, as the numerical details
of extending the approach described herein to include the Fokker-Planck
(FP) collision operator may be considered independently. Interested
readers are referred to e.g., Refs. \citep{Taitano2015,Taitano2016,Taitano2017}
for a compatible fully implicit finite-difference approach.

To solve the Vlasov-Ampère system, there are a variety of possible
approaches including temporally implicit or explicit applications
of particle-in-cell (PIC) methods \citep{Taitano2013,Chen2011,Chen2014,Chen2015},
semi-Lagrangian grid-based methods \citep{Rossmanith2011,Besse2003,Cheng1976},
or Eulerian grid-based approaches \citep{Taitano2015a,Horne2001,Cheng2014}.
In addition, it is possible to utilize semi-implicit strategies \citep{Boscarino2016},
which aim at treating only the stiff physics implicitly, as well as
various asymptotic-preserving (AP) schemes \citep{Degond2017,Chertock2018},
which propose discrete formulations capable of capturing the correct
asymptotic limit when stiff physics are stepped over. In this work,
we apply a temporally fully implicit grid-based Eulerian approach.
An implicit approach has significant advantages over explicit schemes,
particularly for grid-based approaches where the advective Courant-Friedrichs-Lewy
(CFL) time-step limit is determined by the fastest speed on the velocity
grid. With a fully-implicit nonlinear iterative solver, highly temporally
multiscale problems \textendash{} in which the system dynamics are
driven on time-scales much longer than the fastest supported time-scales
\textendash{} may become much more tractable. A classic collisionless
example is the ion-acoustic shock wave, where the dynamic time-scale
is roughly 100 times longer than the inverse electron plasma frequency,
and may be 1000 times (or more) longer than the explicit time-scale
based on the maximum grid velocity \citep{Taitano2015a}. In addition,
a fully implicit iterative scheme exhibits advantages over semi-implicit
and AP schemes primarily in the straightforward application of existing
strategies to enforce discrete conservation. In the current work,
the high-order(HO)/low-order(LO) scheme (HOLO) is used so that the
LO (moment) system of equations accelerates convergence of the HO
(kinetic) system. The LO system consists of the moments of the plasma
species' Vlasov equations coupled to Ampère's equation, while the
HO system consists of the Vlasov equations. HOLO approaches have been
used to solve a variety of systems \citep{Chacon2017}, from neutron
\citep{knoll2011} and thermal radiation transport \citep{Park2012}
to BGK gas-kinetics \citep{Taitano2014}. More importantly, the HOLO
approach has been applied to the solution of collisionless \citep{Taitano2013,Taitano2015a}
and collisional \citep{Taitano2015b} plasma systems. In an earlier
study by Taitano and Chacón \citep{Taitano2015a}, the HOLO approach
was used to accelerate Vlasov-Ampère convergence by using the LO system
to efficiently evaluate the electric field with the higher-order moment
closure provided by the HO system. In this work, we generalize this
study both by applying a non-centered time integration scheme and
by considering the adaptive velocity-space strategy proposed in Ref.
\citep{Taitano2016}.

When solving the Vlasov-Ampère system, a static velocity mesh may
become inefficient in problems where the species temperatures and
bulk velocities exhibit significant temporal and spatial variations.
Specifically, the mesh must be large enough to capture both the shift
in the bulk velocity and temperature evolution at the hottest location
in space and time (the largest thermal speed), while maintaining a
sufficient resolution for the coldest location (the smallest thermal
speed) for all species. In contrast, a mesh that dynamically expands/contracts
in space and time while shifting the center to track changes in their
bulk velocities may efficiently resolve the hottest/coldest regions
of each species. The present work applies an analytic transformation
to the Vlasov equation for each species $\alpha$, scaling it by a
normalizing speed $v_{\alpha}^{*}$ (which is a function of the thermal
speed $v_{th,\alpha}$) and shifting the velocity space by an offset
velocity $u_{\parallel,\alpha}^{*}$ (which is a function of the bulk
velocity $u_{\parallel,\alpha}$). This is similar to the approach
described in Refs. \citep{Taitano2018,Taitano2018a,Larroche2003,Chertock2018,Filbet2013}.

To preserve the numerical accuracy of long-time simulations, we desire
a discretization scheme for which the continuum symmetries of the
governing equations (leading to mass, momentum, and energy conservation)
are preserved in the discrete. Without a discrete conservation principle,
long-term simulations may produce significant violations of the conservation
properties due to accumulated discretization errors, which can manifest
as numerical plasma heating or cooling \citep{Taitano2015b}, or a
departure of the solution from the asymptotic hydrodynamic manifold.
Indeed, as we shall demonstrate later, the failure to ensure discrete
charge conservation (i.e., enforcing the discrete Gauss's law) leads
to catastrophic failure in simulations, with significant departure
from the correct solution. Further, in the case of the Vlasov-Fokker-Planck
equation, Taitano \emph{et al.} \citep{Taitano2018} showed that even
neglecting to ensure discrete momentum and energy conservation relationships
only in the Vlasov equation (while enforcing it in the Fokker-Planck
collision operator) leads to extremely large numerical errors {[}$\sim O\left(1\right)${]}.
Thus, discrete conservation is key for achieving high fidelity and
accuracy.

Broadly, we will distinguish between two different strategies for achieving
discrete conservation. The first, which we term a ``passive'' approach,
relies on specifically chosen discretizations that ``passively'' preserve
the structure of the governing equations. This is a general catch-all
for the symplectic and Hamiltonian-preserving techniques for plasma
physics systems described by Morrison \citep{Morrison2017}. Such
techniques have also been called ``structure-preserving'' \citep{Morrison2017,Shiroto2019},
and are in general very effective at preserving invariants \textendash{}
for the Vlasov-Maxwell system, Shiroto \emph{et al.} \citep{Shiroto2019}
demonstrated conservation errors on the order of machine precision.
However, from our perspective, there are two significant shortcomings
of this approach. The first is that they are generally only possible
through central differencing schemes, which are not monotonic, positivity-preserving,
or non-oscillatory (all of which are desirable properties). The second
is that, in the case of the velocity-space transformed Vlasov equation
used in this work, it is not readily apparent whether such ``structure-preserving''
discretizations are possible. The second strategy, which we term an
``active'' approach, is the strategy we employ in this work to achieve
discrete conservation. To implement this approach, we introduce Lagrange-multiplier-like
constraint functions into the discretized governing equations. These
``nonlinear constraint functions'' are defined so as to \emph{actively
}enforce certain continuum symmetries of the governing equations,
ensuring conservation of e.g., mass, momentum, and energy in the discretized
system. The primary benefit of the ``active'' strategy is that it
permits a choice of arbitrary discretization schemes in both time
(e.g., backward Euler, or BDF2) and phase space (e.g., SMART \citep{Gaskell1988}
or WENO \citep{Jiang1996}).

The present approach is similar to the strategies used in Refs. \citep{Taitano2015a}
and \citep{Taitano2018a}; however, there are some important differences.
In the previous Vlasov-Ampère implementation, the approach relied
on a time-centered Crank-Nicolson integrator to achieve energy conservation.
The current work uses a BDF2 temporal integration scheme (which is
more appropriate for an eventual application to the collisional system),
and can in principle be applied to an arbitrary temporal integration
scheme. Further, the approach for enforcing discrete conservation
with the velocity-space adaptivity (as in Ref. \citep{Taitano2018a})
must be modified because of interaction with the additional constraint
functions. Thus, the current work delivers an implicit algorithm for
the fully kinetic, arbitrary-species 1D-2V Vlasov-Ampère system, which
conserves mass, momentum, and energy to within nonlinear convergence
tolerance. The algorithm is adaptive in the velocity space to ease
meshing requirements due to temporal and spatial variations in the
local bulk velocity and thermal speed of each species, while the nonlinear
constraint functions that ensure conservation also allow substantial
freedom of choice for temporal and advective discretizations.

The rest of this paper is organized as follows. Section \ref{sec:Vlasov-Ampere-System-of}
gives an overview of the governing equations for the Vlasov-Ampère/Poisson
system in 1D-2V, its transformation in the velocity space, and the
continuum-conservation symmetries of the system. Section \ref{sec:Numerical-Implementation}
describes the discretization of the Vlasov system. Section \ref{sec:Discrete-conservation-strategy}
provides details of our strategy for ensuring the continuum conservation
symmetries in the discretized system. In Sec. \ref{sec:Solution-strategy-for}
we present our nonlinear iterative strategy for solving the discretized
system implicitly in time using a HOLO acceleration scheme. We present
numerical results highlighting the accuracy and performance of the
algorithm for several canonical problems of varying difficulty in
Sec. \ref{sec:Numerical-Results}, and provide concluding remarks
in Sec. \ref{sec:Conclusions}.

\input{Methods.tex} \input{Results.tex}\input{Appendices.tex}

\section*{\textemdash \textemdash \textemdash \textemdash \textemdash \textendash{}}

\bibliographystyle{unsrt}
\bibliography{library}

\end{document}

%% file: Methods.tex
\section{Vlasov-Ampère system of equations\label{sec:Vlasov-Ampere-System-of}}

The Vlasov-Ampère/Poisson system may be regarded as a first-principles
representation for a fully ionized electrostatic collisionless plasma.
The governing equations are the Vlasov equations for each species
$\alpha$, 

\begin{equation}
\partial_{t}f_{\alpha}+\bm{\nabla}_{\bm{x}}\cdot\left(\bm{v}f_{\alpha}\right)+\frac{q_{\alpha}}{m_{\alpha}}\bm{E}\cdot\bm{\nabla}_{\bm{v}}\left(f_{\alpha}\right)=0,\label{eq:vlasov}
\end{equation}
which describe the evolution in phase space of distribution functions,
$f_{\alpha}$, and Ampère's equation,

\begin{equation}
\epsilon_{0}\partial_{t}\bm{E}+\sum\limits _{\alpha}q_{\alpha}\bm{\Gamma}_{\alpha}=0,\label{eq:Ampere}
\end{equation}
which describes the evolution of the electric field, $\boldsymbol{E}$.
In Eqs. (\ref{eq:vlasov}) and (\ref{eq:Ampere}), $\bm{v}$ is the
particle velocity, and $q_{\alpha}$ and $m_{\alpha}$ are the particle
charge and mass of species $\alpha$, respectively. We define the
particle flux density to be $\bm{\Gamma}_{\alpha}\equiv n_{\alpha}\bm{u}_{\alpha}$,
where $n_{\alpha}$ is the number density and $\bm{u}_{\alpha}$ the
bulk velocity. The Vlasov-Poisson system instead utilizes Gauss's
law,

\begin{equation}
\epsilon_{0}\bm{\nabla}_{\bm{x}}\cdot\bm{E}-\sum\limits _{\alpha}q_{\alpha}n_{\alpha}=0,\label{eq:Gauss}
\end{equation}
and the electric potential, $\Phi$, defined by $\boldsymbol{\nabla}_{\boldsymbol{x}}\Phi=-\boldsymbol{E}$,
to obtain Poisson's equation,
\begin{equation}
\epsilon_{0}\bm{\nabla}_{\bm{x}}^{2}\Phi+\sum\limits _{\alpha}q_{\alpha}n_{\alpha}=0.\label{eq:Poisson}
\end{equation}
The Vlasov-Poisson and Vlasov-Ampère systems can be shown to be equivalent
through charge conservation, i.e.,
\begin{equation}
0=\boldsymbol{\nabla_{x}}\cdot\left[\epsilon_{0}\partial_{t}\boldsymbol{E}+\sum_{\alpha}q_{\alpha}\boldsymbol{\Gamma}_{\alpha}\right]=\partial_{t}\left[\epsilon_{0}\bm{\nabla}_{\bm{x}}\cdot\bm{E}-\sum\limits _{\alpha}q_{\alpha}n_{\alpha}\right]\Longrightarrow\partial_{t}\sum_{\alpha}q_{\alpha}n_{\alpha}+\boldsymbol{\nabla_{x}}\cdot\sum_{\alpha}q_{\alpha}\bm{\Gamma}_{\alpha}=0.\label{eq:charge-cons-vec}
\end{equation}

In one-dimensional configuration space, Eqs. (\ref{eq:vlasov})\textendash (\ref{eq:Ampere})
may be expressed as

\begin{align}
\partial_{t}f_{\alpha}+\partial_{x}\left(v_{\parallel}f_{\alpha}\right)+\frac{q_{\alpha}}{m_{\alpha}}E_{\parallel}\partial_{v_{\parallel}}\left(f_{\alpha}\right)=0,\label{eq:Vlasov-1D}\\
\epsilon_{0}\partial_{t}E_{\parallel}+\sum\limits _{\alpha}q_{\alpha}\Gamma_{\parallel,\alpha}=\overline{j}_{\parallel},\label{eq:Ampere-1D}
\end{align}
where 
\[
\overline{\Phi}=\frac{1}{x_{\mathrm{max}}-x_{\mathrm{min}}}\int_{x_{\mathrm{min}}}^{x_{\mathrm{max}}}\Phi dx
\]
denotes the spatial average of a quantity $\Phi$. We note that we
have included the spatial average of the current density $\overline{j}_{\parallel}$
in Ampère's equation. This is necessary in 1D periodic systems to
preserve Galilean invariance, and to ensure $\overline{E}_{\parallel}=0$
\citep{Chen2011,Chen2014,Taitano2013}. For details, see \ref{app:Nonlinear-generation-of}.
Without loss of generality, the velocity-space domain may be reduced
to two dimensions by invoking cylindrical symmetry, such that the
velocity coordinates reduce from $(v_{x,}v_{y,}v_{z})$ to $(v_{\parallel},v_{\bot})$.
In Eqs. (\ref{eq:Vlasov-1D})\textendash (\ref{eq:Ampere-1D}), we
use the parallel notation to indicate vector components.

\subsection{Velocity-space coordinate transformation \label{subsec:Velocity-space-coordinate-transf}}

We perform a phase-space coordinate transformation of Eq. (\ref{eq:Vlasov-1D})
as proposed in Refs. \citep{Taitano2018a,Filbet2013,Chertock2018}.
Namely, for each species $\alpha$ we transform the velocity space
(i.e. the velocity coordinate $\bm{v}$) by a translating with a reference
offset velocity $u_{\parallel,\alpha}^{*}(x,t)$ and then normalizing
by a reference speed $v_{\alpha}^{*}(x,t)$. These quantities are
related to each species bulk-flow velocity and thermal speed, respectively,
but are not necessarily equal to them. For each species, we thus define
a transformed velocity coordinate $\bm{c}$ as
\begin{equation}
\bm{c}=\hat{\bm{v}}-\hat{u}_{\parallel,\alpha}^{*}\bm{e}_{\parallel},\label{eq:v-transformed}
\end{equation}
where $\hat{\bm{v}}=\bm{v}/v_{\alpha}^{*}(x,t)$ is the normalized
velocity coordinate, $\hat{u}_{\parallel,\alpha}^{*}=u_{\parallel,\alpha}^{*}/v_{\alpha}^{*}$
is the normalized offset velocity, and $\bm{e}_{\parallel}$ is the
unit vector along $x$. The velocity coordinate $\bm{v}$ may be thus
decomposed using $\bm{c}$, $v_{\alpha}^{*}$, $u_{\parallel,\alpha}^{*}$,
and $\bm{e}_{\parallel}$ as,
\begin{equation}
\bm{v}=v_{\alpha}^{*}(x,t)\bm{c}+u_{\parallel,\alpha}^{*}(x,t)\bm{e}_{\parallel}.\label{eq:v-decomposed}
\end{equation}
In this work, the goal of the velocity transformation is to ensure
that the \emph{computational} velocity-space domain (i.e. the set
of logical velocity space coordinates) is identical for all species.
Thus, regardless of spatio-temporal variations in bulk velocity and
temperature (thermal speed) between species, we are able to use the
same mesh for all species in velocity space.

For full details of the transformation, we refer readers to the work
of Taitano \emph{et al.} \citep{Taitano2018a}. The final form of
the transformed Vlasov equation is thus
\begin{multline}
\partial_{t}\tilde{f}_{\alpha}+\partial_{x}\left(v_{\alpha}^{*}\hat{v}_{\parallel}\tilde{f}_{\alpha}\right)+\frac{q_{\alpha}}{m_{\alpha}v_{\alpha}^{*}}E_{\parallel}\partial_{c_{\parallel}}\tilde{f}_{\alpha}\\
-\frac{1}{v_{\alpha}^{*}}\nabla_{\bm{c}}\cdot\left\{ \left[\partial_{t}\left(\bm{c}v_{\alpha}^{*}+\boldsymbol{e}_{\parallel}u_{\parallel,\alpha}^{*}\right)+\partial_{x}\left(\bm{c}v_{\alpha}^{*}+\boldsymbol{e}_{\parallel}u_{\parallel,\alpha}^{*}\right)\hat{v}_{\parallel}v_{\alpha}^{*}\right]\tilde{f}_{\alpha}\right\} =0,\label{eq:Vlasov-transformed-energy-final}
\end{multline}
where $\tilde{f}_{\alpha}=f_{\alpha}(v_{\alpha}^{*})^{3}$. In what
follows, we will use the shorthand notation 
\begin{equation}
\left\langle \Phi\left(c_{\parallel},c_{\bot}\right),\tilde{f}_{\alpha}\left(c_{\parallel},c_{\bot}\right)\right\rangle _{\bm{c}}\equiv2\pi\int_{-\infty}^{\infty}dc_{\parallel}\int_{0}^{\infty}c_{\bot}dc_{\bot}\Phi\left(c_{\parallel},c_{\bot}\right)\tilde{f}_{\alpha}\left(c_{\parallel},c_{\bot}\right)\label{eq:v-space-moment}
\end{equation}
to denote the velocity-space moment of a function, $\tilde{f}_{\alpha}\left(c_{\parallel},c_{\bot}\right)$,
with the weight, $\Phi\left(c_{\parallel},c_{\bot}\right)$. 

\subsection{Summary of key continuum symmetries \label{subsec:Continuum-conservation-symmetrie}}

Equations (\ref{eq:Ampere-1D}) and (\ref{eq:Vlasov-transformed-energy-final})
conserve mass, momentum, and energy in the continuum.  However, these
continuum properties are not automatically preserved when the governing
equations are discretized, as we shall see in Sec. \ref{sec:Discrete-conservation-strategy}.
In what follows, we will highlight particular symmetries of the governing
equations that lead to the desired conservation properties. Detailed
proofs of the conservation properties including these symmetries can
be found in \ref{app:Deriviation-of-continuum}. 

\subsubsection{Symmetries relating to the Vlasov equation velocity-space transformation\label{subsec:Symmetries-relating-to-1}}

For the transformed Vlasov equation, we recall that our independent
velocity variables have become $\left(c_{\parallel},c_{\bot}\right)$.
However, momentum and energy conservation are still defined in terms
of $v_{\parallel}$ and $v^{2}$ (the original velocity) moments of
$f_{\alpha}$. The important point is that $\boldsymbol{v}$ moments
do not commute with temporal and spatial derivatives in the transformed
space, e.g., 
\begin{align}
\left\langle m_{\alpha}v_{\parallel},\partial_{t}\tilde{f}_{\alpha}\right\rangle _{\bm{c}} & \neq\left\langle 1,\partial_{t}\left(m_{\alpha}v_{\parallel}\tilde{f}_{\alpha}\right)\right\rangle _{\bm{c},}\label{eq:temporal-gradients}\\
\left\langle m_{\alpha}v_{\parallel},\partial_{x}\left(v_{\parallel}\tilde{f}_{\alpha}\right)\right\rangle _{\bm{c}} & \neq\left\langle 1,\partial_{x}\left(m_{\alpha}v_{\parallel}^{2}\tilde{f}_{\alpha}\right)\right\rangle _{\bm{c}}.\label{eq:spatial-gradients}
\end{align}
Thus, to obtain the momentum conservation theorem for the transformed
system, we must utilize integration by parts and the product rule
on the temporal and spatial components of the Vlasov equation to obtain
the identity:
\begin{multline}
\left\langle m_{\alpha}v_{\parallel},\partial_{t}\tilde{f}_{\alpha}\right\rangle _{\bm{c}}+\left\langle m_{\alpha}v_{\parallel},\partial_{x}\left(v_{\parallel}\tilde{f}_{\alpha}\right)\right\rangle _{\bm{c}}-\frac{1}{v_{\alpha}^{*}}\left\langle m_{\alpha}v_{\parallel},\nabla_{\bm{c}}\cdot\left\{ \left[\partial_{t}\left(\boldsymbol{v}\right)+\partial_{x}\left(\boldsymbol{v}\right)\hat{v}_{\parallel}v_{\alpha}^{*}\right]\tilde{f}_{\alpha}\right\} \right\rangle _{\bm{c}}\\
=\left\langle 1,\partial_{t}\left(m_{\alpha}v_{\parallel}\tilde{f}_{\alpha}\right)\right\rangle _{\bm{c}}+\left\langle 1,\partial_{x}\left(m_{\alpha}v_{\parallel}^{2}\tilde{f}_{\alpha}\right)\right\rangle _{\bm{c}},\label{eq:momentum-inertial_continuum}
\end{multline}
which must be satisfied locally for each species. Likewise, for the
energy conservation theorem we must have the identity:
\begin{multline}
\left\langle m_{\alpha}\frac{1}{2}\boldsymbol{v}^{2},\partial_{t}\tilde{f}_{\alpha}\right\rangle _{\bm{c}}+\left\langle m_{\alpha}\frac{1}{2}\boldsymbol{v}^{2},\partial_{x}\left(v_{\parallel}\tilde{f}_{\alpha}\right)\right\rangle _{\bm{c}}-\frac{1}{v_{\alpha}^{*}}\left\langle m_{\alpha}\frac{1}{2}\boldsymbol{v}^{2},\nabla_{\bm{c}}\cdot\left\{ \left[\partial_{t}\left(\boldsymbol{v}\right)+\partial_{x}\left(\boldsymbol{v}\right)v_{\parallel}\right]\tilde{f}_{\alpha}\right\} \right\rangle _{\bm{c}}\\
=\left\langle 1,\partial_{t}\left(m_{\alpha}\frac{1}{2}\boldsymbol{v}^{2}\tilde{f}_{\alpha}\right)\right\rangle _{\bm{c}}+\left\langle 1,\partial_{x}\left(m_{\alpha}\frac{1}{2}\boldsymbol{v}^{2}v_{\parallel}\tilde{f}_{\alpha}\right)\right\rangle _{\bm{c}}.\label{eq:energy-inertial_continuum}
\end{multline}
For an arbitrary discretization, these identities will generally not
be satisfied simultaneously or even independently.

\subsubsection{Symmetries relating to Ampère's equation\label{subsec:Symmetries-relating-to}}

In addition to the preceding symmetries for the velocity-space adaptivity,
there are several which must be satisfied for the Vlasov-Ampère system
as a whole. The first is the equivalence through charge conservation
between Gauss's law, Ampère's equation, and the continuity equation:
\begin{equation}
\partial_{t}\rho_{q}+\partial_{x}j_{\parallel}=\sum\limits _{\alpha}q_{\alpha}\left[\left\langle 1,\partial_{t}\tilde{f}_{\alpha}\right\rangle _{\bm{c}}+\left\langle 1,\partial_{x}\left(v_{\parallel}\tilde{f}_{\alpha}\right)\right\rangle _{\bm{c}}\right].\label{eq:charge-conservation}
\end{equation}
In Eq. (\ref{eq:charge-conservation}), we defined $\rho_{q}\equiv\sum_{\alpha}q_{\alpha}n_{\alpha}$
and $j_{\parallel}\equiv\sum\limits _{\alpha}q_{\alpha}\Gamma_{\parallel,\alpha}$,
and we further define $\left\langle 1,\tilde{f}_{\alpha}\right\rangle _{\bm{c}}=n_{\alpha}$
and $\left\langle v_{\parallel},\tilde{f}_{\alpha}\right\rangle _{\bm{c}}=\Gamma_{\parallel,\alpha}$.
We see that the charge density in Gauss's law and the current in Ampère's
equation must be proportional to the particle number density and flux
in the continuity equation {[}i.e., the $\boldsymbol{v}^{0}$ moment
of the transformed Vlasov equation, Eq. (\ref{eq:Vlasov-transformed-energy-final}){]}.
As we will see shortly, the crux is that the current in Ampère's equation
(which drives the electric field $E_{\parallel}$) must be consistent
with Eq. (\ref{eq:charge-conservation}), thereby ensuring that $\epsilon_{0}\partial_{x}E_{\parallel}=\rho_{q}$
discretely. This equivalence is in some sense a fundamental ``zeroth-order''
requirement for the Vlasov-Ampère system, and, as we shall see shortly,
neglecting it in the discrete will lead to catastrophic errors \citep{Mardahl1997,Marder1987,Villasenor1992,Chen2019}.

The second requirement is that the $m_{\alpha}v_{\parallel}$ moment
of the acceleration operator in Eq. (\ref{eq:Vlasov-transformed-energy-final})
must produce a number density $n_{\alpha}$ that is identical to the
one in Gauss's law. If we sum the moment of this term over all species
we find 
\begin{equation}
\sum_{\alpha}\frac{q_{\alpha}}{m_{\alpha}v_{\alpha}^{*}}E_{\parallel}\left\langle m_{\alpha}v_{\parallel},\partial_{c_{\parallel}}\tilde{f}_{\alpha}\right\rangle _{\bm{c}}=-E_{\parallel}\sum_{\alpha}q_{\alpha}n_{\alpha}=-\partial_{x}\left(\epsilon_{0}\frac{1}{2}E_{\parallel}^{2}\right).\label{eq:momentum-cons-VA}
\end{equation}
We see that this symmetry introduces the divergence of the electrostatic
energy into the momentum equation, which is key to achieve momentum
conservation. 

To arrive at energy conservation in the Vlasov-Ampère system, we again
inspect the acceleration operator. Taking the $m_{\alpha}\frac{1}{2}\boldsymbol{v}^{2}$
moment of the acceleration operator we find: 
\begin{equation}
\sum_{\alpha}\frac{q_{\alpha}}{m_{\alpha}v_{\alpha}^{*}}E_{\parallel}\left\langle m_{\alpha}\frac{1}{2}\boldsymbol{v}^{2},\partial_{c_{\parallel}}\tilde{f}_{\alpha}\right\rangle _{\bm{c}}=-E_{\parallel}\sum_{\alpha}q_{\alpha}\Gamma_{\parallel,\alpha}=E_{\parallel}\left[\epsilon_{0}\partial_{t}\left(E_{\parallel}\right)-\overline{j}_{\parallel}\right]=\partial_{t}\left(\epsilon_{0}\frac{1}{2}E_{\parallel}^{2}\right)-E_{\parallel}\overline{j}_{\parallel}.\label{eq:energy-cons-VA}
\end{equation}
Thus, we see that this moment must produce a particle flux density
that is consistent with the current that appears in Ampère's equation.
Further, we observe that, while the equivalence
\[
E_{\parallel}\partial_{t}E_{\parallel}=\partial_{t}\left(\frac{1}{2}E_{\parallel}^{2}\right)
\]
is true in the continuum, it will not be so for an arbitrary temporal
discretization. 

\subsection{Strategy for enforcing continuum symmetries in the discrete}

To enforce the preceding continuum symmetries in the discretized Vlasov-Ampère
system, we introduce a set of nonlinear constraint functions to the
discrete representation of Eq. (\ref{eq:Vlasov-transformed-energy-final}).
These added nonlinear constraints take the form of Lagrange-multiplier-like
coefficients and associated operators. For the symmetries related
to the velocity-space adaptivity, we introduce the constraint functions
$\gamma_{t}$ and $\gamma_{x}$, which modify the discretized inertial
terms in a manner similar to that of in Refs. \citep{Taitano2016,Taitano2018b}.
In the discretized form of Eq. (\ref{eq:Vlasov-transformed-energy-final}),
these will be highlighted in red. Following a similar approach to
Refs. \citep{Taitano2015a,Taitano2015b} for the Vlasov-Ampère symmetries,
we introduce the constraint functions $\xi,$ $\phi$, and $\gamma_{q}$,
which appear in their own phase-space advection ``pseudo-operators''.
We will depict these highlighted in blue. These constraint functions
all act to expose the underlying continuum symmetries of the governing
equations and eliminate the truncation errors between different discrete
operators, which would break the symmetries. We note that, while having
five distinct nonlinear constraint functions ($\xi$, $\phi$, $\gamma_{q}$,
$\gamma_{t}$, and $\gamma_{x}$) to enforce only three conservation
laws (charge, momentum, and energy conservation) may seem overconstrained,
this is not the case. The critical distinction is that we are\emph{
}not directly enforcing the \emph{conservation laws} themselves, but
rather the \emph{symmetries} that lead to those laws. Details on the
formulation of these nonlinear constraints are discussed in Secs.
\ref{sec:Numerical-Implementation}\textendash \ref{sec:Discrete-conservation-strategy}.

\section{Numerical implementation\label{sec:Numerical-Implementation}}

\subsection{Discretization of the transformed Vlasov equation}

The Vlasov equation, Eq. (\ref{eq:Vlasov-transformed-energy-final}),
is discretized using finite differences in the transformed phase space
as follows. The discrete cylindrical cell volume in the velocity space
for a uniform velocity mesh is 
\begin{equation}
\tilde{\Omega}_{j,k}\equiv2\pi c_{\bot,k}\Delta c_{\bot}\Delta c_{\parallel},\label{eq:phase-space-volume}
\end{equation}
while the total discrete volume including the configuration space
on a uniform mesh is
\begin{equation}
\Delta\tilde{V}_{i,j,k}=\Delta x\tilde{\Omega}_{j,k}.\label{eq:discrete-cell-volume}
\end{equation}
The quantities $\Delta c_{\parallel}$, $\Delta c_{\bot}$, and $\Delta x$
are the mesh spacings for the parallel velocity, perpendicular velocity,
and configuration space, respectively. The domains are defined to
be
\[
x\in\left[x_{min},x_{max}\right],v_{\parallel}\in\left[v_{\parallel,min},v_{\parallel,max}\right],v_{\bot}\in\left[0,v_{\bot,max}\right],
\]
 such that
\[
L_{x}\equiv x_{max}-x_{min},L_{\parallel}\equiv v_{\parallel,max}-v_{\parallel,min},L_{\bot}\equiv v_{\bot,max}.
\]
Thus, for the transformed velocity space the domain becomes
\[
c_{\parallel}\in\left[c_{\parallel,min},c_{\parallel,max}\right],c_{\bot}\in\left[0,c_{\bot,max}\right],
\]
with 
\[
\tilde{L}_{\parallel}=\frac{L_{\parallel}}{v_{\alpha}^{*}}=c_{\parallel,max}-c_{\parallel,min},\tilde{L}_{\bot}=\frac{L_{\bot}}{v_{\alpha}^{*}}=c_{\bot,max}.
\]
The mesh spacings $\Delta x,\Delta c_{\parallel},\Delta c_{\bot}$
are given by
\[
\Delta x=\frac{L_{x}}{N_{x}},\Delta c_{\parallel}=\frac{\tilde{L}_{\parallel}}{N_{\parallel}},\Delta c_{\bot}=\frac{\tilde{L}_{\bot}}{N_{\bot}}.
\]
Here, $N_{x}$, $N_{\parallel}$, and $N_{\bot}$are the numbers of
cells along each coordinate. The coordinates $i,j,k$ are defined
to be at the cell centers, so that the boundary of the domain in each
direction is on cell faces. Thus, a cell-center quantity $\Phi_{j}$
spans $j\in\left[1,N_{\parallel}\right]$, while a cell-face quantity
$\Phi_{j+\frac{1}{2}}$ spans $j\in\left[0,N_{\parallel}\right]$.
Here, we reiterate that the transformed velocity-space domain $c_{\parallel}\in\left[c_{\parallel,min},c_{\parallel,max}\right],c_{\bot}\in\left[0,c_{\bot,max}\right]$
(the computational velocity-space domain) is the same for all species,
and is constant in space and time. The spatio-temporal variations
in bulk velocity and thermal speed (temperature) between species are
dealt with through the reference speed $v_{\alpha}^{*}$ and offset
velocity $u_{\parallel,\alpha}^{*}$.

Discrete moments in the velocity space are computed via a midpoint
quadrature as
\begin{equation}
\left\langle A,B\right\rangle _{\delta\bm{c}}\approx\sum\limits _{j=1}^{N_{\parallel}}\sum\limits _{k=1}^{N_{\bot}}\tilde{\Omega}_{j,k}A_{j,k}B_{j,k}\label{eq:discrete-moment-scalar}
\end{equation}
for scalars defined at the cell centers, and as 
\begin{equation}
\left\langle 1,\bm{A}\cdot\bm{B}\right\rangle _{\delta\bm{c}}\approx\sum\limits _{j=0}^{N_{\parallel}}\sum\limits _{k=1}^{N_{\bot}}\tilde{\Omega}_{j+\frac{1}{2},k}A_{\parallel,j+\frac{1}{2},k}B_{\parallel,j+\frac{1}{2},k}+\sum\limits _{j=1}^{N_{\parallel}}\sum\limits _{k=0}^{N_{\bot}}\tilde{\Omega}_{j,k+\frac{1}{2}}A_{\bot,j,k+\frac{1}{2}}B_{\bot,j,k+\frac{1}{2}}\label{eq:discrete-moment-innerproduct}
\end{equation}
for scalar products of velocity-space vectors defined at the cell
faces. Quantities at half-indices (e.g., $j+\frac{1}{2}$) are at
the cell faces.

A discretization of the Vlasov equation, Eq. (\ref{eq:Vlasov-transformed-energy-final}),
which includes all the relevant `pseudo-operators' and nonlinear constraint
functions to enforce discrete conservation, is then given by
\begin{multline}
\delta_{t}\tilde{f}_{\alpha,i,j,k}^{p+1}+\underbrace{\delta_{x}\left[v_{\parallel,\alpha,j}^{p}\overline{\left(\tilde{f}_{\alpha}^{p+1}\right)}_{j,k}^{v_{\parallel}}\right]_{i}}_{(a)}+\underbrace{\frac{q_{\alpha}}{m_{\alpha}}\frac{E_{\parallel,i}^{p+1}}{v_{\alpha,i}^{*,p}}\delta_{c_{\parallel}}\left[\overline{\left(\tilde{f}_{\alpha}^{p+1}\right)}_{i,k}^{q_{\alpha}E_{\parallel}}\right]_{j}}_{(b)}\\
+\underbrace{\tcboxmath[colback=blue!10!white,colframe=blue]{\delta_{x}\left[\xi_{\alpha}^{p+1}\left|v_{\parallel,\alpha,j}^{p}\right|\overline{\left(\tilde{f}_{\alpha}^{p+1}\right)}_{j,k}^{\xi}\right]_{i}}}_{(c)}+\underbrace{\tcboxmath[colback=blue!10!white,colframe=blue]{\delta_{c_{\parallel}}\left[\phi_{\alpha,i}^{p+1}\overline{\left(\tilde{f}_{\alpha}^{p+1}\right)}_{i,k}^{\phi}\right]_{j}}}_{(d)}+\underbrace{\tcboxmath[colback=blue!10!white,colframe=blue]{\delta_{c_{\parallel}}\left[\gamma_{q,\alpha,i}^{p+1}\overline{\left(\tilde{f}_{\alpha}^{p+1}\right)}_{i,k}^{\gamma_{q}}\right]_{j}}}_{(e)}\\
\underbrace{-\frac{1}{v_{\alpha,i}^{*,p}}\delta_{\bm{c}}\cdot\left[\tcboxmath[colback=red!10!white,colframe=red]{\gamma_{t,\alpha,i}^{p+1}}\delta_{t}\left(\boldsymbol{v}_{\alpha,i}\right)^{p+1}\overline{\left(\tilde{f}_{\alpha}^{p+1}\right)}_{i}^{\delta_{t}\left(\boldsymbol{v}\right)}\right]_{j,k}}_{(f)}\\
\underbrace{-\frac{1}{2v_{\alpha,i}^{*,p}}\delta_{\bm{c}}\cdot\left[\tcboxmath[colback=red!10!white,colframe=red]{\gamma_{x,\alpha,i+\frac{1}{2}}^{p+1}}v_{\alpha,i+\frac{1}{2}}^{*,p}\tcboxmath[colback=blue!10!white,colframe=blue]{\hat{v}_{\parallel,\alpha,\mathrm{eff},i+\frac{1}{2}}^{p+1}}\delta_{x}\left[\boldsymbol{v}_{\alpha}^{p}\right]_{i+\frac{1}{2}}\overline{\left(\tilde{f}_{\alpha}^{p+1}\right)}_{i}^{v_{\parallel,\mathrm{eff}}\delta_{x}\left(\boldsymbol{v}\right)}\right]_{j,k}}_{(g,1)}\\
\underbrace{-\frac{1}{2v_{\alpha,i}^{*,p}}\delta_{\bm{c}}\cdot\left[\tcboxmath[colback=red!10!white,colframe=red]{\gamma_{x,\alpha,i-\frac{1}{2}}^{p+1}}v_{\alpha,i-\frac{1}{2}}^{*,p}\tcboxmath[colback=blue!10!white,colframe=blue]{\hat{v}_{\parallel,\alpha,\mathrm{eff},i-\frac{1}{2}}^{p+1}}\delta_{x}\left[\boldsymbol{v}_{\alpha}^{p}\right]_{i-\frac{1}{2}}\overline{\left(\tilde{f}_{\alpha}^{p+1}\right)}_{i}^{v_{\parallel,\mathrm{eff}}\delta_{x}\left(\boldsymbol{v}\right)}\right]_{j,k}}_{(g,2)}=0.\label{eq:Vlasov-disc-simp}
\end{multline}
Here, we define the following notation
\begin{align}
\delta_{t}\left(F\right)^{p+1} & \equiv\frac{b^{p+1}F^{p+1}+b^{p}F^{p}+b^{p-1}F^{p-1}}{\Delta t^{p}},\label{eq:partial_t_discrete}\\
\delta_{x}F_{i} & \equiv\frac{F_{i+\frac{1}{2}}-F_{i-\frac{1}{2}}}{\Delta x},\label{eq:partial_x_discrete}\\
\delta_{c_{\parallel}}F_{\parallel,j,k} & \equiv\frac{F_{\parallel,j+\frac{1}{2},k}-F_{\parallel,j-\frac{1}{2},k}}{\Delta c_{\parallel}},\label{eq:partial_v_parallel}\\
\delta_{c_{\bot}}F_{\bot,j,k} & \equiv\frac{c_{\bot,k+\frac{1}{2}}F_{\bot,j,k+\frac{1}{2}}-c_{\bot,k-\frac{1}{2}}F_{\bot,j,k-\frac{1}{2}}}{c_{\bot,k}\Delta c_{\bot}},\label{eq:partial_v_perp}\\
\delta_{\bm{c}}\cdot\boldsymbol{\left[F\right]}_{j,k} & \equiv\delta_{c_{\parallel}}F_{\parallel,j,k}+\delta_{c_{\bot}}F_{\bot,j,k},\label{eq:partial_v_discrete}
\end{align}

\begin{equation}
\hat{v}_{\parallel,\alpha,\mathrm{eff},i+\frac{1}{2},j}^{p+1}\equiv\left(c_{\parallel,j}+\hat{u}_{\parallel,\alpha,i+\frac{1}{2}}^{*,p}+\tcboxmath[colback=blue!10!white,colframe=blue]{\xi_{\alpha,i+\frac{1}{2}}^{p+1}}\left|c_{\parallel,j}+\hat{u}_{\parallel,\alpha,i+\frac{1}{2}}^{*,p}\right|\right),\label{eq:v_effective}
\end{equation}
with $v_{\alpha,i+\frac{1}{2}}^{*,p}=\frac{v_{\alpha,i}^{*,p}+v_{\alpha,i+1}^{*,p}}{2}$,
$\hat{u}_{\parallel,\alpha,i+\frac{1}{2}}^{*,p}=\frac{\hat{u}_{\parallel,\alpha,i}^{*,p}+\hat{u}_{\parallel,\alpha,i+1}^{*,p}}{2}$,
and $\bm{v}_{\alpha,i,j,k}^{p}=v_{\alpha,i}^{*,p}\bm{c}_{j,k}+u_{\parallel,\alpha,i}^{*,p}\bm{e}_{\parallel}$.
Note that the nonlinear constraint function $\xi$ is included in
the definition of $\hat{v}_{\parallel,\alpha,\mathrm{eff},i-\frac{1}{2}}^{p+1}$
(boxed in blue). In Eq. (\ref{eq:Vlasov-disc-simp}), we have utilized
a second-order backwards finite difference scheme in time (BDF2, \citep[Chapter III]{Hairer2008}),
with coefficients $b^{p+1}=1.5,$ $b^{p}=-2,$ and $b^{p-1}=0.5$.
The temporal index is $p$. Here, as in Refs. \citep{Taitano2016,Taitano2018a},
we lag the time-level of the reference speed $v_{\alpha,i}^{*,p}$
and offset velocity $\hat{u}_{\parallel,\alpha,i}^{*,p}$ for robustness;
thus, for discrete temporal derivatives involving these quantities,
we have
\begin{align}
\delta_{t}\left(v_{\alpha}^{*}F\right)^{p+1} & =\frac{b^{p+1}v_{\alpha}^{*,p}F^{p+1}+b^{p}v_{\alpha}^{*,p-1}F^{p}+b^{p-1}v_{\alpha}^{*,p-2}F^{p-1}}{\Delta t^{p}},\label{eq:partial_t_v_alpha}\\
\delta_{t}\left(\hat{u}_{\parallel,\alpha}^{*}F\right)^{p+1} & =\frac{b^{p+1}\hat{u}_{\parallel,\alpha}^{*,p}F^{p+1}+b^{p}\hat{u}_{\parallel,\alpha}^{*,p-1}F^{p}+b^{p-1}\hat{u}_{\parallel,\alpha}^{*,p-2}F^{p-1}}{\Delta t^{p}}.\label{eq:partial_t_u_alpha}
\end{align}

For compactness of notation, we define an advective interpolation
operator acting on a scalar $\phi$ at a cell face based on an advection
coefficient $a$: 
\begin{equation}
\overline{\left(\Phi\right)}_{\mathrm{face}}^{a}=\sum\limits _{i'=1}^{N}\omega_{\mathrm{face},i'}(a,\phi)\phi_{i'}.\label{eq:interpolation-discrete}
\end{equation}
In Eq. (\ref{eq:interpolation-discrete}), $\omega_{\mathrm{face},i'}$
are interpolation weights for the $i'$ cells surrounding the cell
face, and $\phi_{i'}$ are the values of the interpolated quantity
in those cells. We note here that the cell-center electric field is
defined as the interpolation of adjacent cell-face electric fields,
\[
E_{\parallel,i}^{p+1}=\frac{E_{\parallel,i+\frac{1}{2}}^{p+1}+E_{\parallel,i-\frac{1}{2}}^{p+1}}{2}.
\]
In Eq. \ref{eq:Vlasov-disc-simp}, terms $(a)$ and $(b)$ represent
the discrete form of the physical configuration-space advection and
velocity-space advection due to the acceleration of the electric field,
respectively. Terms $(c)$, $(d)$, and $(e)$ (boxed in blue) are
the discretized forms of the `pseudo-operators' arising from the nonlinear
constraint coefficients, $\xi$, $\phi$, and $\gamma_{q}$. These
constraint functions are responsible for enforcing the symmetries
discussed in Sec. \ref{subsec:Symmetries-relating-to} relating to
the Vlasov-Ampère coupling in the discretized system. Terms $(f)$
and $(g)$ are the discretized versions of the inertial terms arising
from the respective temporal and spatial gradients in $v_{\alpha}^{*}$
and $\hat{u}_{\parallel,\alpha}^{*}$, which also include two nonlinear
constraint functions, $\gamma_{t}$ and $\gamma_{x}$, (boxed in red).
The constraint functions $\gamma_{t}$ and $\gamma_{x}$ are responsible
for enforcing the conservation symmetries in Sec. \ref{subsec:Symmetries-relating-to-1}
relating to the velocity-space transformation. Note in the velocity-space
inertial terms pertaining to $\gamma_{x}$, the term $\hat{v}_{\parallel,\alpha,\mathrm{eff},i-\frac{1}{2}}^{p+1}$
is boxed in blue to indicate that it also contains the constraint
function $\xi$. A discussion of the function and definition of the
constraint functions and their `pseudo-operators' is presented in
Sec. \ref{sec:Discrete-conservation-strategy}, with more detailed
derivations left to \ref{app:Details-on-the} and \ref{app:Derivation-of-constraint}.

In the present study, several different advective schemes have been
utilized for different terms, in accordance with an empirical hierarchy
of priority based on observed sensitivity and behavior of different
terms. In general, we have observed that the electrons are highly
sensitive to numerical dissipation, particularly with schemes (such
as SMART \citep{Gaskell1988}) that switch between low-order (upwinding)
and higher-order schemes (e.g., QUICK \citep{Leonard1979}). Thus,
for the electron physical configuration space advection {[}term $\left(a\right)${]},
we have chosen a relatively low-dissipation 5th-order WENO scheme
(WENO5 \citep{Jiang1996}). While it does not possess the positivity-
and monotonicity-preserving properties of schemes such as SMART, WENO5
is more robust than a central differencing scheme. The electron physical
velocity space advection term {[}term $\left(b\right)${]} is more
sensitive still to numerical dissipation, and so, while it is less
robust overall than WENO5, a central-differencing scheme is used.
The ions are not as sensitive to dissipation, so the SMART scheme
is used for the configuration-space advection. This is because of
its monotonicity- and positivity-preserving properties, as well as
for being well-posed for nonlinear iterative methods. It is also cheaper
to evaluate than the WENO5 scheme. In the ion electrostatic acceleration
operator, we use WENO5 for increased robustness relative to a central-differencing
scheme. For the velocity-space adaptivity inertial terms $\left(f\right)$
and $\left(g\right)$, we use WENO5 for all species for low dissipation
and greater robustness over central differencing. The choice of discretization
for the `pseudo-operators' $(c)$, $(d)$, and $(e)$ is generally
much less restrictive \textendash{} as we shall see, these terms do
not affect the order of accuracy of the scheme. Thus, for the charge-conservation
pseudo-operator $\left(c\right)$, we use a straightforward upwind
discretization for simplicity and robustness. For the momentum- and
energy-conservation pseudo-operators $\left(d\right)$ and $\left(e\right)$,
we use central differencing primarily for simplicity and to minimize
dissipation in velocity space.

\subsection{Discretization of Ampère's equation, Eq. (\ref{eq:Ampere-1D})}

We follow the approach of Ref. \citep{Taitano2015a} and define the
electric field $E_{\parallel,i+\frac{1}{2}}^{p}$ at cell-faces. The
discrete Ampère equation for the cell-face electric field is thus
\begin{equation}
\epsilon_{0}\delta_{t}E_{\parallel,i+\frac{1}{2}}^{p+1}+\sum\limits _{\alpha}^{N_{sp}}q_{\alpha}\widehat{\Gamma}_{\parallel,\alpha,i+\frac{1}{2}}^{p+1}=\overline{j}_{\parallel}^{p+1}.\label{eq:ampere-discrete}
\end{equation}
 The cell-face particle flux density must be defined to preserve energy
conservation. Thus, instead of being defined based on the ``true''
momentum moment of $\tilde{f}_{\alpha}$,
\begin{equation}
\Gamma_{\parallel,\alpha,i}^{p+1}=\left\langle v_{\parallel},\tilde{f}_{\alpha,i,j,k}^{p+1}\right\rangle _{\delta\bm{c}},\label{eq:momentum_definition}
\end{equation}
$\widehat{\Gamma}_{\parallel,\alpha,i+\frac{1}{2}}^{p+1}$ is defined
from the $\frac{1}{2}m_{\alpha}\boldsymbol{v}^{2}$ moment of the
electrostatic acceleration operator (see the discussion of Eq. (\ref{eq:energy-cons-VA})
in Sec. \ref{subsec:Continuum-conservation-symmetrie}): 
\begin{equation}
\widehat{\Gamma}_{\parallel,\alpha,i}^{p+1}\equiv-\frac{1}{m_{\alpha}}\left\langle \frac{m_{\alpha}\left(\boldsymbol{v}_{\alpha,i,j,k}^{p}\right)^{2}}{2v_{\alpha,i}^{*,p}},\frac{\overline{\left(\tilde{f}_{\alpha}^{p+1}\right)}_{i,j+\frac{1}{2},k}^{q_{\alpha}E_{\parallel}}-\overline{\left(\tilde{f}_{\alpha}^{p+1}\right)}_{i,j-\frac{1}{2},k}^{q_{\alpha}E_{\parallel}}}{\Delta c_{\parallel}}\right\rangle _{\delta\bm{c}},\label{eq:nu-bar}
\end{equation}
\begin{equation}
\widehat{\Gamma}_{\parallel,\alpha,i+\frac{1}{2}}^{p+1}\equiv\frac{\widehat{\Gamma}_{\parallel,\alpha,i}^{p+1}+\widehat{\Gamma}_{\parallel,\alpha,i+1}^{p+1}}{2}.\label{eq:nu-cf-accel}
\end{equation}
We note that, while these momenta are equivalent in the continuum,
choices of discretization and interpolation for the physical acceleration
operator mean this will generally not be so in the discrete. The average
current density $\overline{j}_{\parallel}^{p+1}$ must be based on
this same particle flux density ($\widehat{\Gamma}_{\parallel,\alpha,i+\frac{1}{2}}^{p+1}$)
and is calculated as
\begin{equation}
\overline{j}_{\parallel}^{p+1}\equiv\frac{1}{N_{x}}\sum\limits _{i}^{N_{x}}\left(\sum\limits _{\alpha}^{N_{sp}}q_{\alpha}\widehat{\Gamma}_{\parallel,\alpha,i+\frac{1}{2}}^{p+1}\right).\label{eq:avg-j-discrete-HO}
\end{equation}

\section{Discrete conservation strategy for charge, momentum, and energy\label{sec:Discrete-conservation-strategy}}

As we saw in Sec. \ref{subsec:Continuum-conservation-symmetrie},
there are certain symmetries of the continuum equations that must
be satisfied in order to conserve charge, momentum, and energy. For
an arbitrary discretization, these symmetries will pose conflicting
constraints. As a result, it will not generally be possible to satisfy
\emph{all }of them simultaneously unless we design our discretization
such that it includes elements that ensure these properties. In the
following development, we will present the discrete definitions of
the constraints $\left(\xi,\alpha,\gamma_{q},\gamma_{t},\gamma_{x}\right)$
that will enforce the symmetries discussed in Sec. \ref{subsec:Continuum-conservation-symmetrie}
in the discrete system. For detailed derivations of these constraints,
interested readers are referred to \ref{app:Details-on-the}, \ref{app:Derivation-of-constraint},
and Ref. \citep{Taitano2018a}.

\subsection{Discrete definition of $\gamma_{t}$ and $\gamma_{x}$}

In Sec. \ref{subsec:Continuum-conservation-symmetrie}, we saw that
there are certain continuum identities {[}Eqs. (\ref{eq:momentum-inertial_continuum})
and (\ref{eq:energy-inertial_continuum}){]} that must be satisfied
regarding the inertial terms coming from the velocity-space transformation.
In the discrete, these are used to define the nonlinear constraint
functions $\gamma_{t}$ and $\gamma_{x}$. From Eq. (\ref{eq:momentum-inertial_continuum}),
to obtain discrete momentum conservation, $\gamma_{t}$ must satisfy

\begin{multline}
\left\langle v_{\parallel,\alpha,i,j}^{p},\delta_{t}\tilde{f}_{\alpha,i,j,k}^{p+1}\right\rangle _{\delta\bm{c}}-\left\langle 1,\delta_{t}\left(v_{\parallel,\alpha,i,j}\tilde{f}_{\alpha,i,j,k}\right)^{p+1}\right\rangle _{\delta\bm{c}}\\
-\left\langle v_{\parallel,\alpha,i,j}^{p},\frac{1}{v_{\alpha,i}^{*,p}}\delta_{\bm{c}}\cdot\left[\gamma_{t,\alpha,i}^{p+1}\delta_{t}\left(\boldsymbol{v}_{\alpha,i}\right)^{p+1}\overline{\left(\tilde{f}_{\alpha}^{p+1}\right)}_{i}^{\delta_{t}\left(\boldsymbol{v}\right)}\right]_{j,k}\right\rangle _{\delta\bm{c}}=0,\label{eq:gamma_t-eq1}
\end{multline}
while $\gamma_{x}$ must satisfy 
\begin{multline}
\left\langle v_{\parallel,\alpha,i,j}^{p}-v_{\parallel,\alpha,i+1,j}^{p},\frac{1}{\Delta x}v_{\alpha,i+\frac{1}{2}}^{*,p}\hat{v}_{\parallel,\mathrm{eff},\alpha,i+\frac{1}{2},j}^{p+1}\overline{\left(\tilde{f}_{\alpha}^{p+1}\right)}_{i+\frac{1}{2},j,k}^{\hat{v}_{\parallel,\mathrm{eff}}}\right\rangle _{\delta\bm{c}}\\
-\left\langle v_{\parallel,\alpha,i,j}^{p},\frac{1}{2v_{\alpha,i}^{*,p}}\delta_{\bm{c}}\cdot\left[\gamma_{x,\alpha,i+\frac{1}{2}}^{p+1}v_{\alpha,i+\frac{1}{2}}^{*,p}\hat{v}_{\parallel,\mathrm{eff},i+\frac{1}{2}}\delta_{x}\left(\boldsymbol{v}^{p}\right)_{i+\frac{1}{2}}\overline{\left(\tilde{f}_{\alpha}^{p+1}\right)}_{i}^{v_{\parallel,\mathrm{eff}}\delta_{x}\left(\boldsymbol{v}\right)}\right]_{j,k}\right\rangle _{\delta\bm{c}}\\
-\left\langle v_{\parallel,\alpha,i+1,j}^{p},\frac{1}{2v_{\alpha,i+1}^{*,p}}\delta_{\bm{c}}\cdot\left[\gamma_{x,\alpha,i+\frac{1}{2}}^{p+1}v_{\alpha,i+\frac{1}{2}}^{*,p}\hat{v}_{\parallel,\mathrm{eff},i+\frac{1}{2}}\delta_{x}\left(\boldsymbol{v}^{p}\right)_{i+\frac{1}{2}}\overline{\left(\tilde{f}_{\alpha}^{p+1}\right)}_{i+1}^{v_{\parallel,\mathrm{eff}}\delta_{x}\left(\boldsymbol{v}\right)}\right]_{j,k}\right\rangle _{\delta\bm{c}}=0.\label{eq:gamma_x-eq1}
\end{multline}
To obtain discrete energy conservation, according to Eq. (\ref{eq:energy-inertial_continuum})
$\gamma_{t}$ must satisfy
\begin{multline}
\left\langle m_{\alpha}\frac{1}{2}\left(\boldsymbol{v}_{\alpha,i,j,k}^{p}\right)^{2},\delta_{t}\tilde{f}_{\alpha,i,j,k}^{p+1}\right\rangle _{\delta\bm{c}}-\left\langle 1,\delta_{t}\left(\frac{1}{2}\left(\boldsymbol{v}_{\alpha,i,j,k}\right)^{2}\tilde{f}_{\alpha,i,j,k}\right)^{p+1}\right\rangle _{\delta\bm{c}}\\
-\left\langle m_{\alpha}\frac{1}{2}\left(\boldsymbol{v}_{\alpha,i,j,k}^{p}\right)^{2},\frac{1}{v_{\alpha,i}^{*,p}}\delta_{\bm{c}}\cdot\left[\gamma_{t,\alpha,i}^{p+1}\delta_{t}\left(\boldsymbol{v}_{\alpha,i}\right)^{p+1}\overline{\left(\tilde{f}_{\alpha}^{p+1}\right)}_{i}^{\delta_{t}\left(\boldsymbol{v}\right)}\right]_{j,k}\right\rangle _{\delta\bm{c}}=0,\label{eq:gamma_t-eq2}
\end{multline}
while $\gamma_{x}$ must satisfy 

\begin{multline}
\left\langle \frac{\left(\boldsymbol{v}_{\alpha,i,j,k}^{p}\right)^{2}}{2}-\frac{\left(\boldsymbol{v}_{\alpha,i+1,j,k}^{p}\right)^{2}}{2},\frac{1}{\Delta x}v_{\alpha,i+\frac{1}{2}}^{*,p}\hat{v}_{\parallel,\mathrm{eff},\alpha,i+\frac{1}{2},j}^{p+1}\overline{\left(\tilde{f}_{\alpha}^{p+1}\right)}_{i+\frac{1}{2},j,k}^{\hat{v}_{\parallel,\mathrm{eff}}}\right\rangle _{\delta\bm{c}}\\
-\left\langle \frac{\left(\boldsymbol{v}_{\alpha,i,j,k}^{p}\right)^{2}}{2},\frac{1}{2v_{\alpha,i}^{*,p}}\delta_{\bm{c}}\cdot\left[\gamma_{x,\alpha,i+\frac{1}{2}}^{p+1}v_{\alpha,i+\frac{1}{2}}^{*,p}\hat{v}_{\parallel,\mathrm{eff},i+\frac{1}{2}}\delta_{x}\left(\boldsymbol{v}^{p}\right)_{i+\frac{1}{2}}\overline{\left(\tilde{f}_{\alpha}^{p+1}\right)}_{i}^{v_{\parallel,\mathrm{eff}}\delta_{x}\left(\boldsymbol{v}\right)}\right]_{j,k}\right\rangle _{\delta\bm{c}}\\
-\left\langle \frac{\left(\boldsymbol{v}_{\alpha,i+1,j,k}^{p}\right)^{2}}{2},\frac{1}{2v_{\alpha,i+1}^{*,p}}\delta_{\bm{c}}\cdot\left[\gamma_{x,\alpha,i+\frac{1}{2}}^{p+1}v_{\alpha,i+\frac{1}{2}}^{*,p}\hat{v}_{\parallel,\mathrm{eff},i+\frac{1}{2}}\delta_{x}\left(\boldsymbol{v}^{p}\right)_{i+\frac{1}{2}}\overline{\left(\tilde{f}_{\alpha}^{p+1}\right)}_{i+1}^{v_{\parallel,\mathrm{eff}}\delta_{x}\left(\boldsymbol{v}\right)}\right]_{j,k}\right\rangle _{\delta\bm{c}}=0.\label{eq:gamma_x-eq2}
\end{multline}
The constraint functions $\gamma_{t}$ and $\gamma_{x}$ are expanded
in velocity-space basis functions, e.g.,
\[
\gamma_{t}=1+\sum_{r=0}^{P_{\parallel}}\sum_{s=0}^{P_{\bot}}C_{r,s}B_{\parallel,r}\left(c_{\parallel}\right)B_{\bot,s}\left(c_{\bot}\right),
\]
where the $B_{\parallel,r}\left(c_{\parallel}\right),B_{\bot,s}\left(c_{\bot}\right)$
are the $r^{th}$ and $s^{th}$ velocity-space functions in $c_{\parallel}$
and $c_{\bot}$, respectively, in some chosen basis (in this work,
we use a Fourier representation as in \citep{Taitano2018a}). The
$C_{r,s}$ are the corresponding coefficient weights, which are obtained
via the solution of a constrained-minimization problem using Eqs.
(\ref{eq:momentum-inertial_continuum})\textendash (\ref{eq:energy-inertial_continuum}).
More details on the approach can be found in Ref. \citep{Taitano2018a}.
We note here that the solution is relatively inexpensive, involving
the computation of the discrete moments in Eqs. (\ref{eq:momentum-inertial_continuum})\textendash (\ref{eq:energy-inertial_continuum}),
and the solution of a straightforward linear system of the size $\mathcal{O}\left(P_{\parallel}P_{\bot}\right)\sim\mathcal{O}\left(10\right)-\mathcal{O}\left(100\right)$.
Note that $\gamma_{x}$ depends on $\xi$ through $\hat{v}_{\parallel,\alpha,\mathrm{eff}}$,
and so must be calculated after $\xi$ is obtained. Together, these
constraint functions ensure that the integration by parts and product
rule that produce Eqs. (\ref{eq:momentum-inertial_continuum})\textendash (\ref{eq:energy-inertial_continuum})
are upheld discretely.

\subsection{Discrete definition of $\xi$, $\phi$, and $\gamma_{q}$}

The constraint function $\xi$ is defined by 
\begin{equation}
\xi_{\alpha,i+\frac{1}{2}}^{p+1}=\frac{\widehat{\Gamma}_{\parallel,\alpha,i+\frac{1}{2}}^{p+1}-\widetilde{\Gamma}_{\parallel,\alpha,i+\frac{1}{2}}^{p+1}}{\Pi_{\xi,\parallel,\alpha,i+\frac{1}{2}}^{p+1}}.\label{eq:xi-definition}
\end{equation}
The quantity $\widehat{\Gamma}_{\parallel,\alpha,i+\frac{1}{2}}^{p+1}$
is defined in Eqs. (\ref{eq:nu-bar})-(\ref{eq:nu-cf-accel}), and
$\widetilde{\Gamma}_{\parallel,\alpha,i+\frac{1}{2}}^{p+1}$ and $\Pi_{\xi,\parallel,\alpha,i+\frac{1}{2}}^{p+1}$
are given by:

\begin{align}
\widetilde{\Gamma}_{\parallel,\alpha,i+\frac{1}{2}}^{p+1} & \equiv\frac{1}{m_{\alpha}}\left\langle m_{\alpha},v_{\alpha,i+\frac{1}{2}}^{*,p}\left(c_{\parallel,j}+\hat{u}_{\parallel,\alpha,i+\frac{1}{2}}^{*,p}\right)\overline{\left(\tilde{f}_{\alpha}^{p+1}\right)}_{i+\frac{1}{2},j,k}^{v_{\parallel,i+\frac{1}{2},j}^{p}}\right\rangle _{\delta\bm{c}},\label{eq:nu-tilde}\\
\Pi_{\xi,\parallel,\alpha,i+\frac{1}{2}}^{p+1} & \equiv\frac{1}{m_{\alpha}}\left\langle m_{\alpha},v_{\alpha,i+\frac{1}{2}}^{*,p}\left|c_{\parallel,j}+\hat{u}_{\parallel,\alpha,i+\frac{1}{2}}^{*,p}\right|\overline{\left(\tilde{f}_{\alpha}^{p+1}\right)}_{i+\frac{1}{2},j,k}^{\xi_{\alpha,i+\frac{1}{2}}^{p+1}}\right\rangle _{\delta\bm{c}},\label{eq:gamma-xi}
\end{align}
where the integrals in the scalar product are obtained from the fluxes
in terms $\left(a\right)$ and $\left(c\right)$ in Eq. (\ref{eq:Vlasov-disc-simp}).
The action of the constraint function $\xi$ is to ensure that Gauss's
law is upheld {[}see Eq. (\ref{eq:charge-conservation}){]}, and is
generally described as the ``charge-conserving'' constraint.

The constraint functions $\phi$ and $\gamma_{q}$ are split in velocity
space using the following convention:
\begin{align}
\phi_{\alpha,i,j+\frac{1}{2}}^{p+1} & =\begin{cases}
\phi_{\alpha,i}^{+,p+1} & \text{if }v_{\parallel,\alpha,i,j+\frac{1}{2}}\geq u_{\parallel,\alpha,i}^{p}\\
1 & \text{otherwise }
\end{cases},\nonumber \\
\gamma_{q,\alpha,i,j+\frac{1}{2}}^{p+1} & =\begin{cases}
1 & \text{if }v_{\parallel,\alpha,i,j+\frac{1}{2}}\geq u_{\parallel,\alpha,i}^{p}\\
\gamma_{q,\alpha,i}^{-,p+1} & \text{otherwise}
\end{cases},\label{eq:phi-gammaq-split-1}
\end{align}
where 
\begin{equation}
u_{\parallel,\alpha,i}^{p}=\frac{\left\langle v_{\parallel},\tilde{f}_{\alpha,i,j,k}^{p+1}\right\rangle _{\delta\bm{c}}}{\left\langle 1,\tilde{f}_{\alpha,i,j,k}^{p+1}\right\rangle _{\delta\bm{c}}}\label{eq:bulk-velocity}
\end{equation}
is the bulk velocity of species $\alpha$ from the previous timestep.
The quantities $\phi_{\alpha,i}^{+,p+1}$ and $\gamma_{q,\alpha,i}^{-,p+1}$
are thus coupled by the $2\times2$ linear system
\begin{equation}
\left[\begin{array}{cc}
n_{\alpha,i}^{+,p+1} & n_{\alpha,i}^{-,p+1}\\
\Gamma_{\parallel,\alpha,i}^{+,p+1} & \Gamma_{\parallel,\alpha,i}^{-,p+1}
\end{array}\right]\left[\begin{array}{c}
\phi_{\alpha,i}^{+,p+1}+1\\
\gamma_{q,\alpha,i}^{-,p+1}+1
\end{array}\right]=\left[\begin{array}{c}
\left(n_{\alpha,i}^{p+1}-\widehat{n}_{\alpha,i}^{p+1}\right)E_{\parallel,i}^{p+1}\frac{q_{\alpha}}{m_{\alpha}}\\
\frac{\epsilon_{0}}{m_{\alpha}N_{sp}}\left[E_{\parallel,i+\frac{1}{2}}^{p+1}\delta_{t}E_{\parallel,i+\frac{1}{2}}^{p+1}-\delta_{t}\left(\frac{1}{2}E_{\parallel,i+\frac{1}{2}}^{2}\right)^{p+1}\right]
\end{array}\right],\label{eq:phi-alpha-def}
\end{equation}
which is well-posed (see \ref{app:Well-posedness-of-}) and may be
easily inverted analytically to calculate $\phi$ and $\gamma_{q}$.
In the preceding, we defined the discrete number densities
\begin{equation}
n_{\alpha,i}^{p+1}\equiv\left\langle 1,\tilde{f}_{\alpha,i,j,k}^{p+1}\right\rangle _{\delta\bm{c}},\label{eq:n-def}
\end{equation}
\begin{equation}
\widehat{n}_{\alpha,i}^{p+1}\equiv-\left\langle v_{\parallel,\alpha,i,j}^{p},\frac{\overline{\left(\tilde{f}_{\alpha}^{p+1}\right)}_{i,j+\frac{1}{2},k}^{q_{\alpha}E_{\parallel}}-\overline{\left(\tilde{f}_{\alpha}^{p+1}\right)}_{i,j-\frac{1}{2},k}^{q_{\alpha}E_{\parallel}}}{v_{\alpha,i}^{*,p}\Delta c_{\parallel}}\right\rangle _{\delta\bm{c}},\label{eq:n-bar}
\end{equation}
which come from the direct $v_{\parallel}^{0}$ moment of $\tilde{f}_{\alpha}$,
and the $v_{\parallel}$ moment of the electrostatic acceleration
operator, respectively. We also defined the ``upper'' and ``lower''
densities 
\begin{align}
n_{\alpha,i}^{+,p+1} & \equiv-\left\langle v_{\parallel,\alpha,i,j}^{p},\frac{1}{v_{\alpha,i}^{*,p}}\frac{\overline{\left(\tilde{f}_{\alpha}^{p+1}\right)}_{i,j+\frac{1}{2},k}^{\mathrm{central}}-\overline{\left(\tilde{f}_{\alpha}^{p+1}\right)}_{i,j-\frac{1}{2},k}^{\mathrm{central}}}{\Delta c_{\parallel}}\right\rangle _{\delta\bm{c}\text{ where }v_{\parallel,\alpha,i,j+\frac{1}{2}}\geq u_{\parallel,\alpha,i}^{p}},\label{eq:n-plus}\\
n_{\alpha,i}^{-,p+1} & \equiv-\left\langle v_{\parallel,\alpha,i,j}^{p},\frac{1}{v_{\alpha,i}^{*,p}}\frac{\overline{\left(\tilde{f}_{\alpha}^{p+1}\right)}_{i,j+\frac{1}{2},k}^{\mathrm{central}}-\overline{\left(\tilde{f}_{\alpha}^{p+1}\right)}_{i,j-\frac{1}{2},k}^{\mathrm{central}}}{\Delta c_{\parallel}}\right\rangle _{\delta\bm{c}\text{ where }v_{\parallel,\alpha,i,j+\frac{1}{2}}<u_{\parallel,\alpha,i}^{p}},\label{eq:n-minus}
\end{align}
which come from the appropriate $m_{\alpha}v_{\parallel}$ ``half
moments'' of the pseudo-operators associated with $\phi$ and $\gamma_{q}$,
as well as the upper and lower momenta
\begin{align}
\Gamma_{\parallel,\alpha,i}^{+,p+1} & \equiv-\left\langle \frac{1}{2}\left(\boldsymbol{v}_{\alpha,i,j,k}^{p}\right)^{2},\frac{1}{v_{\alpha,i}^{*,p}}\frac{\overline{\left(\tilde{f}_{\alpha}^{p+1}\right)}_{i,j+\frac{1}{2},k}^{\mathrm{central}}-\overline{\left(\tilde{f}_{\alpha}^{p+1}\right)}_{i,j-\frac{1}{2},k}^{\mathrm{central}}}{\Delta c_{\parallel}}\right\rangle _{\delta\bm{c}\text{ where }v_{\parallel,\alpha,i,j+\frac{1}{2}}\geq u_{\parallel,\alpha,i}^{p}},\label{eq:nu-plus}\\
\Gamma_{\parallel,\alpha,i}^{-,p+1} & \equiv-\left\langle \frac{1}{2}\left(\boldsymbol{v}_{\alpha,i,j,k}^{p}\right)^{2},\frac{1}{v_{\alpha,i}^{*,p}}\frac{\overline{\left(\tilde{f}_{\alpha}^{p+1}\right)}_{i,j+\frac{1}{2},k}^{\mathrm{central}}-\overline{\left(\tilde{f}_{\alpha}^{p+1}\right)}_{i,j-\frac{1}{2},k}^{\mathrm{central}}}{\Delta c_{\parallel}}\right\rangle _{\delta\bm{c}\text{ where }v_{\parallel,\alpha,i,j+\frac{1}{2}}<u_{\parallel,\alpha,i}^{p}},\label{eq:nu-minus}
\end{align}
from the $\frac{1}{2}m_{\alpha}\boldsymbol{v}^{2}$ half moments of
the same operators. The constraint functions $\phi$ and $\gamma_{q}$
together act to enforce the symmetries in Eqs. (\ref{eq:momentum-cons-VA})
and (\ref{eq:energy-cons-VA}), which lead to momentum and energy
conservation.

The nonlinear constraint function approach has been employed previously
for actively enforcing conservation for the Vlasov-Ampère system \citep{Taitano2015a,Taitano2015b}
with a symplectic time-integration scheme as well as for the Vlasov-Fokker-Planck
system with a velocity-space adaptive transformation \citep{Taitano2016,Taitano2018a}.
Here, this approach has been applied to the velocity-space transformed
Vlasov-Ampère system with BDF2 temporal discretization, though it
may in principle be applied to a wide variety of temporal discretizations.
Here, we further point out that all the constraint functions $\left(\gamma_{t},\gamma_{x},\xi,\phi,\gamma_{q}\right)$
are calculated locally in configuration space, and are almost entirely
independent of one another. The single exception is the dependence
of $\gamma_{x}$ on $\xi$ through the quantity $v_{\parallel,\mathrm{eff}}$,
which is satisfied simply by calculating $\xi$ before $\gamma_{x}$.

\section{Solving the discretized Vlasov-Ampère system\label{sec:Solution-strategy-for}}

To solve the discretized Vlasov-Ampère system, we use the high-order/low-order
(HOLO) nonlinear acceleration iterative strategy \citep{Chacon2017}.
HOLO accelerates the nonlinear convergence of the temporally implicit
high-order (HO) Vlasov-Ampère system through a low-order (LO) representation,
which efficiently exposes the stiff physics. The LO system is obtained
from the velocity-space moments of the HO system. This approach has
been successfully employed to solve the Vlasov\textendash Ampère and
Vlasov-Fokker-Planck\textendash Ampère systems among many other problems
\citep{Taitano2015a,Taitano2015b,Taitano2013,Park2012,knoll2011,Taitano2014,Chacon2017}.

\subsection{Formulation, discretization, and solution of the LO system\label{subsec:Discretization-of-LO}}

In our context, the LO equations (moments of the species' Vlasov equations)
are used to provide a well-informed guess for the electric field $E_{\parallel}$,
which results in fast nonlinear convergence of the original Vlasov-Ampère
system. The LO moment-Ampère system allows for the stiff time-scales
associated with collective physics (e.g., plasma waves) to be efficiently
captured in a lower-dimensional system. A key component of the strategy
is the enslavement of the discretization error and any missing physics
in the LO system through discrete consistency terms. This ensures
that the LO and HO moments agree exactly upon nonlinear convergence. 

To obtain the LO quantities and their respective equations, we take
the appropriate velocity-space moments of the Vlasov equation in the
continuum:

\begin{equation}
\left\langle \cdot,\cdot\right\rangle _{\bm{c}}\equiv2\pi\int_{-\infty}^{\infty}dc_{\parallel}\int_{0}^{\infty}c_{\bot}dc_{\bot}(\cdot*\cdot).\label{eq:moment-def}
\end{equation}
The evolution of $n_{\alpha}$ and $\Gamma_{\parallel,\alpha}$ are
described by the corresponding moments of the Vlasov equation, Eq.(\ref{eq:Vlasov-transformed-energy-final}):

\begin{eqnarray}
\left\langle 1,Vlasov\right\rangle _{\bm{c}}= & \partial_{t}n_{\alpha}+\partial_{x}\left(\Gamma_{\parallel,\alpha}\right)= & 0,\label{eq:continuity-def}\\
\left\langle v_{\parallel},Vlasov\right\rangle _{\bm{c}}= & \partial_{t}\left(\Gamma_{\parallel,\alpha}\right)+\partial_{x}S_{\parallel\parallel,\alpha}^{(2)}-\frac{q_{\alpha}}{m_{\alpha}}n_{\alpha}E_{\parallel}= & 0.\label{eq:momentum-cons-def}
\end{eqnarray}
In Eq. (\ref{eq:momentum-cons-def}), the quantity $S_{\parallel\parallel,\alpha}^{(2)}$
is the $v_{\parallel}^{2}$ moment of $\tilde{f}_{\alpha}$.

Thus, the LO system consists of the moment equations together with
Ampère's equation:

\begin{eqnarray}
\epsilon_{0}\partial_{t}E_{\parallel}^{LO}+\sum\limits _{\alpha}q_{\alpha}\Gamma_{\parallel,\alpha}^{LO} & = & \overline{j}_{\parallel},\label{eq:ampere-1D_LO}\\
\partial_{t}n_{\alpha}^{LO}+\partial_{x}\left(\Gamma_{\parallel,\alpha}^{LO}\right) & = & \eta_{n_{\alpha}}^{HO},\label{eq:continuity_LO}\\
\partial_{t}\left(\Gamma_{\parallel,\alpha}^{LO}\right)+\partial_{x}\left(n_{\alpha}^{LO}\widetilde{S^{(2)}}_{\parallel\parallel,\alpha}^{HO}\right)-\frac{q_{\alpha}}{m_{\alpha}}n_{\alpha}^{LO}E_{\parallel}^{LO} & = & \eta_{nu_{\parallel,\alpha}}^{HO}.\label{eq:momentum_LO}
\end{eqnarray}
In Eqs. (\ref{eq:continuity_LO}) and (\ref{eq:momentum_LO}) we have
introduced the HO consistency terms, $\eta_{n_{\alpha}}^{HO}$ and
$\eta_{nu_{\parallel,\alpha}}^{HO}$, which enslave the truncation
error (and any missing physics) of the LO system to the HO system
(to be explicitly defined later). Note that, to provide a closure
for the higher-order moments to the LO system, we use in Eq. (\ref{eq:momentum_LO})
the density-normalized total stress tensor from the HO system,

\begin{equation}
\widetilde{S^{(2)}}_{\parallel\parallel,\alpha}^{HO}=\frac{\left\langle v_{\parallel}^{2},f_{\alpha}\right\rangle _{\bm{c}}}{\left\langle 1,f_{\alpha}\right\rangle _{\bm{c}}}.\label{eq:HO-stress-norm}
\end{equation}
The presence of the LO density $n_{\alpha}^{LO}$ in the LO momentum
advection exposes the stiff isothermal wave in the LO system \citep{Taitano2015a,Taitano2013}. 

The LO system is discretized on a staggered finite-difference grid,
where we define the density, $n_{\alpha,i}^{LO}$, at cell centers
and the particle number density flux, $nu_{\parallel,\alpha,i+\frac{1}{2}}^{LO}$,
and electric field, $E_{\parallel,i+\frac{1}{2}}^{LO}$, at cell faces.
The discrete form of the LO system is

\begin{equation}
R_{n_{\alpha},i}^{l}\equiv\frac{b^{p+1}n_{\alpha,i}^{LO,p+1,l}+b^{p}n_{\alpha,i}^{HO,p}+b^{p-1}n_{\alpha,i}^{HO,p-1}}{\Delta t^{p}}+\frac{\Gamma_{\parallel,\alpha,i+\frac{1}{2}}^{LO,p+1,l}-\Gamma_{\parallel,\alpha,i-\frac{1}{2}}^{LO,p+1,l}}{\Delta x}-\eta_{n_{\alpha},i}^{l-1},\label{eq:continuity-semidisc-LO}
\end{equation}
\begin{multline}
R_{nu_{\parallel,\alpha},i+\frac{1}{2}}^{l}\equiv\frac{b^{p+1}\Gamma_{\parallel,\alpha,i+\frac{1}{2}}^{LO,p+1,l}+b^{p}\Gamma_{\parallel,\alpha,i+\frac{1}{2}}^{HO,p}+b^{p-1}\Gamma_{\parallel,\alpha,i+\frac{1}{2}}^{HO,p-1}}{\Delta t^{p}}\\
+\frac{n_{\alpha,i+1}^{LO,p+1,l}\widetilde{S^{(2)}}_{\parallel\parallel,\alpha,i+1}^{HO,p+1,l-1}-n_{\alpha,i}^{LO,p+1,l}\widetilde{S^{(2)}}_{\parallel\parallel,\alpha,i}^{HO,p+1,l-1}}{\Delta x}-\frac{q_{\alpha}}{m_{\alpha}}n_{\alpha,i+\frac{1}{2}}^{LO,p+1,l}E_{\parallel,i+\frac{1}{2}}^{LO,p+1,l}-\eta_{nu_{\parallel,\alpha}i+\frac{1}{2}}^{l-1},\label{eq:momentum-semidisc-LO}
\end{multline}
\begin{equation}
R_{E_{\parallel},i+\frac{1}{2}}^{l}\equiv\epsilon_{0}\frac{b^{p+1}E_{\parallel,i+\frac{1}{2}}^{LO,p+1,l}+b^{p}E_{\parallel,i+\frac{1}{2}}^{LO,p}+b^{p-1}E_{\parallel,i+\frac{1}{2}}^{LO,p-1}}{\Delta t^{p}}+\sum\limits _{\alpha}q_{\alpha}\Gamma_{\parallel,\alpha,i+\frac{1}{2}}^{LO,p+1,l}-\overline{j}_{\parallel}^{p+1,l}.\label{eq:Ampere-semidisc-LO}
\end{equation}
In Eqs. (\ref{eq:continuity-semidisc-LO})\textendash (\ref{eq:Ampere-semidisc-LO}),
the quantities $R_{n_{\alpha},i}$, $R_{nu_{\parallel,\alpha},i+\frac{1}{2}}$,
and $R_{E_{\parallel},i+\frac{1}{2}}$ are the nonlinear residuals
for the corresponding LO quantities (which should be converged to
zero), $i$ is the configuration space index, and $l$ is the HOLO
iteration index. The quantity $n_{\alpha,i+\frac{1}{2}}^{LO,p+1,l}$
is the cell-centered density linearly interpolated to cell faces.
$\widetilde{S^{(2)}}_{\parallel\parallel,\alpha,i}^{HO}$ is defined
at cell centers. The HO quantities are generally defined as the corresponding
direct moments of the distribution $\tilde{f}_{\alpha,i,j,k}^{p+1,l}$
\begin{align}
n_{\alpha,i}^{HO,p+1,l} & =\left\langle 1,\tilde{f}_{\alpha,i}^{p+1,l}\right\rangle _{\delta\bm{c}}\label{eq:n_HO}\\
\widetilde{S^{(2)}}_{\parallel\parallel,\alpha}^{HO,p+1,l} & =\frac{\left\langle v_{\parallel,j}^{2},\tilde{f}_{\alpha,i}^{p+1,l}\right\rangle _{\delta\bm{c}}}{\left\langle 1,\tilde{f}_{\alpha,i}^{p+1,l}\right\rangle _{\delta\bm{c}}}\label{eq:S2_HO}
\end{align}
The exception is the cell-face HO particle flux density used in the
LO system,
\[
\Gamma_{\parallel,\alpha,i+\frac{1}{2}}^{HO,p+1,l}=\widehat{\Gamma}_{\parallel,\alpha,i+\frac{1}{2}}^{p+1,l},
\]
which is taken to be the same as the flux that forms the current for
Ampère's equation {[}$\widehat{\Gamma}_{\parallel,\alpha,i+\frac{1}{2}}^{p+1,l}$
is defined as in Eqs. (\ref{eq:nu-bar})-(\ref{eq:nu-cf-accel}){]}.
We note here that, for the HOLO system, we use a slightly different
definition for the discrete averaged current, $\overline{j}_{\parallel}^{p+1,l}$,
than in Eq. (\ref{eq:avg-j-discrete-HO}), based on the cell-face
parallel LO particle flux density: 
\begin{equation}
\overline{j}_{\parallel}^{p+1,l}\equiv\frac{1}{N_{x}}\sum\limits _{i}^{N_{x}}\left(\sum\limits _{\alpha}^{N_{sp}}q_{\alpha}\Gamma_{\parallel,\alpha,i+\frac{1}{2}}^{LO,p+1,l}\right).\label{eq:avg-j-discrete-LO}
\end{equation}
 The discrete consistency terms, $\eta_{n_{\alpha},i}^{l}$ and $\eta_{nu_{\parallel,\alpha},i+\frac{1}{2}}^{l}$,
are defined by introducing HO moments into the LO equations as

\begin{equation}
\eta_{n_{\alpha},i}^{l}\equiv\frac{b^{p+1}n_{\alpha,i}^{HO,p+1,l}+b^{p}n_{\alpha,i}^{HO,p}+b^{p-1}n_{\alpha,i}^{HO,p-1}}{\Delta t^{p}}+\frac{\Gamma_{\parallel,\alpha,i+\frac{1}{2}}^{HO,p+1,l}-\Gamma_{\parallel,\alpha,i-\frac{1}{2}}^{HO,p+1,l}}{\Delta x}-\left\langle 1,R_{\tilde{f},\alpha,i}^{l}\right\rangle _{\delta\bm{c}},\label{eq:gamma-discrete-density}
\end{equation}
\begin{multline}
\eta_{nu_{\parallel,\alpha},i+\frac{1}{2}}^{l}\equiv\frac{b^{p+1}\Gamma_{\parallel,\alpha,i+\frac{1}{2}}^{HO,p+1,l}+b^{p}\Gamma_{\parallel,\alpha,i+\frac{1}{2}}^{HO,p}+b^{p-1}\Gamma_{\parallel,\alpha,i+\frac{1}{2}}^{HO,p-1}}{\Delta t^{p}}+\frac{n_{\alpha,i+1}^{HO,p+1,l}\widetilde{S^{(2)}}_{\parallel\parallel,\alpha,i+1}^{HO,p+1,l}-n_{\alpha,i}^{HO}\widetilde{S^{(2)}}_{\parallel\parallel,\alpha,i}^{HO,p+1,l}}{\Delta x}\\
-\frac{q_{\alpha}}{m_{\alpha}}n_{\alpha,i+\frac{1}{2}}^{HO,p+1,l}E_{\parallel,i+\frac{1}{2}}^{LO,p+1,l-1}-\frac{\left(\left\langle v_{\parallel,j},R_{\tilde{f},\alpha,i}^{l}\right\rangle _{\delta\bm{c}}+\left\langle v_{\parallel,j},R_{\tilde{f},\alpha,i+1}^{l}\right\rangle _{\delta\bm{c}}\right)}{2}.\label{eq:gamma-discrete-momentum}
\end{multline}
The consistency terms will converge towards zero along with the residuals
as the HOLO iteration progresses. In Eqs. (\ref{eq:gamma-discrete-density})\textendash (\ref{eq:gamma-discrete-momentum}),
$\left\langle 1,R_{\tilde{f},\alpha,i}^{l}\right\rangle _{\delta\bm{c}}$
and $\left\langle v_{\parallel,j},R_{\tilde{f},\alpha,i}^{l}\right\rangle _{\delta\bm{c}}$
indicate moments of the HO system residual, Eq. (\ref{eq:HO-residual}),
which will be discussed in detail in Sec. \ref{subsec:Discretization-and-solution}.

The coupled LO system is solved with an Anderson-accelerated nonlinear
quasi-Newton iteration \citep{Walker2011,Anderson1965}. The quasi-Newton
iteration is preconditioned with a direct solution of the linearized
moment equations:

\begin{equation}
\frac{b^{p+1}\delta n_{\alpha,i}}{\Delta t^{p}}+\frac{\delta\Gamma_{\parallel,\alpha,i+\frac{1}{2}}-\delta\Gamma_{\parallel,\alpha,i-\frac{1}{2}}}{\Delta x}=-R_{n_{\alpha},i}^{l}\,,\label{eq:continuity-linearized}
\end{equation}
\begin{multline}
\frac{b^{p+1}\delta\Gamma_{\parallel,\alpha,i+\frac{1}{2}}}{\Delta t^{p}}+\frac{\delta n_{\alpha,i+1}\widetilde{S^{(2)}}_{\parallel\parallel,\alpha,i+1}^{HO,p+1,l-1}-\delta n_{\alpha,i}\widetilde{S^{(2)}}_{\parallel\parallel,\alpha,i}^{HO,p+1,l-1}}{\Delta x}\\
-\frac{q_{\alpha}}{m_{\alpha}}\left(\delta n_{\alpha,i+\frac{1}{2}}E_{\parallel,i+\frac{1}{2}}^{LO,p+1,l}+n_{\alpha,i+\frac{1}{2}}^{LO,p+1,l}\delta E_{\parallel,i+\frac{1}{2}}\right)=-R_{nu_{\parallel,\alpha},i+\frac{1}{2}}^{l},\label{eq:momentum-linearized}
\end{multline}
\begin{equation}
\epsilon_{0}\frac{b^{p+1}\delta E_{\parallel,i+\frac{1}{2}}}{\Delta t^{p}}+\sum\limits _{\alpha}q_{\alpha}\delta\Gamma_{\parallel,\alpha,i+\frac{1}{2}}=-R_{E_{\parallel},i+\frac{1}{2}}^{l}.\label{eq:Ampere-linearized}
\end{equation}
After substitution of Eqs. (\ref{eq:continuity-linearized}) and (\ref{eq:Ampere-linearized})
into Eq. (\ref{eq:momentum-linearized}) for $\delta n_{\alpha}$
and $\delta E$, respectively, the system reduces to a single equation
for $\delta\Gamma_{\parallel,\alpha,i+\frac{1}{2}}$ at cell faces,
coupled in space and across species. After linear inversion for $\delta\Gamma_{\parallel,\alpha,i+\frac{1}{2}}$,
$\delta n_{\alpha,i}$ and $\delta E_{\parallel,i+\frac{1}{2}}$ are
found directly from Eqs. (\ref{eq:continuity-linearized}) and (\ref{eq:Ampere-linearized}).
To accelerate HOLO convergence further, an additional layer of Anderson
acceleration is wrapped around the outer HOLO iteration driven by
to the LO solution, similarly to what was considered in Ref. \citep{Willert2014}.

\subsection{Discretization and solution of the HO system\label{subsec:Discretization-and-solution}}

The HO system is discretized essentially as presented in Sec. \ref{sec:Numerical-Implementation},
with the result reproduced here to highlight the coupling with the
LO system:

\begin{multline}
R_{\tilde{f},\alpha,i,j,k}^{l}\equiv\Bigg\{\delta_{t}\tilde{f}_{\alpha,i,j,k}^{p+1,l}+\delta_{x}\left[v_{\parallel,\alpha,j}^{p}\overline{\left(\tilde{f}_{\alpha}^{p+1,l}\right)}_{j,k}^{v_{\parallel}}\right]_{i}+\frac{q_{\alpha}}{m_{\alpha}}\frac{E_{\parallel,i}^{LO,p+1,l-1}}{v_{\alpha,i}^{*,p}}\delta_{c_{\parallel}}\left[\overline{\left(\tilde{f}_{\alpha}^{p+1,l}\right)}_{i,k}^{q_{\alpha}E_{\parallel}}\right]_{j}\\
+\delta_{x}\left[\xi_{\alpha}^{p+1,l}\left|v_{\parallel,\alpha,j}^{p}\right|\overline{\left(\tilde{f}_{\alpha}^{p+1,l}\right)}_{j,k}^{\xi}\right]_{i}+\delta_{c_{\parallel}}\left[\phi_{\alpha,i}^{p+1,l}\overline{\left(\tilde{f}_{\alpha}^{p+1,l}\right)}_{i,k}^{\phi}\right]_{j}+\delta_{c_{\parallel}}\left[\gamma_{q,\alpha,i}^{p+1,l}\overline{\left(\tilde{f}_{\alpha}^{p+1,l}\right)}_{i,k}^{\gamma_{q}}\right]_{j}\\
-\frac{1}{v_{\alpha,i}^{*,p}}\delta_{\bm{c}}\cdot\left[\gamma_{t,\alpha,i}^{p+1,l}\delta_{t}\left(\boldsymbol{v}_{\alpha,i}\right)\overline{\left(\tilde{f}_{\alpha}^{p+1,l}\right)}_{i}^{\delta_{t}\left(\boldsymbol{v}\right)}\right]_{j,k}\\
-\frac{1}{2v_{\alpha,i}^{*,p}}\delta_{\bm{c}}\cdot\left[\gamma_{x,\alpha,i+\frac{1}{2}}^{p+1,l}v_{\alpha,i+\frac{1}{2}}^{*,p}\hat{v}_{\parallel,\alpha,\mathrm{eff},i+\frac{1}{2}}^{p+1,l}\delta_{x}\left[\boldsymbol{v}_{\alpha}^{p}\right]_{i+\frac{1}{2}}\overline{\left(\tilde{f}_{\alpha}^{p+1,l}\right)}_{i}^{v_{\parallel,\mathrm{eff}}\delta_{x}\left(\boldsymbol{v}\right)}\right]_{j,k}\\
-\frac{1}{2v_{\alpha,i}^{*,p}}\delta_{\bm{c}}\cdot\left[\gamma_{x,\alpha,i-\frac{1}{2}}^{p+1,l}v_{\alpha,i-\frac{1}{2}}^{*,p}\hat{v}_{\parallel,\alpha,\mathrm{eff},i-\frac{1}{2}}^{p+1,l}\delta_{x}\left[\boldsymbol{v}_{\alpha}^{p}\right]_{i-\frac{1}{2}}\overline{\left(\tilde{f}_{\alpha}^{p+1,l}\right)}_{i}^{v_{\parallel,\mathrm{eff}}\delta_{x}\left(\boldsymbol{v}\right)}\right]_{j,k}\Bigg\},\label{eq:HO-residual}
\end{multline}
where 
\[
\hat{v}_{\parallel,\alpha,\mathrm{eff},i+\frac{1}{2},j}^{p+1,l}=\left(c_{\parallel,j}+\hat{u}_{\parallel,\alpha,i+\frac{1}{2}}^{*,p}\right)+\xi_{\alpha,i+\frac{1}{2}}^{p+1,l}\left|c_{\parallel,j}+\hat{u}_{\parallel,\alpha,i+\frac{1}{2}}^{*,p}\right|.
\]
The quantity $R_{\tilde{f},\alpha,i,j,k}^{l}$ is the HO system residual.
Note that we have included the superscript $l$ for the HOLO iteration
index. Observe that the electric field, $E_{\parallel}^{LO}$, in
Eq. (\ref{eq:HO-residual}) is obtained from the solution of the LO
system, which effectively Picard-linearizes the individual Vlasov
equations (HO system) in $\tilde{f}_{\alpha}$ and is key for effective
nonlinear convergence acceleration. However, the discretization scheme
employed in Eq. (\ref{eq:HO-residual}) may still include significant
nonlinearities in the advective terms. Thus, the HO system is also
solved with Anderson acceleration. For preconditioning, Eq. (\ref{eq:HO-residual})
is linearized in $\delta f_{\alpha,i}$, with a linear upwind discretization
for all operators. The system is then solved with the multigrid-preconditioned
Flexible Generalized Minimal RESiduals (FGMRES) method \citep{Saad1993}.

\subsection{HOLO solution algorithm\label{subsec:HOLO-acceleration-strategy}}

Thus, the coupled HOLO system is represented by 1) the HO system,
which consists a system of the species' Vlasov equations, Eq. (\ref{eq:HO-residual}),
and 2) the LO system, which consists of the moment equations for each
species' mass and momentum and Ampère's equation, Eqs. (\ref{eq:continuity-semidisc-LO})\textendash (\ref{eq:Ampere-semidisc-LO}).
Algorithm \ref{alg:HOLO} depicts the HOLO-accelerated iteration.
\begin{algorithm}[t]
\caption{HOLO solution for VA system} 
\label{alg:HOLO} 
Initialize HOLO iteration index $\left(l=0\right)$\;
Initialize outer HO residual $\left(R^{0}_{\tilde{f}}\right)$\;
Initialize consistency terms $\left(\eta^{l}_{n_\alpha}, \eta^{l}_{nu_{\parallel,\alpha}}\right)$\;
\While{HOLO not converged}{
	Solve LO system for $E^{LO,l}_{\parallel}$ from Eqs. \eqref{eq:continuity-semidisc-LO}--\eqref{eq:Ampere-semidisc-LO}\;
	Solve HO system for $\tilde{f}^{l}_{\alpha}$ from Eq. \eqref{eq:HO-residual}\;
	Compute consistency terms $\left(\eta^{l}_{n_\alpha}, \eta^{l}_{nu_{\parallel,\alpha}}\right)$ from Eqs. \eqref{eq:gamma-discrete-density}--\eqref{eq:gamma-discrete-momentum}\;
	Increment HOLO iteration $\left(l=l+1\right)$\;
	Check HOLO convergence $\left(\left|R^{l}_{\tilde{f}}\right|_{\mathrm{rms}}<\epsilon\right)$\;
}
Save $\tilde{f}_{\alpha}^{p+1}$, $E_{\parallel}^{LO,p+1}$\;
\end{algorithm}Convergence is measured through the root-mean-square (rms) of the
HO residual vector,
\[
\left|R_{\tilde{f}}^{l}\right|_{\mathrm{rms}}=\sqrt{\frac{1}{N_{\alpha}N_{x}N_{y}N_{z}}\sum_{\alpha=1}^{N_{sp}}\sum_{i=1}^{N_{x}}\sum_{j=1}^{N_{y}}\sum_{k=1}^{N_{z}}\left(R_{\tilde{f},\alpha,i,j,k}^{l}\right)^{2}}.
\]
Here, the convergence tolerance, $\epsilon$, is defined as
\begin{equation}
\epsilon=\epsilon_{a}+\epsilon_{r}\left|R_{\tilde{f}}^{0}\right|_{\mathrm{rms}},\label{eq:tolerance_HOLO}
\end{equation}
where $\epsilon_{a}$ is an absolute tolerance and $\epsilon_{r}$
is the relative tolerance.

%% file: Results.tex
\section{Numerical results\label{sec:Numerical-Results}}

In this section, we demonstrate the accuracy, convergence, and conservation
properties of the proposed numerical scheme. We do so using several
canonical collisionless problems of increasing complexity, ranging
from the linear Landau damping to an ion-acoustic shock wave. Unless
specified otherwise, for all the problems presented we normalize the
particle mass and charge to the electron mass, $m_{e}$, and proton
charge, $q_{p}$, while normalizing the temperature, density, velocity,
and time to the reference temperature, $T_{0,}$ density, $n_{0,}$
speed, $v_{0}=\sqrt{(T_{0}/m_{e})},$ and time-scale, $\tau_{0}=\omega_{p,e}^{-1}$
(where $\omega_{p,e}$ is the electron plasma frequency). The initial
velocity distributions for each species, $\alpha,$ are assumed to
be normalized Maxwellians

\begin{equation}
\tilde{f}_{M,\alpha}=\frac{n_{\alpha}}{\pi^{3/2}}\left(\frac{v_{\alpha}^{*}}{v_{th,\alpha}}\right)^{3}\exp\left[-\frac{1}{v_{th,\alpha}^{2}}\left(v_{\alpha}^{*}\left(\bm{c}+\hat{u}_{\parallel,\alpha}^{*}\boldsymbol{e}_{\parallel}\right)-u_{\parallel,\alpha}\boldsymbol{e}_{\parallel}\right)^{2}\right],\label{eq:maxwell-norm}
\end{equation}
where $v_{th,\alpha}\equiv\sqrt{\frac{2T_{\alpha}}{m_{\alpha}}}$.
Unless stated otherwise, the velocity-space adaptivity metrics (i.e.,
the offset velocity $u_{\parallel,\alpha}^{*}$ and reference speed
$v_{\alpha}^{*}$) are initialized and spatio-temporally adapted using
each species' initial bulk velocity $u_{\parallel,\alpha}$ and thermal
speed $v_{\alpha}^{*}$, respectively. For robustness, some smoothing
and limiting strategies are applied to avoid large spatial or temporal
gradients in the metrics. See Ref. \citep{Taitano2018a} for specific
details. The initial electric field, $E_{\parallel,}$is determined
from the solution of Poisson's equation driven by the initial charge
density:

\[
-\epsilon_{0}\frac{\partial^{2}\Phi_{0}}{\partial x^{2}}=\sum_{\alpha}^{N_{s}}q_{\alpha}n_{\alpha},
\]

\[
E_{\parallel,0}=-\frac{\partial\Phi_{0}}{\partial x}.
\]
A realistic proton-electron mass ratio $m_{i}/m_{e}=1836$ is used
for all cases. Unless otherwise specified, the relative nonlinear
convergence tolerance is $\epsilon_{r}=10^{-4}$, while the absolute
tolerance is set to a low value ($\epsilon_{a}=10^{-14}$) to avoid
interference with the relative convergence (see Sec. \ref{subsec:HOLO-acceleration-strategy}).

\subsection{Landau damping}

The linear and nonlinear electron Landau damping tests show the ability
of the solver to capture fine collisionless features in phase space.
For this problem, the rate of oscillation and decay of the electric
field energy is determined by the dispersion relation,

\begin{equation}
1+\frac{1}{k^{2}}\left[1+\frac{\omega}{\sqrt{2}k}Z\left(\frac{\omega}{\sqrt{2}k}\right)\right]=0,\label{eq:Landau-dispersion}
\end{equation}
which determines the complex frequency $\omega=\omega_{r}-i\gamma$
for a given wavenumber $k$. The function $Z$ is the plasma dispersion
function for a Maxwellian \citep{Fried1961}.

\subsubsection{Linear Landau damping}

\begin{figure}[h]
\begin{centering}
\includegraphics[scale=0.4]{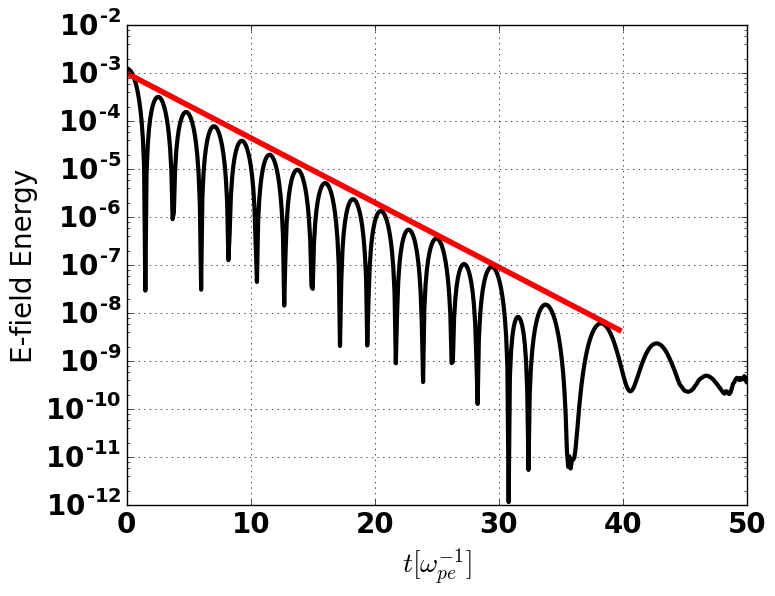}
\par\end{centering}
\caption{Decay ($\gamma=-0.155$) of the electric field energy for the linear
Landau damping test.\label{fig:Linear-Landau-Damping}}
\end{figure}

To initiate the linear Landau damping in an electron-proton plasma,
we impose a sinusoidal density perturbation on electrons with wavenumber
$k\lambda_{D}=0.5$ and amplitude $\delta n=0.01$. We expect it to
decay at a linear rate of $\gamma=-0.155$ \citep{Taitano2015a,Watanabe2005}.
Both species have the same initial temperature $T_{0}=1$, bulk velocity
$u_{\parallel,0}=0$, and unperturbed density $n_{0}=1$. The simulation
is performed on the mesh $N_{x}=32$, $N_{v_{\parallel}}=512$, $N_{v_{\bot}}=32$,
with an average time-step of $\Delta t=0.1\omega_{pe}^{-1}$. The
velocity-space domain is $c_{\parallel}\in\left[-6,+6\right]$, $c_{\bot}\in\left[0,5\right]$.
The offset velocity for both species is $u_{\parallel}^{*}=0$, while
the species normalization reference speeds are $v_{e,0}^{*}=\sqrt{\frac{2T_{0}}{m_{e}}},$
$v_{i,0}^{*}=\sqrt{\frac{2T_{0}}{m_{i}}}$. The configuration space
domain size is $L_{x}=4\pi$. From Fig. \ref{fig:Linear-Landau-Damping},
we see that the rate of decay matches the linear theory well. 

\subsubsection{Numerical convergence of the method}

To demonstrate that the full set of discrete governing equations achieve
our desired level of accuracy, we perform a convergence study in time
and space. Convergence is measured by computing the $L_{2}$ norm,
$L_{2}^{E,\Delta}$, of the difference in the electric field for each
solution relative to a reference solution $E_{\parallel}^{\Delta,\mathrm{ref}}$
obtained with a small time-step or with high resolution in the configuration
or velocity space,
\[
L_{2}^{E,\Delta}\equiv\sqrt{\sum\limits _{i}^{N_{x}}\Delta x_{i}\left[\left(E_{\parallel,i}-E_{\parallel,i}^{\Delta,\mathrm{ref}}\right)\right]^{2}}.
\]
For the convergence studies, a relative nonlinear convergence tolerance
of $\epsilon_{r}=10^{-12}$ is used to resolve the difference in truncation
error at small time-steps and fine-grid resolutions.
\begin{figure}[h]
\centering{}%
\noindent\begin{minipage}[t]{1\columnwidth}%
\begin{center}
\includegraphics[scale=0.25]{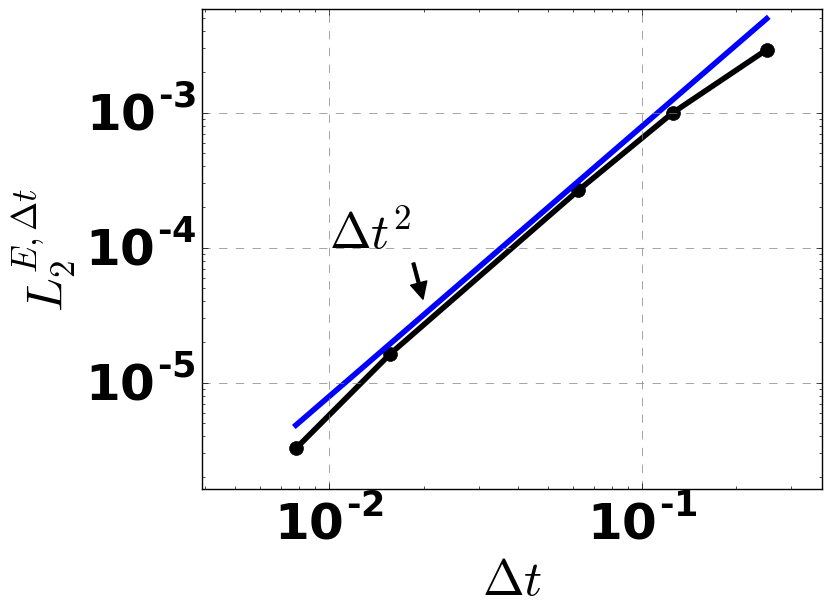}\hfill{}\includegraphics[scale=0.25]{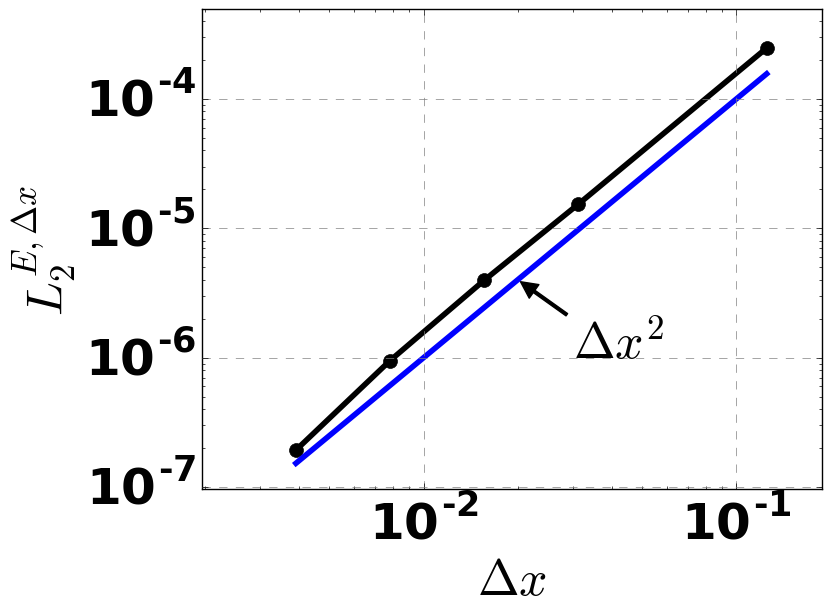}\hfill{}\includegraphics[scale=0.25]{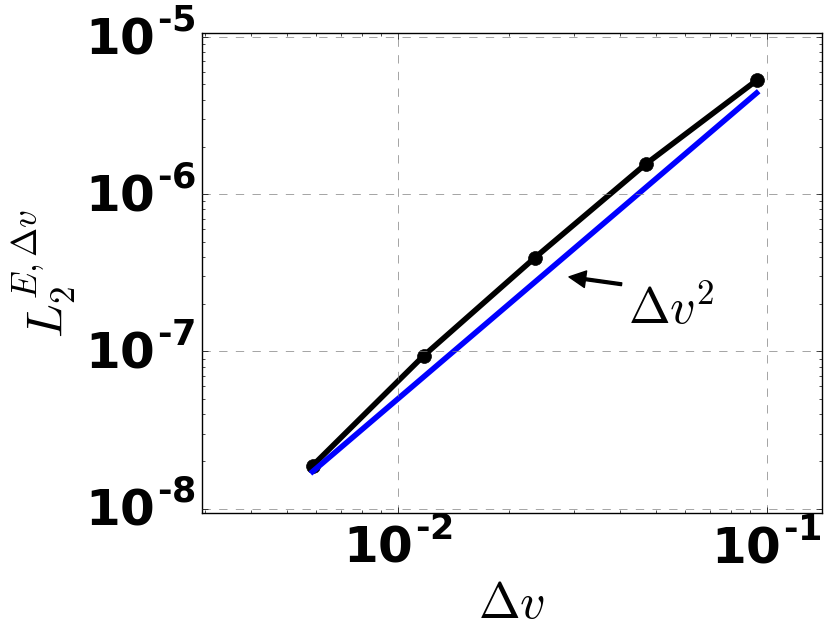}
\par\end{center}%
\end{minipage}\caption{Numerical convergence studies for temporal (left), configuration space
(center), and velocity space (right) resolutions.\label{fig:Linear-Landau-grid-convergence}}
\end{figure}
Figure \ref{fig:Linear-Landau-grid-convergence} shows convergence
with temporal resolution (left, with a reference time-step of $\Delta t=3.90625\times10^{-3}$,
and a mesh of $N_{x}=32$, $N_{v_{\parallel}}=512$, $N_{v_{\bot}}=32$),
configuration space resolution (center, with a reference mesh of $N_{x}=2048$,
using a time-step $\Delta t=0.1$, with a velocity-space mesh of $N_{v_{\parallel}}=512$,
$N_{v_{\bot}}=32$), and velocity space resolution (right, with a
reference mesh of $N_{v_{\parallel}}=4096$, $N_{v_{\bot}}=2048$,
using a time-step of $\Delta t=0.1$, and a configuration-space mesh
of $N_{x}=32$). The maximum simulation time is $t_{max}=1$ for configuration
and velocity-space convergence, and $t_{max}=10$ for the temporal
convergence. As can be seen, second-order convergence rates are observed
with respect to all the independent variables.

\subsubsection{Nonlinear Landau damping}

To simulate nonlinear (strong) Landau damping, we again choose $k\lambda_{D}=0.5,$
but increase the electron density perturbation magnitude to $\delta n=0.5.$
The simulation is performed on a mesh of $N_{x}=256$, $N_{v_{\parallel}}=512$,
$N_{v_{\bot}}=32$, and the initialization is otherwise identical
to the linear Landau damping case. According to the literature, this
should produce an initial decay and a subsequent recurrence with rates
of $\gamma_{1}=-0.292$ and $\gamma_{2}=0.0815$, respectively. 
\begin{figure}[h]
\centering{}\includegraphics[scale=0.4]{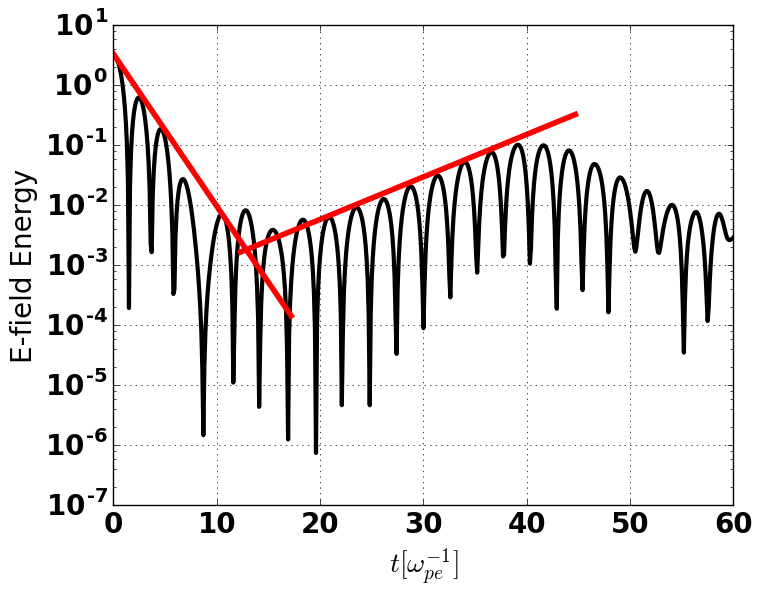}\caption{Initial decay ($\gamma_{1}=-0.292$) and a subsequent recurrence growth
($\gamma_{2}=0.0815$) of the electric field energy for the nonlinear
Landau damping test.\label{fig:Nonlinear-Landau-Damping}}
\end{figure}
 As we see in Fig. (\ref{fig:Nonlinear-Landau-Damping}), the decay
and growth of the electric field energy show excellent agreement with
other published results for this problem \citep{Taitano2015a,Rossmanith2011}.

\subsection{Two-stream instability\label{subsec:Two-stream-instability}}

The electron-electron two-stream instability simulation \citep{Chen1989}
is initialized as two relatively cold, counterstreaming Maxwellian
electron beams, each with the bulk velocity $\pm v_{b},$ and with
the thermal speed $v_{th,b}<<v_{b}$, against a neutralizing background
of stationary ions. The dispersion relation for this problem is
\begin{equation}
1+\frac{\omega_{p,b}^{2}}{k^{2}v_{th,b}^{2}}\left[2+\zeta_{+}Z\left(\zeta_{+}\right)+\zeta_{-}Z\left(\zeta_{-}\right)\right]=0,\label{eq:dispersion-two-stream-general}
\end{equation}
where
\[
\zeta_{\pm}\equiv\frac{\omega\mp kv_{b}}{kv_{th,b}},
\]
and $\omega_{p,b}$ is the beam plasma frequency. In the limit of
$v_{th,b}\rightarrow0$, Eq. (\ref{eq:dispersion-two-stream-general})
becomes

\begin{equation}
1-\frac{1}{\left(\omega+v_{b}k\right)^{2}}-\frac{1}{\left(\omega-v_{b}k\right)^{2}}=0.\label{eq:dispersion-two-stream-delta}
\end{equation}

\begin{figure}[h]
\centering{}%
\begin{minipage}[t]{0.49\columnwidth}%
\begin{center}
\includegraphics[scale=0.4]{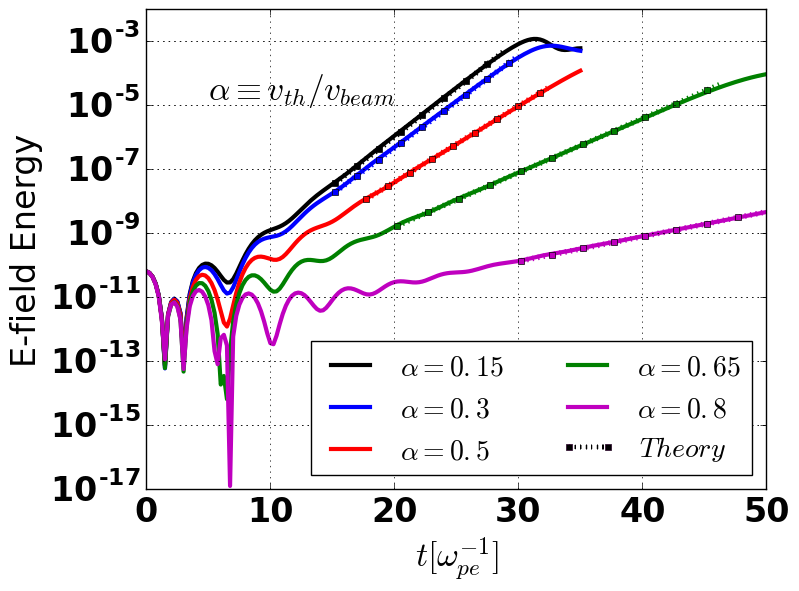}\caption{Growth of the electric field energy due to the two-stream instability
for various ratios of the beam thermal speed $v_{th,b}$ to the offset
velocity $v_{b}$.\label{fig:Two-Stream-Instability-alphasweep}}
\par\end{center}%
\end{minipage}\hfill{}%
\begin{minipage}[t]{0.49\columnwidth}%
\begin{center}
\includegraphics[scale=0.4]{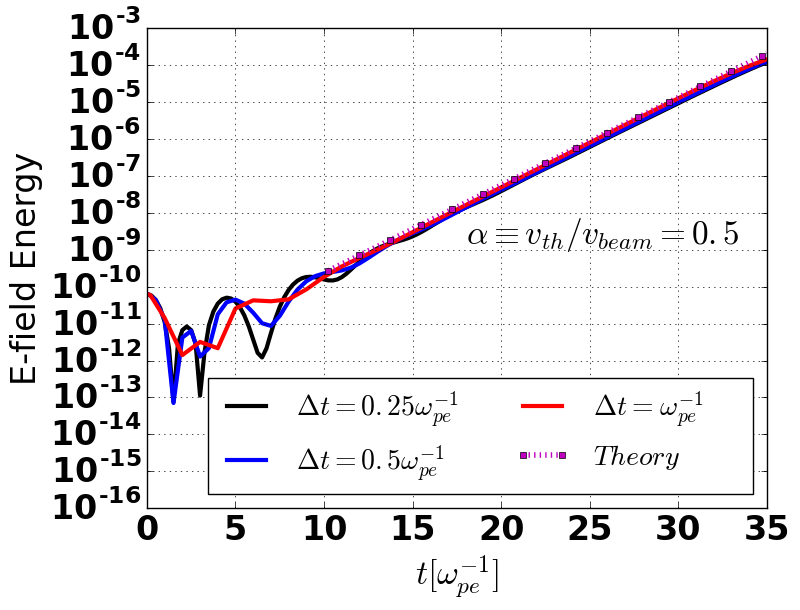}\caption{Growth of the electric field energy due to the two stream instability
for various time-steps at a beam thermal velocity ratio of $v_{th,b}/v_{b}=0.5$.\label{fig:Two-Stream-Instability-delta-t}}
\par\end{center}%
\end{minipage}
\end{figure}

For our simulation, we use electron beam densities of $n_{0}=0.5$,
beam velocities $v_{b}=\pm0.1$ and beam thermal velocities $v_{th,b}/v_{b}=[0.15,0.3,0.5,0.65,0.8]$.
The electron-beam densities are perturbed sinusoidally with wavenumber
$k=2\pi/L_{x}$ and magnitude $\delta n=0.00005.$ The domain size
is $L_{x}=1$. The velocity-space domain is $c_{\parallel}\in\left[-4,+4\right]$
for $v_{th,b}/v_{b}=0.15$, $c_{\parallel}\in\left[-5,+5\right]$
for $v_{th,b}/v_{b}=0.3$, and $c_{\parallel}\in\left[-6,+6\right]$
for $v_{th,b}/v_{b}\geq0.5$, with $c_{\bot}\in\left[0,5\right]$.
The electrons have an initial offset velocity of $u_{\parallel,e}^{*}=0$
and normalizing speed of $v_{\alpha,e}^{*}=\left[0.083,0.087,0.096,0.10,0.114\right]$.
The mesh is $N_{x}=128$, $N_{v_{\parallel}}=512$, $N_{v_{\bot}}=32$.
In Fig. \ref{fig:Two-Stream-Instability-alphasweep}, we perform a
sweep in beam thermal velocity ratio with $\Delta t=0.25\omega_{pe}^{-1},$
while in Fig. \ref{fig:Two-Stream-Instability-delta-t} we choose
$v_{th,b}/v_{b}=0.5$ and use $\Delta t=[0.25,1.0,2.0]\omega_{pe}^{-1}$.

Based on the delta-function dispersion relation, Eq. (\ref{eq:dispersion-two-stream-delta}),
the growth rate of electric field energy is $\gamma=0.353\omega_{p,b}$.
However, for thermalized beams there will be some deviation, and we
expect that as the ratio $v_{th,b}/v_{b}$ increases the system to
become less unstable (i.e., $\gamma$ will decrease). Indeed, in Fig.
\ref{fig:Two-Stream-Instability-alphasweep} we see that if we increase
$v_{th,b}/v_{b}$ towards some critical ratio near unity, the growth
rate decreases precipitously. As shown in Table \ref{tab:two-stream},
growth rates calculated from simulations ($\gamma_{sim}$) agree very
well with the growth rates obtained from a numerical solution of the
dispersion relation for thermalized beams ($\gamma_{\mathrm{num}}$)
\textendash{} details of this analysis may be found in \ref{app:Two-stream-instability-for}.
\begin{table}[h]
\centering{}%
\begin{tabular}{|c|c|c|c|c|c|}
\hline 
$v_{th,b}/v_{b}$ & $0.15$ & $0.3$ & $0.5$ & $0.65$ & $0.8$\tabularnewline
\hline 
\hline 
$\gamma_{\mathrm{num}}/\omega_{p,b}$ & $0.3488$ & $0.3318$ & $0.2734$ & $0.1953$ & $0.08911$\tabularnewline
\hline 
$\gamma_{\mathrm{sim}}/\omega_{p,b}$ & $0.3459$ & $0.3291$ & $0.2722$ & $0.1927$ & $0.08745$\tabularnewline
\hline 
$\frac{\gamma_{\mathrm{sim}}-\gamma_{\mathrm{num}}}{\gamma_{\mathrm{num}}}\times100\%$ & $-0.81\%$ & $-0.81\%$ & $-0.44\%$ & $-1.3\%$ & $-1.9\%$\tabularnewline
\hline 
\end{tabular}\caption{Numerical solution of two-stream instability growth rate\label{tab:two-stream}}
\end{table}
In Fig. \ref{fig:Two-Stream-Instability-delta-t}, we show that there
is little change in the simulated electric-field growth rate as we
vary the time-step size.

\subsection{\label{subsec:Ion-acoustic-shockwave}Ion-acoustic shock wave}

The final test is the ion-acoustic shock wave (IASW) \citep{Shay2007}.
This problem is an excellent test of the scheme because it exhibits
strongly nonlinear multi-scale behavior. In this problem, the dynamical
time-scale of the system is orders of magnitude larger than the inverse
electron plasma frequency, and so the simulation provides a stringent
test of the HOLO algorithm to step over $\omega_{pe}^{-1}$ time-scales,
which do not significantly contribute to the system evolution (since
the evolution is largely ambipolar). 

\begin{figure}[h]
\centering{}%
\noindent\begin{minipage}[t]{1\columnwidth}%
\begin{center}
\includegraphics[scale=0.4]{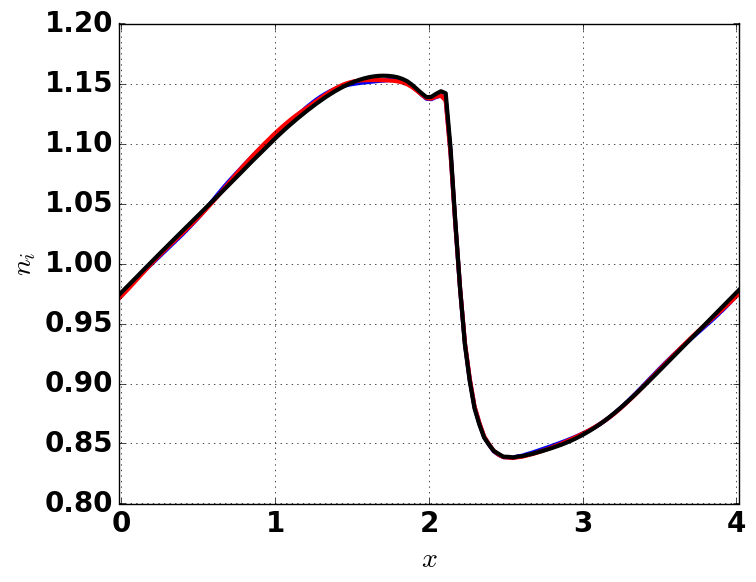}\includegraphics[scale=0.4]{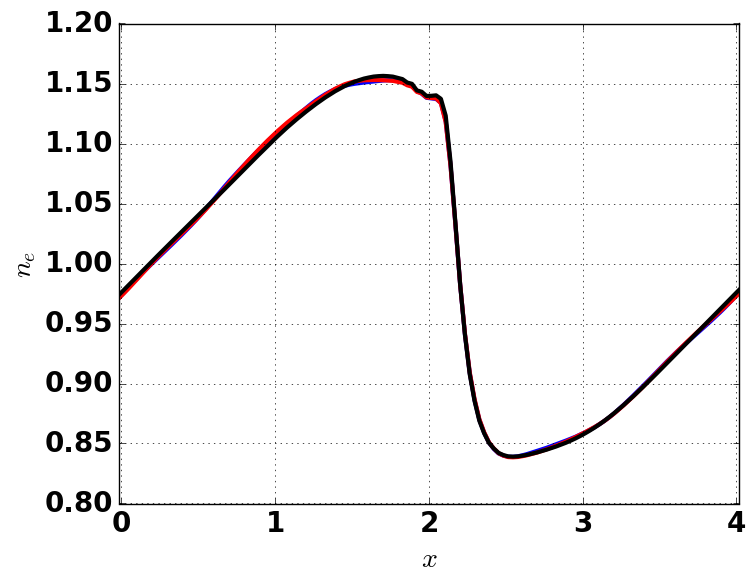}
\par\end{center}%
\end{minipage}\vfill{}
\noindent\begin{minipage}[t]{1\columnwidth}%
\begin{center}
\includegraphics[scale=0.4]{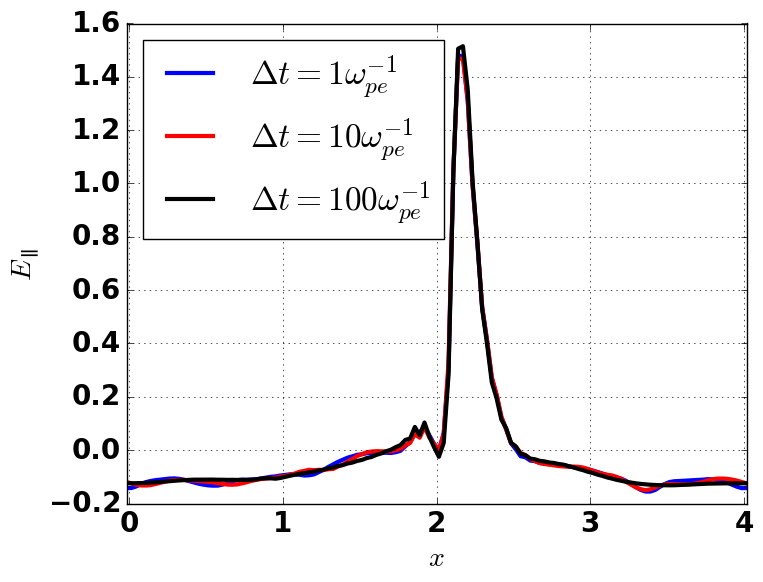}\includegraphics[scale=0.4]{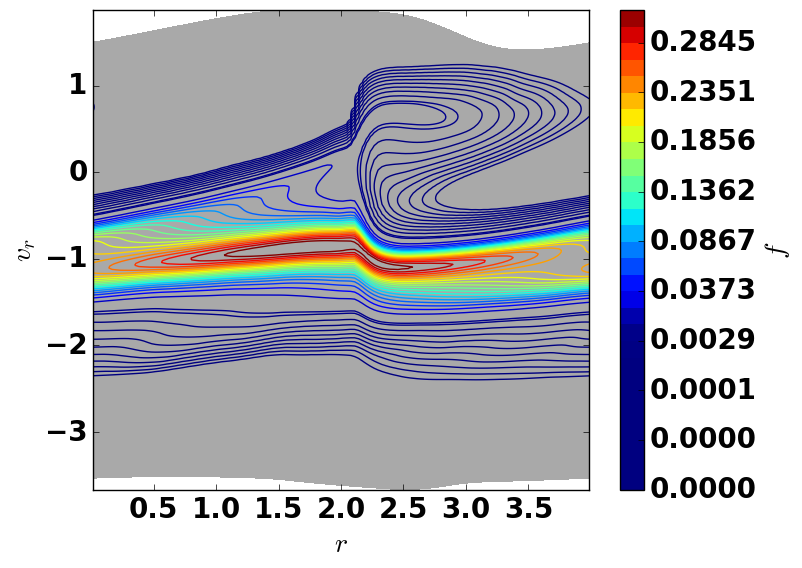}
\par\end{center}%
\end{minipage}\caption{ion-acoustic shock wave solution for the ion number density (top left),
electron number density (top right), and electric field (bottom left),
at $t\approx5000\omega_{pe}^{-1}$ for various time-step sizes. Also
given is a contour plot of the ion distribution, $f_{i}$, (bottom
right) showing the wave-breaking in phase-space and the velocity-space
domain adaptivity.\label{fig:Ion-acoustic-shockwave}}
\end{figure}

For this problem, we normalize particle mass to the proton mass, with
the electron mass $m_{e}=1/1836$. We take the Debye length as $\lambda_{D}=1/36,$
with the system length $L_{x}=144\lambda_{D}$. The problem is initialized
with sinusoidally perturbed ion and electron density profiles
\[
n_{0,i}=1+0.2\mathrm{sin}\left(kx\right),
\]

\[
n_{0,e}=1+0.2\left(1-k^{2}\lambda_{D}^{2}\right)\mathrm{sin}\left(kx\right),
\]
and with the same sinusoidal bulk velocity profiles for both species.
The velocity is chosen such that the simulation proceeds in the frame
of the shock:
\[
u_{\parallel,0}=-1+0.2\mathrm{sin}\left(kx\right).
\]
 The species temperatures are initially $T_{0,i}=0.05$, $T_{0,e}=1$,
with the large temperature ratio chosen to avoid electron Landau damping
\citep{Shay2007}. As a consequence, the ion-acoustic time- and length-scales
of the problem are much longer than the inverse plasma frequency $\omega_{pe}^{-1}$
and the Debye length \citep{Chen2011}. The wave number is $k=2\pi/L_{x}$.
The simulation is performed with a velocity-space domain $c_{\parallel}\in\left[-8,+8\right]$,
$c_{\bot}\in\left[0,5\right]$, on a mesh of $N_{x}=128$, $N_{v_{\parallel}}=256$,
$N_{v_{\bot}}=64$. The initial offset velocity of each species is
set equal to the bulk velocity, $u_{\parallel,0}^{*}=u_{\parallel,0}$,
with the initial normalizing speed equal to $v_{0,i}^{*}=\sqrt{2T_{0,i}}$,
$v_{0,e}^{*}=\sqrt{\frac{2T_{0,e}}{m_{e}}}$. In this problem, the
offset velocity is set to track the quantity $u_{\parallel,\alpha}+\Delta w_{\parallel,\alpha}$,
where the normalized heat flux, $\Delta w_{\parallel,\alpha}=\frac{\left\langle \frac{1}{2}m_{\alpha}\left(v_{\parallel}-u_{\parallel}\right)\left(\bm{v}-\bm{u}\right)^{2},\tilde{f}_{\alpha}\right\rangle _{\bm{c}}}{\frac{3}{2}n_{\alpha}T_{\alpha}}$,
has been included to aid in capturing the significant non-Maxwellian
wave-breaking feature in the ion distribution function at late times
(see Fig. \ref{fig:Ion-acoustic-shockwave}, lower right).

The first set of results consider a varying time-step size: $\Delta t=\left[1,10,100\right]\omega_{pe}^{-1}$.
The timesteps were chosen in such a way that the smallest time-step
is of the order of the stiff time-scale (i.e. the inverse plasma frequency)
while the largest time-step is of the order of the dynamical time-scale
(i.e., approximately the ion acoustic wave CFL: $100\omega_{pe}^{-1}/\Delta t_{\mathrm{CFL}}\sim2$).
For this problem, a relative nonlinear convergence tolerance of $\epsilon_{r}=10^{-6}$
was used. At the largest $\Delta t$, we are stepping over many plasma
periods. However, since this problem is not driven by the physics
on this time-scale, we do not need to resolve it to capture the solution
correctly.
Figure \ref{fig:Ion-acoustic-shockwave} shows spatial profiles for
the number density of ions and electrons, and the electric field at
$t\approx5000\omega_{pe}^{-1}$ for varying time-step sizes. We observe
that the solution quality is not significantly affected even at time-steps
far larger than the inverse electron plasma frequency (which is itself
much larger than the explicit CFL). Table \ref{tab:HOLO-solver-statistics}
shows the solver statistics for the simulations at each time-step
size (obtained by averaging the number of HOLO iterations for each
time-step over the simulation duration), indicating excellent performance
even at large $\Delta t$. Here, we estimate the explicit time-step
size as $\Delta t_{\mathrm{explicit}}=\frac{\Delta x}{v_{\parallel,e,\mathrm{max}}}$.
To demonstrate the capabilities of the velocity-space adaptive scheme,
we also include in Fig. \ref{fig:Ion-acoustic-shockwave} a contour
plot of the perpendicular velocity-integrated unnormalized ion distribution
function, $f_{i,\parallel}\equiv\left(v_{\alpha,i}^{*}\right)^{-1}\int_{v_{\bot,\mathrm{min}}}^{v_{\bot,\mathrm{max}}}\tilde{f}_{i}\left(x,c_{\parallel},c_{\bot}\right)c_{\bot}dc_{\bot}$,
at $t\approx5000\omega_{pe}^{-1}$ for the case with $\Delta t=\omega_{pe}^{-1}$.
The velocity-space boundary adapts to variations in the ion bulk velocity,
$u_{\parallel}$, and thermal speed, $v_{th}=\sqrt{\frac{2T}{m}}$,
as can be clearly seen by the gray background fill of the simulation
$\left(x,v_{\parallel}\right)$ domain. We can also clearly see the
characteristic `wave-breaking' feature of the ion distribution in
the velocity space.
\begin{table}[h]
\begin{centering}
\begin{tabular}{|c|c|c|c|}
\hline 
$\Delta t/\omega_{pe}^{-1}$ & $1$ & $10$ & $100$\tabularnewline
\hline 
\hline 
$\Delta t/\Delta t_{\mathrm{explicit}}$ & $1.01\mathrm{E+1}$ & $1.01\mathrm{E+2}$ & $1.01\mathrm{E+3}$\tabularnewline
\hline 
HOLO iters & $3.5$ & $6.2$ & $11.3$\tabularnewline
\hline 
\end{tabular}\caption{HOLO solver statistics for the ion-acoustic shock wave at various
time-step sizes\label{tab:HOLO-solver-statistics}}
\par\end{centering}
\end{table}

\begin{figure}[h]
\noindent\begin{minipage}[t]{1\columnwidth}%
\begin{center}
\includegraphics[scale=0.4]{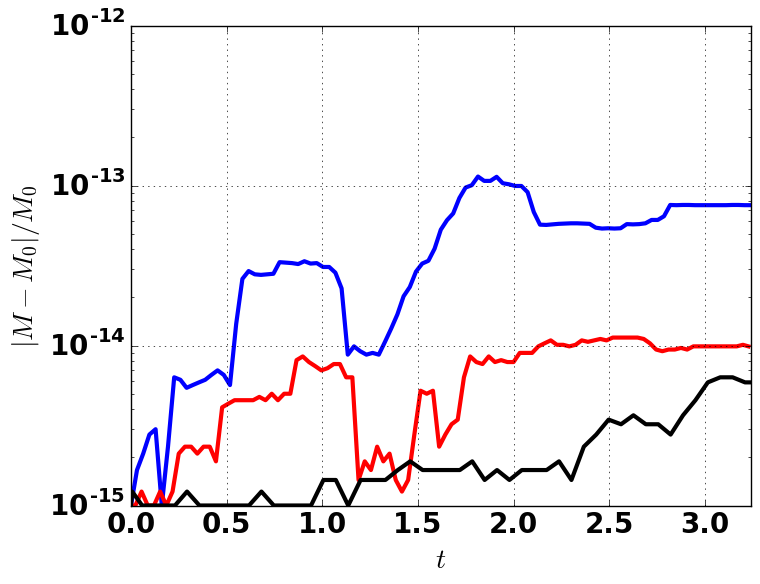}\includegraphics[scale=0.4]{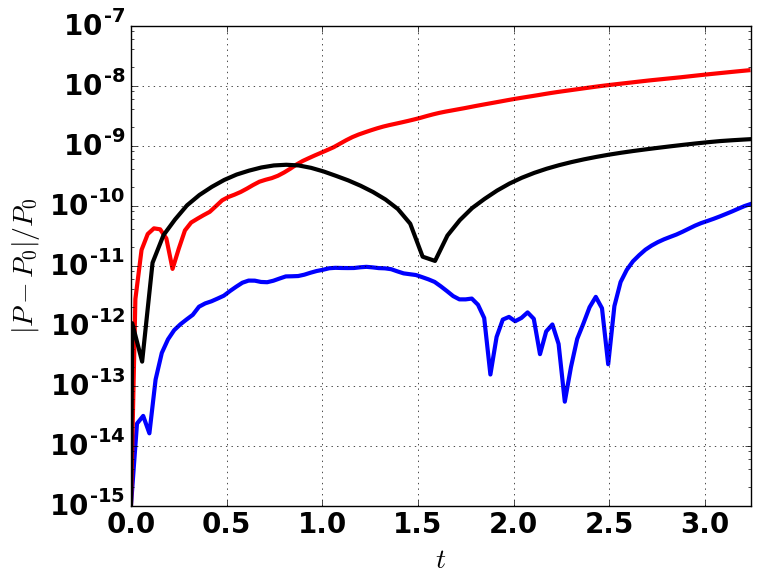}
\par\end{center}%
\end{minipage}\vfill{}
\noindent\begin{minipage}[t]{1\columnwidth}%
\begin{center}
\includegraphics[scale=0.4]{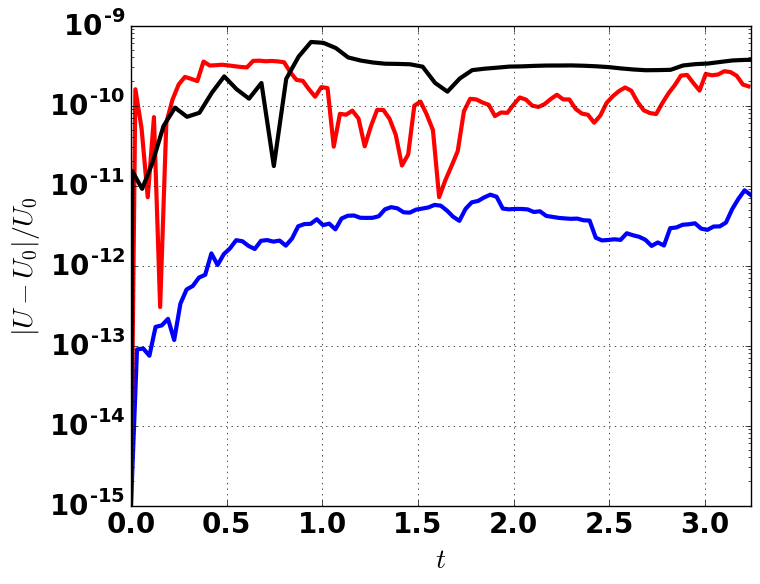}\includegraphics[scale=0.4]{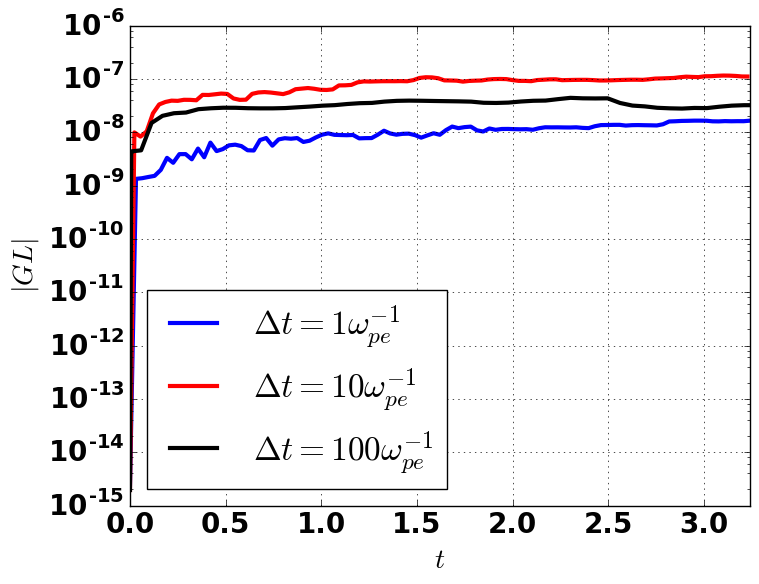}
\par\end{center}%
\end{minipage}\caption{Conservation errors for the ion-acoustic shock wave simulations using
various time-step sizes for the total mass (top left), total momentum
(top right), total energy (bottom left), and charge from Gauss' law
(bottom right).\label{fig:Conservation-properties-for}}
\end{figure}

In Sec. \ref{sec:Discrete-conservation-strategy}, we discussed the
need for enforcing discrete conservation properties. Figure \ref{fig:Conservation-properties-for}
shows the error in mass, momentum, and energy conservation and Gauss's
law for the IASW for various time-step sizes. The error is measured
as the absolute value of the difference in a quantity at a given time
relative to the initial value, 
\begin{equation}
\mathrm{\left|\frac{\Phi^{p}-\Phi^{0}}{\Phi^{0}}\right|},\label{eq:Conservation-error}
\end{equation}
where $\Phi$ is the total mass $\mathrm{TM}$, total momentum $\mathrm{TP}$,
or total energy $\mathrm{TE}$. Additionally, we calculate the $L^{1}$
norm of the error in the discrete form of Gauss's law:
\begin{equation}
\left|GL\right|^{p}=\sum\limits _{i}^{N_{x}}\left|\epsilon_{0}\frac{\left(E_{\parallel,i+\frac{1}{2}}^{p}-E_{\parallel,i-\frac{1}{2}}^{p}\right)}{\Delta x_{i}}-\sum\limits _{\alpha}^{N_{sp}}q_{\alpha}n_{\alpha,i}^{p}\right|.\label{eq:Gauss-error}
\end{equation}
 We see that, in all cases, the conservation error is affected by
the different time-step sizes, but is kept well within acceptable
levels. 

Next, we investigate the effects of the relative nonlinear convergence
tolerance, $\epsilon_{r}$, on the magnitude of the conservation error.
Here, we use a time-step of $\Delta t=\omega_{pe}^{-1}$, and vary
the tolerance: $\epsilon_{r}=\left[10^{-4},10^{-6},10^{-8}\right]$.
\begin{figure}[h]
\noindent\begin{minipage}[t]{1\columnwidth}%
\begin{center}
\includegraphics[scale=0.4]{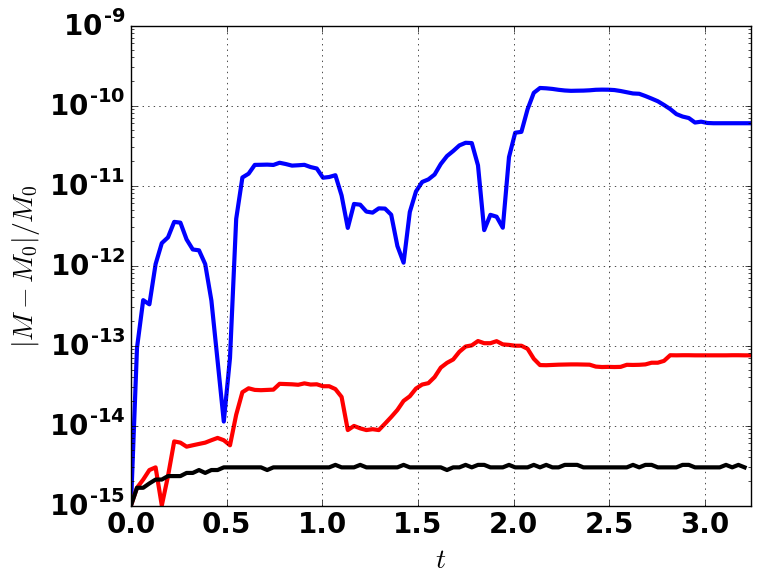}\includegraphics[scale=0.4]{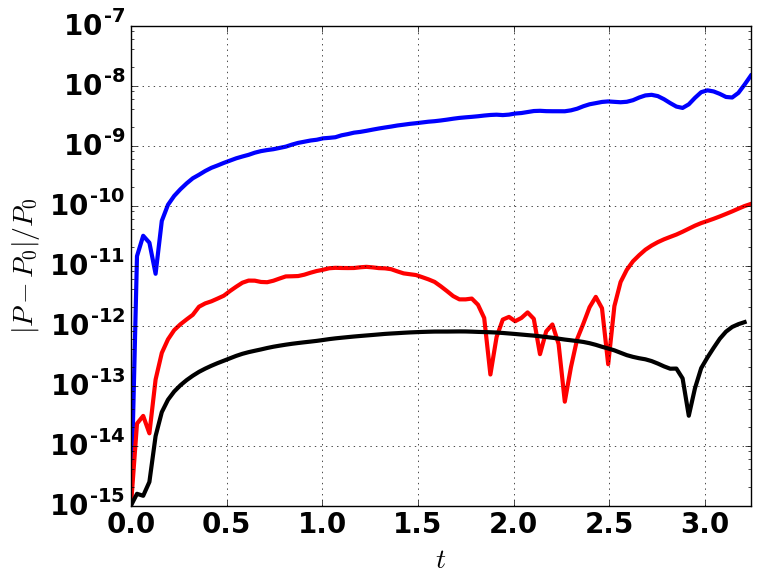}
\par\end{center}%
\end{minipage}\vfill{}
\noindent\begin{minipage}[t]{1\columnwidth}%
\begin{center}
\includegraphics[scale=0.4]{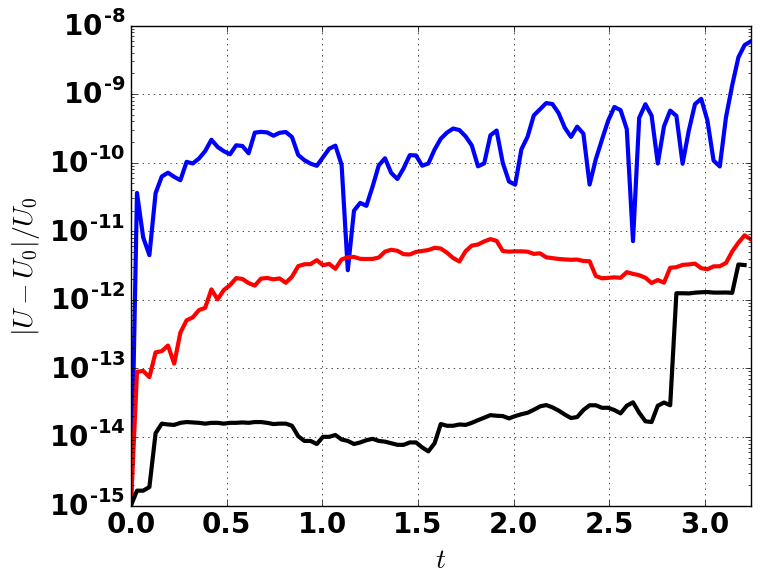}\includegraphics[scale=0.4]{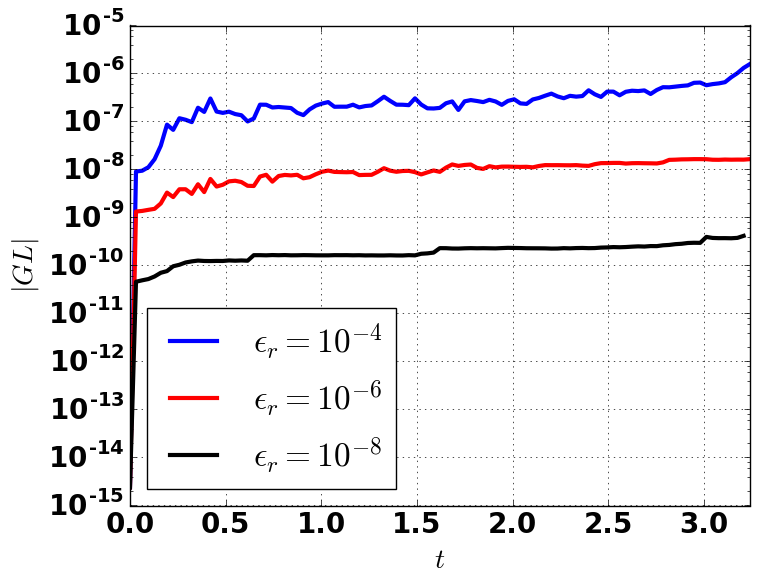}
\par\end{center}%
\end{minipage}\caption{Conservation errors for the ion-acoustic shock wave simulations for
various nonlinear convergence tolerances for the total mass (top left),
total momentum (top right), total energy (bottom left), and charge
from Gauss' law (bottom right).\label{fig:Conservation-properties_eps_r}}
\end{figure}
As we can clearly see in Fig. \ref{fig:Conservation-properties_eps_r},
as the tolerance is tightened the conservation error decreases. Further
decrease in $\epsilon_{r}$ will eventually push the error to machine
roundoff. 

\subsubsection{Importance of discrete conservation}

\begin{figure}[h]
\noindent\begin{minipage}[t]{1\columnwidth}%
\begin{center}
\includegraphics[scale=0.4]{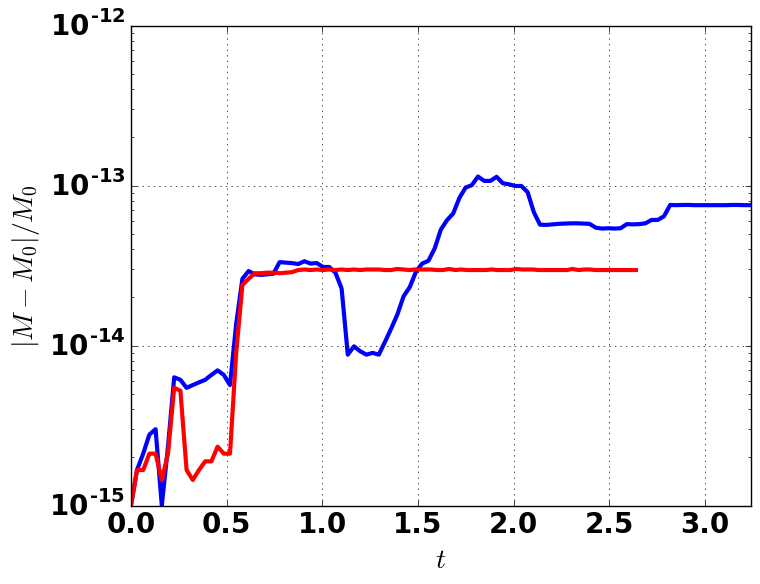}\includegraphics[scale=0.4]{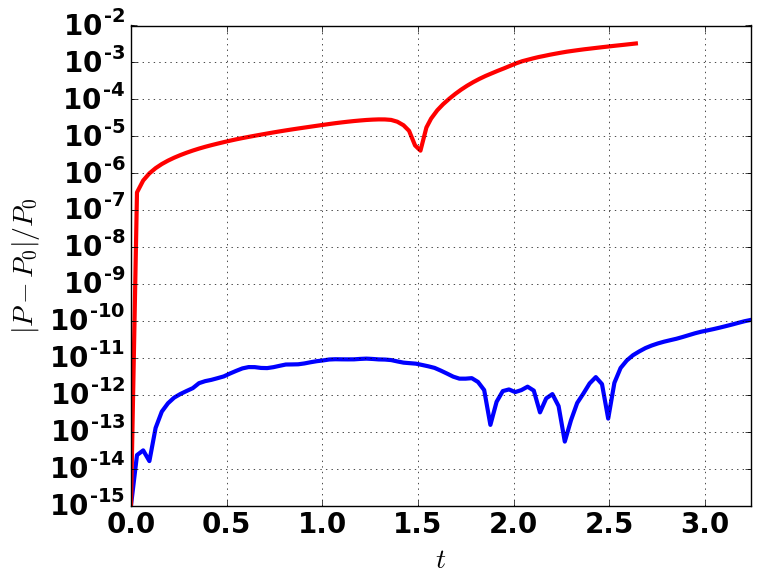}
\par\end{center}%
\end{minipage}\vfill{}
\noindent\begin{minipage}[t]{1\columnwidth}%
\begin{center}
\includegraphics[scale=0.4]{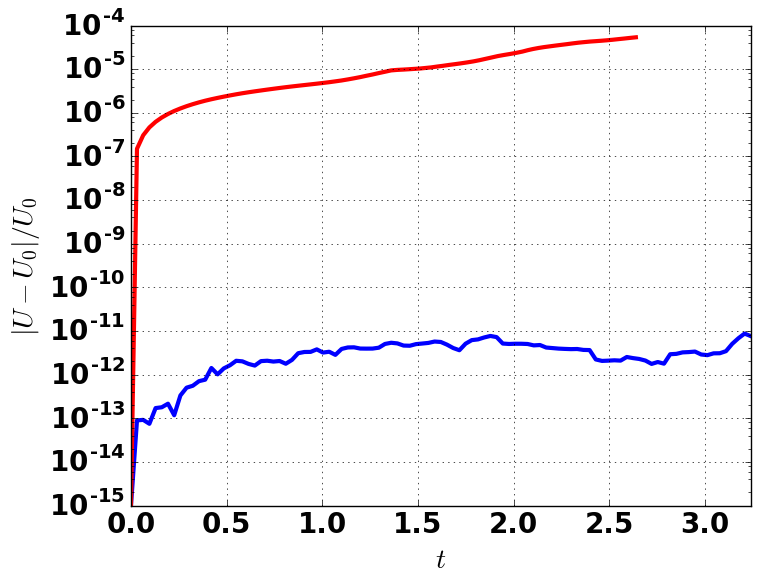}\includegraphics[scale=0.4]{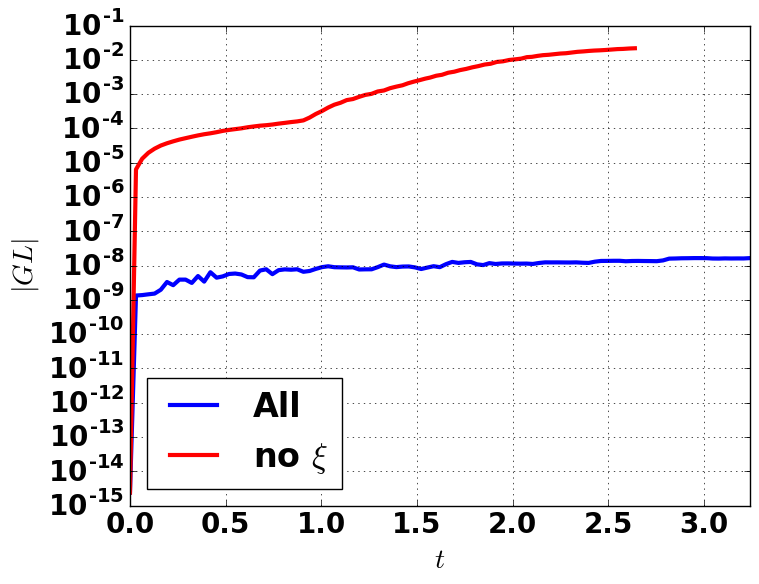}
\par\end{center}%
\end{minipage}\caption{Conservation errors with and without charge conservation for the ion-acoustic
shock wave simulations with for $\Delta t=\omega_{pe}^{-1}$, $\epsilon_{r}=10^{-6}$,
comparing the solution with (``All'', in blue) and without the charge-conserving
nonlinear constraint function $\xi$ active (``no $\xi$'', in red)
for the total mass (top left), total momentum (top right), total energy
(bottom left), and charge from Gauss' law (bottom right).\label{fig:Conservation-properties_cons-noxi}}
\end{figure}

\begin{figure}[h]
\centering{}%
\noindent\begin{minipage}[t]{1\columnwidth}%
\begin{center}
\includegraphics[scale=0.4]{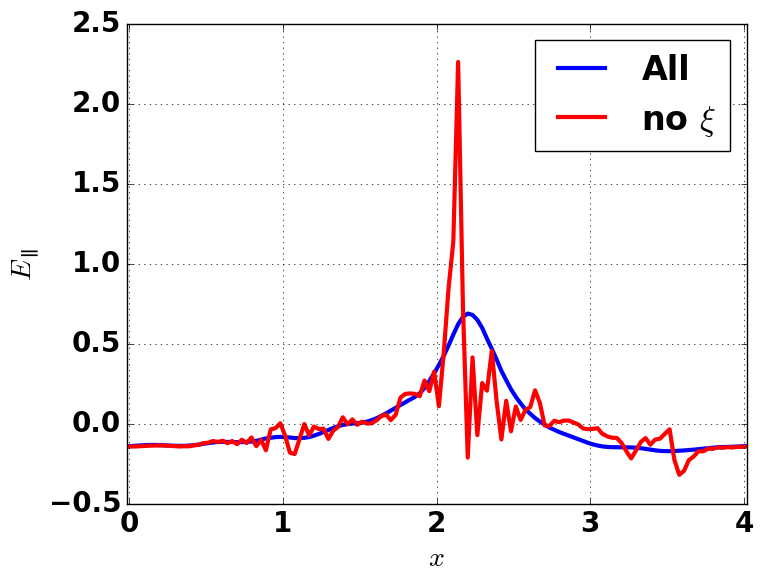}
\par\end{center}%
\end{minipage}\caption{ion-acoustic shock wave solution for electric field $E_{\parallel}$,
at $t\approx3100\omega_{pe}^{-1}$ for $\Delta t=\omega_{pe}^{-1}$
with (``All'', in blue) and without the charge-conserving nonlinear
constraint function $\xi$ active (``no $\xi$'', in red).\label{fig:Ion-acoustic-shockwave_cons-noxi}}
\end{figure}

Here, we present IASW simulation results emphasizing the necessity
of the developments presented in Sec. \ref{sec:Discrete-conservation-strategy}.
As we stated previously, in the Vlasov-Ampère system the most critical
element of the discrete conservation strategy is ensuring that Gauss'
law is satisfied. Here, we compare the case with $\Delta t=\omega_{pe}^{-1}$,
$\epsilon_{r}=10^{-6}$ from Fig. \ref{fig:Conservation-properties-for}
with an identical case without the charge-conserving constraint function
$\xi$ (and its associated pseudo-operator). The result is shown in
Fig. \ref{fig:Conservation-properties_cons-noxi}. With all constraints
\emph{except }$\xi$ active, we see that the charge conservation error
increases monotonically as expected. However, we also observe that
the momentum and energy conservation errors experience a significant
increase. The reason is that violations of charge conservation give
rise to (significant) error in the electric field $E_{\parallel}$
\citep{Mardahl1997}, which results in unphysical acceleration in
the Vlasov equation. This, in turn, results in violations of momentum
and energy conservation. 
Figure \ref{fig:Ion-acoustic-shockwave_cons-noxi} depicts the electric
field $E_{\parallel}$ at $t\approx3100\omega_{pe}^{-1}$. The $E_{\parallel}$
for the simulation lacking charge conservation has at this stage accumulated
significant error. Though not shown here, the solutions for other
moment quantities (e.g., number density, temperature) also show significant
errors. 

\begin{figure}[h]
\noindent\begin{minipage}[t]{1\columnwidth}%
\begin{center}
\includegraphics[scale=0.4]{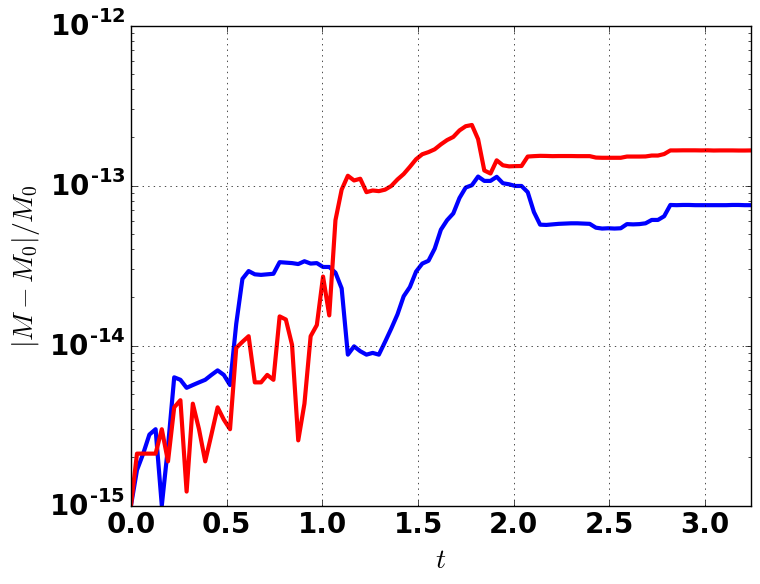}\includegraphics[scale=0.4]{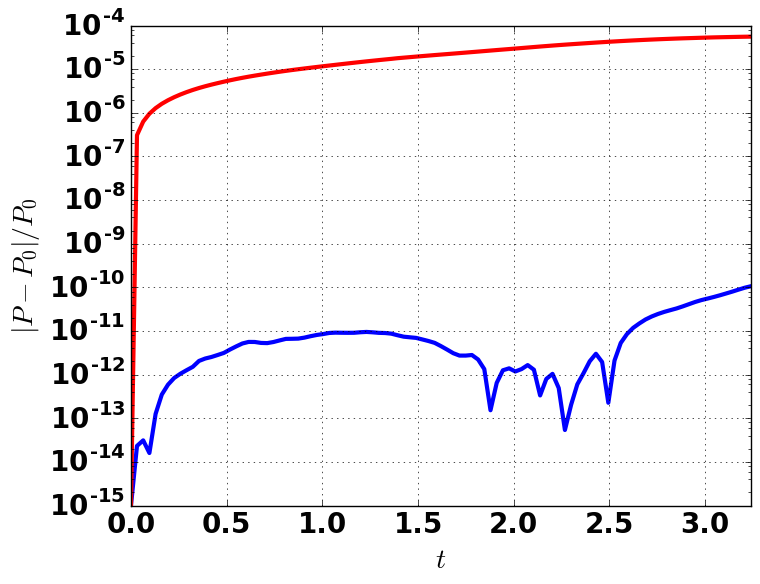}
\par\end{center}%
\end{minipage}\vfill{}
\noindent\begin{minipage}[t]{1\columnwidth}%
\begin{center}
\includegraphics[scale=0.4]{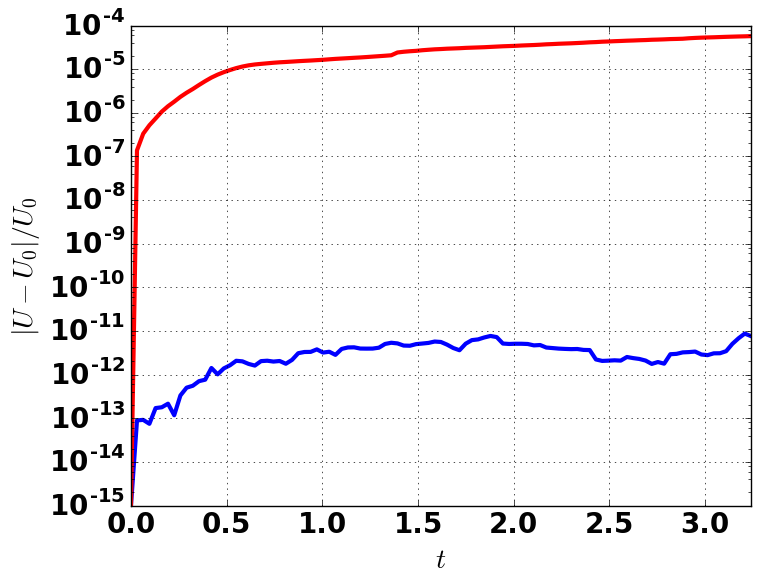}\includegraphics[scale=0.4]{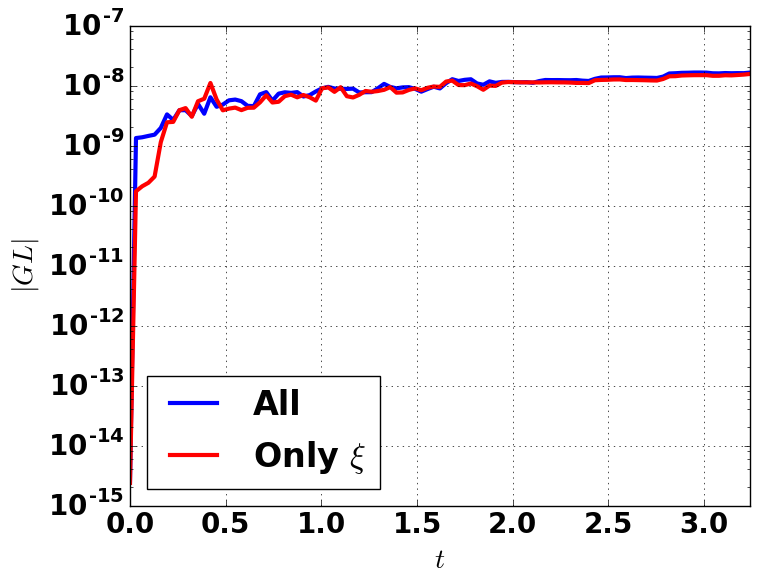}
\par\end{center}%
\end{minipage}\caption{Conservation errors for the ion-acoustic shock wave with all nonlinear
constraint functions active (``All'', in blue) and with only the
charge-conserving constraint function $\xi$ active (``Only $\xi$'',
in red). The total mass (top left), total momentum (top right), total
energy (bottom left), and charge from Gauss' law (bottom right) are
depicted.\label{fig:Conservation-properties_onlyxi}}
\end{figure}

We also investigate the effect of neglecting the momentum- and energy-conserving
constraints (while maintaining the charge conserving constraint $\xi$).
Here, we again compare the case with $\Delta t=\omega_{pe}^{-1}$,
$\epsilon_{r}=10^{-6}$ from Fig. \ref{fig:Conservation-properties-for}
to a case with the same parameters, but which only conserves charge
(thereby failing to conserve momentum and energy). Figure \ref{fig:Conservation-properties_onlyxi}
shows that indeed the simulation fails to conserve momentum and energy
(albeit Gauss's law is still maintained), but without catastrophic
failure. However, based on earlier studies \citep{Taitano2018}, we
do expect catastrophic failure when we couple our Vlasov solver with
the Fokker-Planck collision operator.

\subsubsection{Importance of keeping the discrete averaged current in Ampère's equation}

\begin{figure}[h]
\noindent\begin{minipage}[t]{1\columnwidth}%
\begin{center}
\includegraphics[scale=0.4]{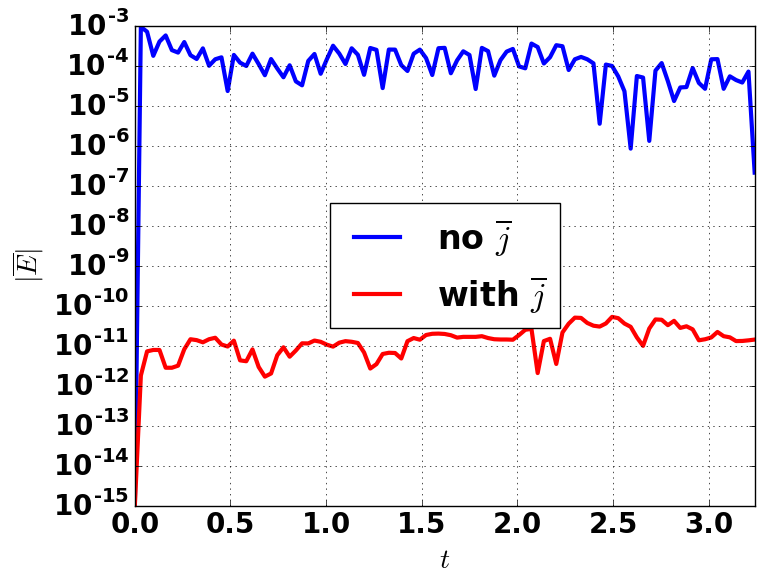}
\par\end{center}%
\end{minipage}\caption{The spatially-averaged electric field $\overline{E}_{\parallel}$
vs. time with (``with $\overline{j}$'' red) and without (``no
$\overline{j}$'', in blue) the spatially-averaged current $\overline{j}_{\parallel}$
included in Ampère's equation.\label{fig:avgE}}
\end{figure}

In Sec. \ref{sec:Vlasov-Ampere-System-of}, we presented Ampère's
equation with the spatially-averaged current, $\overline{j}_{\parallel}$,
which is a necessary solvability constraint to preserve Galilean invariance
in a periodic system. In this work, none of the problems presented
possess an applied electric field, and thus the spatially averaged
electric field, $\overline{E}_{\parallel}$, must be identically zero
at all times. This is clearly seen from the relationship $\boldsymbol{\nabla}_{\boldsymbol{x}}\Phi=-\boldsymbol{E}$,
which relates the electric field to the electrostatic potential. However,
as stated in Sec. \ref{sec:Vlasov-Ampere-System-of}, it is possible
for a finite average current density $\overline{j}_{\parallel}$ to
exist even when $\overline{E}_{\parallel}=0$ due to nonlinear effects
(see \ref{app:Nonlinear-generation-of}). Thus, to ensure $\overline{E}_{\parallel}=0$,
it is necessary to include $\overline{j}_{\parallel}$ in Ampère's
equation. In Fig. \ref{fig:avgE} we compare time traces of the spatially
averaged electric field $\overline{E}_{\parallel}$,
\[
\overline{E}_{\parallel}\equiv\frac{1}{N_{x}}\sum\limits _{i}^{N_{x}}E_{\parallel,i+\frac{1}{2}},
\]
for two simulations with and without $\overline{j}_{\parallel}$ in
Ampère's equation. All other parameters are identical to the $\Delta t=\omega_{pe}^{-1}$
case in Figs. \ref{fig:Ion-acoustic-shockwave} and \ref{fig:Conservation-properties-for}.
We see that there is a significant error increase {[}$\mathcal{O}\left(10^{8}\right)${]}
on the average electric field $\overline{E_{\parallel}}$ when the
average current is not included.

\section{Conclusions\label{sec:Conclusions}}

We have presented a fully conserving, adaptive algorithm for numerically
integrating the 1D-2V multi-species Vlasov-Ampère system. The algorithm
is applicable for the fully kinetic Vlasov system with an arbitrary
number of species of arbitrary mass ratio. The velocity-space adaptivity
scheme allows each species' velocity-space mesh to evolve according
to variations in their bulk velocity and temperature resulting in
very efficient velocity-space meshing. Conservation of the total mass,
momentum, and energy, as well as Gauss's law, are enforced through
the introduction of several nonlinear constraint functions, which
eliminate the truncation error of the conservation properties. We
emphasize that discrete conservation in our algorithm can be achieved
with significant flexibility in temporal and spatial discretizations.
The nonlinear scheme is efficiently accelerated via a HOLO algorithm,
where a LO fluid representation is used to accelerated convergence
of the HO kinetic system. 

The present algorithm is tested with the linear and nonlinear Landau
damping, as well as the two-stream instability and ion-acoustic shock
wave. For the Landau damping and two-stream instability tests, we
achieve excellent agreement with analytical and previously published
growth/decay rates of the electric field energy. For the linear Landau
damping test, we demonstrate that the algorithm achieves second-order
convergence in time, configuration space, and velocity space. For
the ion-acoustic shock wave test, we demonstrate that the algorithm
remains stable when taking time-steps much larger than stiff time-scales
(such as the inverse plasma frequency $\omega_{pe}^{-1}$) without
affecting quality of the solution and while maintaining discrete
conservation. We also demonstrate a commensurate decrease in the discrete
conservation error with decreasing nonlinear convergence tolerance,
in principle allowing us to drive the error to machine precision if
desired. Further, we demonstrate that without enforcing discrete conservation
(particularly for charge conservation), the solution can degrade significantly.

\section*{Acknowledgments}

This work was supported by the Thermonuclear Burn Initiative of the
Advanced Simulation and Computing Program, used resources provided
by the Institutional Computing Program at Los Alamos National Laboratory,
and was performed under the auspices of the National Nuclear Security
Administration of the U.S. Department of Energy at Los Alamos National
Laboratory, managed by Triad National Security, LLC under contract
89233218CNA000001.

%% file: Appendices.tex
\appendix

\section{Nonlinear generation of finite average current density\label{app:Nonlinear-generation-of}}

As presented in Sec. \ref{sec:Vlasov-Ampere-System-of}, in one-dimensional
configuration space the Vlasov-Ampère system of equations becomes
\begin{align*}
\partial_{t}f_{\alpha}+\partial_{x}\left(v_{\parallel}f_{\alpha}\right)+\frac{q_{\alpha}}{m_{\alpha}}E_{\parallel}\partial_{v_{\parallel}}\left(f_{\alpha}\right)=0, & \tag{\ref{eq:Vlasov-1D}}\\
\epsilon_{0}\partial_{t}E_{\parallel}+\sum\limits _{\alpha}q_{\alpha}\Gamma_{\parallel,\alpha}=\overline{j}_{\parallel}. & \tag{\ref{eq:Ampere-1D}}
\end{align*}
We note again the presence of the average current density $\overline{j}_{\parallel}$,
which is necessary in 1D periodic systems. Here $\overline{j}_{\parallel}$
is taken to be the average of the time-dependent current density,
$\overline{j}_{\parallel}\left(t\right)$, rather than the initial
current density, $\overline{j}_{\parallel}\left(t=0\right)$. This
is because a finite average current density may be generated due to
nonlinear effects even in systems where the electric field is a gradient
(i.e. $\overline{E}_{\parallel}=0$). This may be demonstrated as
follows.

The $v_{\parallel}^{1}$ moment of Eq. \ref{eq:Vlasov-1D} will produce
the governing equation for particle flux density $\Gamma_{\parallel,\alpha}$
\begin{equation}
\partial_{t}\Gamma_{\parallel,\alpha}+\partial_{x}S_{\parallel\parallel,\alpha}^{(2)}-\frac{q_{\alpha}}{m_{\alpha}}n_{\alpha}E_{\parallel}=0.\label{eq:momentum}
\end{equation}
Recalling the definition of current density, $j_{\parallel}=\sum_{\alpha}q_{\alpha}\Gamma_{\parallel,\alpha}$,
(\ref{eq:momentum}) becomes the equation for current conservation,
\begin{equation}
\partial_{t}j_{\parallel}+\partial_{x}\left(\sum_{\alpha}q_{\alpha}S_{\parallel\parallel,\alpha}^{(2)}\right)-\sum_{\alpha}\frac{q_{\alpha}^{2}}{m_{\alpha}}n_{\alpha}E_{\parallel}=0.\label{eq:j-cons}
\end{equation}
Taking the spatial average of Eq. (\ref{eq:j-cons}), we obtain
\begin{equation}
\partial_{t}\overline{j}_{\parallel}-\sum_{\alpha}\frac{q_{\alpha}^{2}}{m_{\alpha}}\overline{\left(n_{\alpha}E_{\parallel}\right)}=0.\label{eq:dt_j}
\end{equation}
If we make a substitution for $n_{\alpha}=n_{\alpha}^{*}+\overline{n}_{\alpha}$,
where $\overline{n}_{\alpha}$ is the spatial average of $n_{\alpha}$
(i.e., a constant) and $n_{\alpha}^{*}$ is spatially varying (with
$\overline{n_{\alpha}^{*}}=0$), we see that Eq. (\ref{eq:dt_j})
becomes
\[
\partial_{t}\overline{j}_{\parallel}=\sum_{\alpha}\frac{q_{\alpha}^{2}}{m_{\alpha}}\left[\overline{\left(n_{\alpha}^{*}E_{\parallel}\right)}\right],
\]
which is in general non-zero, even though $\overline{n}_{\alpha}^{*}=0$
and $\overline{E}_{\parallel}=0$. 

\section{Derivation of continuum symmetries for conservation\label{app:Deriviation-of-continuum}}

\subsection{Mass \& charge conservation\label{subsec:Charge-=000026-mass-conservation}}

Mass conservation is demonstrated by taking the $m_{\alpha}v^{0}$
moment of Eq. (\ref{eq:Vlasov-transformed-energy-final}):
\begin{multline}
\left\langle m_{\alpha},\partial_{t}\tilde{f}_{\alpha}\right\rangle _{\bm{c}}+\left\langle m_{\alpha},\partial_{x}\left(v_{\alpha}^{*}\hat{v}_{\parallel}\tilde{f}_{\alpha}\right)\right\rangle _{\bm{c}}+\frac{q_{\alpha}}{m_{\alpha}v_{\alpha}^{*}}E_{\parallel}\left\langle m_{\alpha},\partial_{c_{\parallel}}\tilde{f}_{\alpha}\right\rangle _{\bm{c}}\\
-\frac{1}{v_{\alpha}^{*}}\left\langle m_{\alpha},\nabla_{\bm{c}}\cdot\left\{ \left[\partial_{t}\left(\boldsymbol{v}\right)+\partial_{x}\left(\boldsymbol{v}\right)\hat{v}_{\parallel}v_{\alpha}^{*}\right]\tilde{f}_{\alpha}\right\} \right\rangle _{\bm{c}}=0.\label{eq:vlasov-0th-moment-1}
\end{multline}
Observing that the first two terms produce the mass density $\left\langle m_{\alpha},\tilde{f}_{\alpha}\right\rangle _{\bm{c}}=m_{\alpha}n_{\alpha}$
and the parallel momentum density $\left\langle m_{\alpha},v_{\alpha}^{*}\hat{v}_{\parallel}\tilde{f}_{\alpha}\right\rangle _{\bm{c}}=m_{\alpha}\Gamma_{\parallel,\alpha}$
while the last two terms are zeros, we obtain the mass conservation
equation,
\begin{equation}
m_{\alpha}\left[\partial_{t}n_{\alpha}+\partial_{x}\Gamma_{\parallel,\alpha}\right]=0.\label{eq:mass-cons}
\end{equation}
This is straightforward to satisfy in the discrete with appropriate
boundary conditions on the distribution $f_{\alpha}$.

However, we must also be cognizant of the symmetry between Ampère's
and Gauss's laws and the above statement of mass conservation. If
we take Eq. (\ref{eq:charge-cons-vec}) in one dimension we find 
\begin{equation}
\partial_{t}\left(\sum\limits _{\alpha}\rho_{q,\alpha}\right)+\partial_{x}\left(\sum\limits _{\alpha}j_{\parallel,\alpha}\right)=0,\label{eq:charge-cons}
\end{equation}
 where $\rho_{q,\alpha}\equiv q_{\alpha}n_{\alpha}$ and $j_{\parallel,\alpha}\equiv q_{\alpha}\Gamma_{\parallel,\alpha}$.
Of course, underlying Eq. (\ref{eq:charge-cons}) is simply the summation
over all species of Eq. (\ref{eq:mass-cons}):
\begin{equation}
\partial_{t}\left(\sum\limits _{\alpha}\rho_{q,\alpha}\right)+\partial_{x}\left(\sum\limits _{\alpha}j_{\parallel,\alpha}\right)=\sum\limits _{\alpha}q_{\alpha}\left(\partial_{t}n_{\alpha}+\partial_{x}\Gamma_{\parallel,\alpha}\right).\label{eq:charge-and-mass-cons}
\end{equation}
Thus, we see that the charge density in Gauss's law and the current
in Ampère's equation must be proportional to the particle number density
and flux in Eq. (\ref{eq:mass-cons}).

\subsection{Momentum conservation\label{subsec:Momentum-conservation}}

Momentum conservation is demonstrated by taking the $m_{\alpha}v_{\parallel}$
moment of Eq. (\ref{eq:Vlasov-transformed-energy-final}):

\begin{multline}
\tcboxmath[colback=blue!10!white,colframe=blue]{\left\langle m_{\alpha}v_{\parallel},\partial_{t}\tilde{f}_{\alpha}\right\rangle _{\bm{c}}}+\tcboxmath[colback=red!10!white,colframe=red]{\left\langle m_{\alpha}v_{\parallel},\partial_{x}\left(v_{\alpha}^{*}\hat{v}_{\parallel}\tilde{f}_{\alpha}\right)\right\rangle _{\bm{c}}}+\frac{q_{\alpha}}{m_{\alpha}v_{\alpha}^{*}}E_{\parallel}\left\langle m_{\alpha}v_{\parallel},\partial_{c_{\parallel}}\tilde{f}_{\alpha}\right\rangle _{\bm{c}}\\
-\frac{1}{v_{\alpha}^{*}}\left\langle m_{\alpha}v_{\parallel},\nabla_{\bm{c}}\cdot\left\{ \left[\tcboxmath[colback=blue!10!white,colframe=blue]{\partial_{t}\left(\boldsymbol{v}\right)}+\tcboxmath[colback=red!10!white,colframe=red]{\partial_{x}\left(\boldsymbol{v}\right)}\hat{v}_{\parallel}v_{\alpha}^{*}\right]\tilde{f}_{\alpha}\right\} \right\rangle _{\bm{c}}=0.\label{eq:vlasov-1st-moment-1}
\end{multline}
First, we take note of the terms involving temporal derivatives (boxed
in blue in Eq. (\ref{eq:vlasov-1st-moment-1})) , and apply the chain
rule with integration by parts (once again applying $\lim\limits _{\bm{c}\rightarrow\pm\infty}f_{\alpha}=0$):
\begin{multline}
\left\langle m_{\alpha}v_{\parallel},\partial_{t}\tilde{f}_{\alpha}\right\rangle _{\bm{c}}-\frac{1}{v_{\alpha}^{*}}\left\langle m_{\alpha}v_{\parallel},\nabla_{\bm{c}}\cdot\left\{ \partial_{t}\left(\boldsymbol{v}\right)\tilde{f}_{\alpha}\right\} \right\rangle _{\bm{c}}=\left\langle 1,\partial_{t}\left(m_{\alpha}v_{\parallel}\tilde{f}_{\alpha}\right)\right\rangle _{\bm{c}}-\left\langle m_{\alpha}\tilde{f}_{\alpha},\partial_{t}v_{\parallel}\right\rangle _{\bm{c}}\\
+\frac{1}{v_{\alpha}^{*}}\left\langle 1,m_{\alpha}\partial_{t}\left(\boldsymbol{v}\right)\tilde{f}_{\alpha}\cdot\nabla_{\bm{c}}v_{\parallel}\right\rangle _{\bm{c}}.\label{eq:vlasov-1st-moment-temporal}
\end{multline}
In a similar manner, we may inspect the terms in Eq. (\ref{eq:vlasov-1st-moment-1})
involving spatial derivatives (boxed in red)
\begin{multline}
\left\langle m_{\alpha}v_{\parallel},\partial_{x}\left(v_{\alpha}^{*}\hat{v}_{\parallel}\tilde{f}_{\alpha}\right)\right\rangle _{\bm{c}}-\frac{1}{v_{\alpha}^{*}}\left\langle m_{\alpha}v_{\parallel},\nabla_{\bm{c}}\cdot\left\{ \partial_{x}\left(\boldsymbol{v}\right)\hat{v}_{\parallel}v_{\alpha}^{*}\tilde{f}_{\alpha}\right\} \right\rangle _{\bm{c}}=\left\langle 1,\partial_{x}\left(m_{\alpha}v_{\parallel}^{2}\tilde{f}_{\alpha}\right)\right\rangle _{\bm{c}}-\left\langle m_{\alpha}v_{\parallel}\tilde{f}_{\alpha},\partial_{x}v_{\parallel}\right\rangle _{\bm{c}}\\
+\frac{1}{v_{\alpha}^{*}}\left\langle 1,m_{\alpha}\partial_{x}\left(\boldsymbol{v}\right)\hat{v}_{\parallel}v_{\alpha}^{*}\tilde{f}_{\alpha}\cdot\nabla_{\bm{c}}v_{\parallel}\right\rangle _{\bm{c}}.\label{eq:vlasov-1st-moment-spatial}
\end{multline}
Observing that $\nabla_{\bm{c}}v_{\parallel}=v_{\alpha}^{*}\boldsymbol{e}_{\parallel}$,
we see that the last two terms cancel in both Eq. (\ref{eq:vlasov-1st-moment-temporal})
and Eq. (\ref{eq:vlasov-1st-moment-spatial}). Together these equations
become
\begin{multline}
\left\langle m_{\alpha}v_{\parallel},\partial_{t}\tilde{f}_{\alpha}\right\rangle _{\bm{c}}+\left\langle m_{\alpha}v_{\parallel},\partial_{x}\left(v_{\alpha}^{*}\hat{v}_{\parallel}\tilde{f}_{\alpha}\right)\right\rangle _{\bm{c}}-\frac{1}{v_{\alpha}^{*}}\left\langle m_{\alpha}v_{\parallel},\nabla_{\bm{c}}\cdot\left\{ \left[\partial_{t}\left(\boldsymbol{v}\right)+\partial_{x}\left(\boldsymbol{v}\right)\hat{v}_{\parallel}v_{\alpha}^{*}\right]\tilde{f}_{\alpha}\right\} \right\rangle _{\bm{c}}\\
=\left\langle 1,\partial_{t}\left(m_{\alpha}v_{\parallel}\tilde{f}_{\alpha}\right)\right\rangle _{\bm{c}}+\left\langle 1,\partial_{x}\left(m_{\alpha}v_{\parallel}^{2}\tilde{f}_{\alpha}\right)\right\rangle _{\bm{c}}.\label{eq:vlasov-1st-moment-intermediate}
\end{multline}
If we inspect the acceleration term in Eq. (\ref{eq:vlasov-1st-moment-1})
and observe that $\partial_{c_{\parallel}}v_{\parallel}=v_{\alpha}^{*}$,
we find 
\begin{equation}
\frac{q_{\alpha}}{m_{\alpha}v_{\alpha}^{*}}E_{\parallel}\left\langle m_{\alpha}v_{\parallel},\partial_{c_{\parallel}}\tilde{f}_{\alpha}\right\rangle _{\bm{c}}=-\frac{q_{\alpha}}{m_{\alpha}v_{\alpha}^{*}}E_{\parallel}\left\langle m_{\alpha}\tilde{f}_{\alpha},\partial_{c_{\parallel}}v_{\parallel}\right\rangle _{\bm{c}}=-\frac{q_{\alpha}}{m_{\alpha}}E_{\parallel}\left\langle 1,m_{\alpha}\tilde{f}_{\alpha}\right\rangle _{\bm{c}}.\label{eq:vlasov-1st-moment-accel}
\end{equation}

Thus, if we sum over all species, Eq. (\ref{eq:vlasov-1st-moment-1})
becomes 
\begin{equation}
\sum\limits _{\alpha}\left[\left\langle 1,\partial_{t}\left(m_{\alpha}v_{\parallel}\tilde{f}_{\alpha}\right)\right\rangle _{\bm{c}}+\left\langle 1,\partial_{x}\left(m_{\alpha}v_{\parallel}^{2}\tilde{f}_{\alpha}\right)\right\rangle _{\bm{c}}-\frac{q_{\alpha}}{m_{\alpha}}E_{\parallel}\left\langle 1,m_{\alpha}\tilde{f}_{\alpha}\right\rangle _{\bm{c}}\right]=0.\label{eq:vlasov-1st-moment-2}
\end{equation}
 We may now define $\left\langle 1,m_{\alpha}v_{\parallel}^{2}\tilde{f}_{\alpha}\right\rangle _{\bm{c}}=S_{2,\parallel\parallel,\alpha}$.
If we recall Gauss's law, Eq. (\ref{eq:Gauss}), we may make a substitution
in the acceleration term:
\begin{equation}
\partial_{t}P_{\parallel}+\partial_{x}S_{2,\parallel\parallel}-E_{\parallel}\epsilon_{0}\partial_{x}E_{\parallel}=\partial_{t}P_{\parallel}+\partial_{x}\left[S_{2,\parallel\parallel}-\frac{1}{2}\epsilon_{0}E_{\parallel}^{2}\right]=0,\label{eq:momentum-cons}
\end{equation}
where $P_{\parallel}=\sum\limits _{\alpha}m_{\alpha}\Gamma_{\parallel,\alpha}$
is the total (parallel) momentum density, $S_{2,\parallel\parallel}$
is the total (fluid) stress and $\frac{1}{2}\epsilon_{0}E_{\parallel}^{2}$
is the electrostatic stress. Equation (\ref{eq:momentum-cons}) is
a succinct statement of total momentum conservation \textendash{}
when we integrate over a periodic domain $\frac{1}{L}\int_{0}^{L}dx$
to obtain the total system momentum we obtain 
\[
\partial_{t}\overline{P}_{\parallel}=0,
\]
where $\overline{P}_{\parallel}=\frac{1}{L}\int_{0}^{L}P_{\parallel}dx$.
The key symmetries here are 1) the equivalences in Eqs. (\ref{eq:vlasov-1st-moment-temporal})
and (\ref{eq:vlasov-1st-moment-spatial}), and 2) the equivalence
in Eq. (\ref{eq:vlasov-1st-moment-accel}) of the density arising
from the acceleration term to that which appears in the temporal term
\emph{\textendash{} }i.e., through Gauss's law in Eq. (\ref{eq:charge-cons}).
In general, none of these symmetries are guaranteed in the discrete
system. Indeed, we see that the second symmetry here involving Gauss's
law and the acceleration operator may directly contradict the charge-conservation
requirement leading to Eq. (\ref{eq:charge-cons}).

\subsection{Energy conservation\label{subsec:Energy-conservation}}

Energy conservation is demonstrated by taking the $m_{\alpha}\frac{\boldsymbol{v}^{2}}{2}$
moment of Eq. (\ref{eq:Vlasov-transformed-energy-final}):

\begin{multline}
\left\langle m_{\alpha}\frac{1}{2}\boldsymbol{v}^{2},\partial_{t}\tilde{f}_{\alpha}\right\rangle _{\bm{c}}+\left\langle m_{\alpha}\frac{1}{2}\boldsymbol{v}^{2},\partial_{x}\left(v_{\alpha}^{*}\hat{v}_{\parallel}\tilde{f}_{\alpha}\right)\right\rangle _{\bm{c}}+\frac{q_{\alpha}}{m_{\alpha}v_{\alpha}^{*}}E_{\parallel}\left\langle m_{\alpha}\frac{1}{2}\boldsymbol{v}^{2},\partial_{c_{\parallel}}\tilde{f}_{\alpha}\right\rangle _{\bm{c}}\\
-\frac{1}{v_{\alpha}^{*}}\left\langle m_{\alpha}\frac{1}{2}\boldsymbol{v}^{2},\nabla_{\bm{c}}\cdot\left\{ \left[\partial_{t}\left(\boldsymbol{v}\right)+\partial_{x}\left(\boldsymbol{v}\right)\hat{v}_{\parallel}v_{\alpha}^{*}\right]\tilde{f}_{\alpha}\right\} \right\rangle _{\bm{c}}=0.\label{eq:vlasov-2nd-moment-1}
\end{multline}
As with momentum conservation, we investigate the terms with temporal
and spatial derivatives separately in order to expose their respective
conservation symmetries. Integrating by parts and assuming no boundary
contributions gives, for the temporal terms:
\begin{multline}
\left\langle m_{\alpha}\frac{1}{2}\boldsymbol{v}^{2},\partial_{t}\tilde{f}_{\alpha}\right\rangle _{\bm{c}}-\frac{1}{v_{\alpha}^{*}}\left\langle m_{\alpha}\frac{1}{2}\boldsymbol{v}^{2},\nabla_{\bm{c}}\cdot\left\{ \partial_{t}\left(\boldsymbol{v}\right)\tilde{f}_{\alpha}\right\} \right\rangle _{\bm{c}}=\\
\left\langle 1,\partial_{t}\left(m_{\alpha}\frac{1}{2}\boldsymbol{v}^{2}\tilde{f}_{\alpha}\right)\right\rangle _{\bm{c}}-\left\langle m_{\alpha}\tilde{f}_{\alpha},\partial_{t}\left(\frac{1}{2}\boldsymbol{v}^{2}\right)\right\rangle _{\bm{c}}+\frac{1}{v_{\alpha}^{*}}\left\langle 1,m_{\alpha}\partial_{t}\left(\boldsymbol{v}\right)\tilde{f}_{\alpha}\cdot\nabla_{\bm{c}}\left(\frac{1}{2}\boldsymbol{v}^{2}\right)\right\rangle _{\bm{c}},\label{eq:vlasov-2nd-moment-temporal}
\end{multline}
and for the spatial terms:
\begin{multline}
\left\langle m_{\alpha}\frac{1}{2}\boldsymbol{v}^{2},\partial_{x}\left(v_{\parallel}\tilde{f}_{\alpha}\right)\right\rangle _{\bm{c}}-\frac{1}{v_{\alpha}^{*}}\left\langle m_{\alpha}\frac{1}{2}\boldsymbol{v}^{2},\nabla_{\bm{c}}\cdot\left\{ \partial_{x}\left(\boldsymbol{v}\right)v_{\parallel}\tilde{f}_{\alpha}\right\} \right\rangle _{\bm{c}}=\\
\left\langle 1,\partial_{x}\left(m_{\alpha}\frac{1}{2}\boldsymbol{v}^{2}v_{\parallel}\tilde{f}_{\alpha}\right)\right\rangle _{\bm{c}}-\left\langle m_{\alpha}v_{\parallel}\tilde{f}_{\alpha},\partial_{x}\left(\frac{1}{2}\boldsymbol{v}^{2}\right)\right\rangle _{\bm{c}}+\frac{1}{v_{\alpha}^{*}}\left\langle 1,m_{\alpha}\partial_{x}\left(\boldsymbol{v}\right)v_{\parallel}\tilde{f}_{\alpha}\cdot\nabla_{\bm{c}}\left(\frac{1}{2}\boldsymbol{v}^{2}\right)\right\rangle _{\bm{c}}.\label{eq:vlasov-2nd-moment-spatial}
\end{multline}
Here, we observe that $\nabla_{\bm{c}}\left(\frac{1}{2}\boldsymbol{v}^{2}\right)=\boldsymbol{v}v_{\alpha}^{*}$,
$\partial_{x}\left(\frac{1}{2}\boldsymbol{v}^{2}\right)=\boldsymbol{v}\cdot\partial_{x}\boldsymbol{v}$,
and $\partial_{t}\left(\frac{1}{2}\boldsymbol{v}^{2}\right)=\boldsymbol{v}\cdot\partial_{t}\boldsymbol{v}$.
Thus, the last two terms on the right-hand sides of Eqs. (\ref{eq:vlasov-2nd-moment-temporal})
and (\ref{eq:vlasov-2nd-moment-spatial}) cancel, and together these
equations become
\begin{multline}
\left\langle m_{\alpha}\frac{1}{2}\boldsymbol{v}^{2},\partial_{t}\tilde{f}_{\alpha}\right\rangle _{\bm{c}}+\left\langle m_{\alpha}\frac{1}{2}\boldsymbol{v}^{2},\partial_{x}\left(v_{\parallel}\tilde{f}_{\alpha}\right)\right\rangle _{\bm{c}}-\frac{1}{v_{\alpha}^{*}}\left\langle m_{\alpha}\frac{1}{2}\boldsymbol{v}^{2},\nabla_{\bm{c}}\cdot\left\{ \left[\partial_{t}\left(\boldsymbol{v}\right)+\partial_{x}\left(\boldsymbol{v}\right)v_{\parallel}\right]\tilde{f}_{\alpha}\right\} \right\rangle _{\bm{c}}\\
=\left\langle 1,\partial_{t}\left(m_{\alpha}\frac{1}{2}\boldsymbol{v}^{2}\tilde{f}_{\alpha}\right)\right\rangle _{\bm{c}}+\left\langle 1,\partial_{x}\left(m_{\alpha}\frac{1}{2}\boldsymbol{v}^{2}v_{\parallel}\tilde{f}_{\alpha}\right)\right\rangle _{\bm{c}}.\label{eq:vlasov-2nd-moment-intermediate}
\end{multline}
Returning to the acceleration term in Eq. (\ref{eq:vlasov-2nd-moment-1})
and integrating by parts gives
\begin{equation}
\frac{q_{\alpha}}{m_{\alpha}v_{\alpha}^{*}}E_{\parallel}\left\langle m_{\alpha}\frac{1}{2}\boldsymbol{v}^{2},\partial_{c_{\parallel}}\tilde{f}_{\alpha}\right\rangle _{\bm{c}}=-\frac{q_{\alpha}}{m_{\alpha}v_{\alpha}^{*}}E_{\parallel}\left\langle m_{\alpha}\tilde{f}_{\alpha},\partial_{c_{\parallel}}\left(\frac{1}{2}\boldsymbol{v}^{2}\right)\right\rangle _{\bm{c}}=-\frac{q_{\alpha}}{m_{\alpha}}E_{\parallel}\left\langle 1,m_{\alpha}v_{\parallel}\tilde{f}_{\alpha}\right\rangle _{\bm{c}},\label{eq:vlasov-2nd-moment-accel}
\end{equation}
where we utilized the relationship $\partial_{c_{\parallel}}\left(\frac{1}{2}\boldsymbol{v}^{2}\right)=v_{\parallel}v_{\alpha}^{*}$. 

Combining the preceding results and summing over all species $\alpha$,
Eq. (\ref{eq:vlasov-2nd-moment-1}) becomes
\begin{equation}
\sum\limits _{\alpha}\left[\left\langle 1,\partial_{t}\left(m_{\alpha}\frac{1}{2}\boldsymbol{v}^{2}\tilde{f}_{\alpha}\right)\right\rangle _{\bm{c}}+\left\langle 1,\partial_{x}m_{\alpha}\frac{1}{2}\boldsymbol{v}^{2}v_{\parallel}\tilde{f}_{\alpha}\right\rangle _{\bm{c}}-\frac{q_{\alpha}}{m_{\alpha}}E_{\parallel}\left\langle 1,m_{\alpha}v_{\parallel}\tilde{f}_{\alpha}\right\rangle _{\bm{c}}\right]=0.\label{eq:vlasov-2nd-moment-2}
\end{equation}
We note the definitions $\left\langle 1,m_{\alpha}\frac{1}{2}\boldsymbol{v}^{2}\tilde{f}_{\alpha}\right\rangle _{\bm{c}}=\varepsilon_{\alpha}$
and $\left\langle 1,m_{\alpha}\frac{1}{2}\boldsymbol{v}^{2}v_{\parallel}\tilde{f}_{\alpha}\right\rangle _{\bm{c}}=S_{3,\parallel,\alpha}$
and, recalling Ampère's equation, we introduce it in the acceleration
term to find
\begin{equation}
\partial_{t}U+\partial_{x}S_{3,\parallel}+E_{\parallel}\epsilon_{0}\partial_{t}E_{\parallel}=\partial_{t}\left[U+\frac{1}{2}\epsilon_{0}E_{\parallel}^{2}\right]+\partial_{x}S_{3,\parallel}-E_{\parallel}\overline{j}_{\parallel}=0.\label{eq:energy-cons}
\end{equation}
Here, $U$ is the total (fluid) energy density, $\frac{1}{2}\epsilon_{0}E_{\parallel}^{2}$
is the electrostatic energy density, and $S_{3,\parallel}$ is the
total energy flux. Equation (\ref{eq:energy-cons}) expresses conservation
of the total energy density of the system. Integrating over the periodic
domain gives the total energy conservation: 
\[
\partial_{t}\overline{U}_{tot}-\overline{E}_{\parallel}\overline{j}_{\parallel}=\partial_{t}\overline{U}_{tot}=0.
\]
The $\overline{E}_{\parallel}\overline{j}_{\parallel}$ term vanishes
in a periodic system with no external electric field. Again, the symmetry
of the moments of the temporal/spatial inertial terms is a key point.
We also note that, in the discrete, the requirements for energy conservation
with the inertial terms are not guaranteed to be compatible with those
for momentum conservation. Further, we again observe that the second
moment of the acceleration term must correspond to the current in
Ampère's equation, which, as we saw in \ref{subsec:Charge-=000026-mass-conservation}
must also correspond to the zeroth moment of the advective flux. This
once again presents apparently conflicting requirements for discrete
conservation.

\section{Detailed definitions of discrete fluxes\label{app:Detailed-definitions-of}}

As we saw in section \ref{sec:Numerical-Implementation}, the transformed
Vlasov equation is discretized conservatively as

\begin{multline}
\delta_{t}\tilde{f}_{\alpha,i,j,k}+\underbrace{\delta_{x}\left[v_{\parallel,\alpha,j}^{p}\overline{\left(\tilde{f}_{\alpha}^{p+1}\right)}_{j,k}^{v_{\parallel}}\right]_{i}}_{(a)}+\underbrace{\frac{q_{\alpha}}{m_{\alpha}}\frac{E_{\parallel,i}^{p+1}}{v_{\alpha,i}^{*,p}}\delta_{c_{\parallel}}\left[\overline{\left(\tilde{f}_{\alpha}^{p+1}\right)}_{i,k}^{q_{\alpha}E_{\parallel}}\right]_{j}}_{(b)}\\
+\underbrace{\tcboxmath[colback=blue!10!white,colframe=blue]{\delta_{x}\left[\xi_{\alpha}^{p+1}\left|v_{\parallel,\alpha,j}^{p}\right|\overline{\left(\tilde{f}_{\alpha}^{p+1}\right)}_{j,k}^{\xi}\right]_{i}}}_{(c)}+\underbrace{\tcboxmath[colback=blue!10!white,colframe=blue]{\delta_{c_{\parallel}}\left[\phi_{\alpha,i}^{p+1}\overline{\left(\tilde{f}_{\alpha}^{p+1}\right)}_{i,k}^{\phi}\right]_{j}}}_{(d)}+\underbrace{\tcboxmath[colback=blue!10!white,colframe=blue]{\delta_{c_{\parallel}}\left[\gamma_{q,\alpha,i}^{p+1}\overline{\left(\tilde{f}_{\alpha}^{p+1}\right)}_{i,k}^{\gamma_{q}}\right]_{j}}}_{(e)}\\
\underbrace{-\frac{1}{v_{\alpha,i}^{*,p}}\delta_{\bm{c}}\cdot\left[\tcboxmath[colback=red!10!white,colframe=red]{\gamma_{t,\alpha,i}^{p+1}}\delta_{t}\left(\boldsymbol{v}_{\alpha,i}\right)\overline{\left(\tilde{f}_{\alpha}^{p+1}\right)}_{i}^{\delta_{t}\left(\boldsymbol{v}\right)}\right]_{j,k}}_{(f)}\\
\underbrace{-\frac{1}{2v_{\alpha,i}^{*,p}}\delta_{\bm{c}}\cdot\left[\tcboxmath[colback=red!10!white,colframe=red]{\gamma_{x,\alpha,i+\frac{1}{2}}^{p+1}}v_{\alpha,i+\frac{1}{2}}^{*,p}\tcboxmath[colback=blue!10!white,colframe=blue]{\hat{v}_{\parallel,\alpha,\mathrm{eff},i+\frac{1}{2}}^{p+1}}\delta_{x}\left[\boldsymbol{v}_{\alpha}^{p}\right]_{i+\frac{1}{2}}\overline{\left(\tilde{f}_{\alpha}^{p+1}\right)}_{i}^{v_{\parallel,\mathrm{eff}}\delta_{x}\left(\boldsymbol{v}\right)}\right]_{j,k}}_{(g,1)}\\
\underbrace{-\frac{1}{2v_{\alpha,i}^{*,p}}\delta_{\bm{c}}\cdot\left[\tcboxmath[colback=red!10!white,colframe=red]{\gamma_{x,\alpha,i-\frac{1}{2}}^{p+1}}v_{\alpha,i-\frac{1}{2}}^{*,p}\tcboxmath[colback=blue!10!white,colframe=blue]{\hat{v}_{\parallel,\alpha,\mathrm{eff},i-\frac{1}{2}}^{p+1}}\delta_{x}\left[\boldsymbol{v}_{\alpha}^{p}\right]_{i-\frac{1}{2}}\overline{\left(\tilde{f}_{\alpha}^{p+1}\right)}_{i}^{v_{\parallel,\mathrm{eff}}\delta_{x}\left(\boldsymbol{v}\right)}\right]_{j,k}}_{(g,2)}=0.\label{eq:Vlasov-disc-simp-app}
\end{multline}
The various fluxes in Eq. (\ref{eq:Vlasov-disc-simp-app}) are defined
as follows. The physical configuration-space advection, $(a)$, is
defined as
\begin{equation}
\left[v_{\parallel,\alpha,j}^{p}\overline{\left(\tilde{f}_{\alpha}^{p+1}\right)}_{j,k}^{v_{\parallel}}\right]_{i+\frac{1}{2}}=v_{\alpha,i+\frac{1}{2}}^{*,p}\left(c_{\parallel,j}+\hat{u}_{\parallel,\alpha,i+\frac{1}{2}}^{*,p}\right)\mathrm{Interp}\left(c_{\parallel,j}+\hat{u}_{\parallel,\alpha,i+\frac{1}{2}}^{*,p},\tilde{f}_{\alpha}^{p+1}\right)_{i+\frac{1}{2},j,k}.\label{eq:discrete-advection-physical}
\end{equation}
Term $(b)$ \textendash{} the velocity-space advection operator due
to electric field acceleration \textendash{} is defined as
\begin{equation}
\left[\overline{\left(\tilde{f}_{\alpha}^{p+1}\right)}_{i,k}^{q_{\alpha}E_{\parallel}}\right]_{j+\frac{1}{2}}=\frac{q_{\alpha}}{m_{\alpha}}\frac{E_{\parallel,i}^{p+1}}{v_{\alpha,i}^{*,p}}\mathrm{Interp}\left(q_{\alpha}E_{\parallel,i}^{p+1},\tilde{f}_{\alpha}^{p+1}\right)_{i,j+\frac{1}{2},k}.\label{eq:discrete-acceleration-physical}
\end{equation}
Terms $(c)$, $(d)$, and $(e)$ are the `pseudo-operators' introduced
by the inclusion of nonlinear constraint functions, $\xi_{\alpha}$,
$\phi_{\alpha}$, and $\gamma_{q,\alpha}$, which act to enforce the
conservation symmetries discussed in Sec. \ref{subsec:Continuum-conservation-symmetrie}
(see Ref. \citep{Taitano2015a}). Here, we will only discuss their
discrete appearance in the numerical implementation of the governing
equation. The nature and definitions of these constraint functions
and their respective pseudo-operators are discussed \ref{app:Derivation-of-constraint}
and Sec. \ref{sec:Discrete-conservation-strategy}. Term $(c)$ is
the pseudo-advection operator arising due to the discrete nonlinear
constraint function $\xi_{\alpha}$ and is defined to be
\begin{equation}
\left[\xi_{\alpha}^{p+1}\left|v_{\parallel,\alpha,j}^{p}\right|\overline{\left(\tilde{f}_{\alpha}^{p+1}\right)}_{j,k}^{\xi}\right]_{i+\frac{1}{2}}=v_{\alpha,i+\frac{1}{2}}^{*,p}\xi_{\alpha,i+\frac{1}{2}}^{p+1}\left|c_{\parallel,j}+\hat{u}_{\parallel,\alpha,i+\frac{1}{2}}^{*,p}\right|\mathrm{Upw}\left(\xi_{\alpha,i+\frac{1}{2}}^{p+1},\tilde{f}_{\alpha}^{p+1}\right)_{i+\frac{1}{2},j,k},\label{eq:discrete-advection-xi}
\end{equation}
where $\mathrm{Upw}$ denotes the use of straightforward upwinding
based on the sign of $\xi_{\alpha,i+\frac{1}{2}}^{p+1}$. The discretization
of terms $(d)$ and $(e)$ is given as
\begin{equation}
\left[\phi_{\alpha,i}^{p+1}\overline{\left(\tilde{f}_{\alpha}^{p+1}\right)}_{i,k}^{\phi}\right]_{j+\frac{1}{2}}=\phi_{\alpha,i,j+\frac{1}{2}}^{p+1}\frac{1}{v_{\alpha,i}^{*,p}}\frac{\tilde{f}_{\alpha,i,j-\frac{1}{2},k}^{p+1}+\tilde{f}_{\alpha,i,j+\frac{1}{2},k}^{p+1}}{2},\label{eq:discrete-acceleration-phi}
\end{equation}
\begin{equation}
\left[\gamma_{q,\alpha,i}^{p+1}\overline{\left(\tilde{f}_{\alpha}^{p+1}\right)}_{i,k}^{\gamma_{q}}\right]_{j+\frac{1}{2}}=\gamma_{q,\alpha,i,j+\frac{1}{2}}^{p+1}\frac{1}{v_{\alpha,i}^{*,p}}\frac{\tilde{f}_{\alpha,i,j-\frac{1}{2},k}^{p+1}+\tilde{f}_{\alpha,i,j+\frac{1}{2},k}^{p+1}}{2},\label{eq:discrete-acceleration-gammaq}
\end{equation}
with a straighforward central differencing of $\tilde{f}_{\alpha,i,j,k}^{p+1}$
in $c_{\parallel}$. Recall that $\phi_{\alpha}$ and $\gamma_{q,\alpha}$
have a dependence on the parallel velocity space; the details of this
dependence are given in Sec. \ref{sec:Discrete-conservation-strategy}. 

The inertial terms $(f)$ and $(g)$ arise due to the velocity-coordinate
transformation. Similar to terms $(c)$, $(d)$, and $(e)$, they
contain additional nonlinear constraint functions $\gamma_{t}$ and
$\gamma_{x}$, which also act so as to enforce the continuum conservation
symmetries discussed previously. The specific definitions and action
of $\gamma_{t,\alpha}$ and $\gamma_{x,\alpha}$ are discussed in
more detail in Sec. \ref{sec:Discrete-conservation-strategy}, \ref{app:Details-on-the},
and Ref. \citep{Taitano2018a}. The parallel-velocity flux of the
temporal inertial term, $(f)$, is defined as
\begin{multline}
-\left[\gamma_{t,\alpha,i}^{p+1}\delta_{t}\left(v_{\parallel,\alpha,i}^{p}\right)\overline{\left(\tilde{f}_{\alpha}^{p+1}\right)}_{i}^{\delta_{t}\left(\boldsymbol{v}\right)}\right]_{j+\frac{1}{2},k}=\\
-\gamma_{t,\alpha,i,j+\frac{1}{2},k}^{p+1}\frac{1}{v_{\alpha,i}^{*,p}}\delta_{t}\left(v_{\parallel,\alpha,i,j+\frac{1}{2}}^{p}\right)\mathrm{\mathrm{Interp}}\left(-\delta_{t}\left(v_{\parallel,\alpha,i,j+\frac{1}{2}}^{p}\right),\tilde{f}_{\alpha}^{p+1}\right)_{i,j+\frac{1}{2},k}.\label{eq:discrete-vsamr-t-parallel}
\end{multline}
 The perpendicular-velocity flux of the temporal inertial term is
defined similarly:
\begin{multline}
-\left[\gamma_{t,\alpha,i}^{p+1}\delta_{t}\left(v_{\bot,\alpha,i}^{p}\right)\overline{\left(\tilde{f}_{\alpha}^{p+1}\right)}_{i}^{\delta_{t}\left(\boldsymbol{v}\right)}\right]_{j,k+\frac{1}{2}}=\\
-\gamma_{t,\alpha,i,j,k+\frac{1}{2}}^{p+1}\frac{1}{v_{\alpha,i}^{*,p}}\delta_{t}\left(v_{\bot,\alpha,i,k+\frac{1}{2}}^{p}\right)\mathrm{\mathrm{Interp}}\left(-\delta_{t}\left(v_{\bot,\alpha,i,k+\frac{1}{2}}^{p}\right),\tilde{f}_{\alpha}^{p+1}\right)_{i,j,k+\frac{1}{2}}.\label{eq:discrete-vsamr-t-perpendicular}
\end{multline}

For the spatial inertial terms, $(g)$, the parallel-velocity flux
is defined by
\begin{multline}
-\left[\gamma_{x,\alpha,i+\frac{1}{2}}^{p+1}v_{\alpha,i+\frac{1}{2}}^{*,p}\hat{v}_{\parallel,\alpha,\mathrm{eff},i+\frac{1}{2}}^{p+1}\delta_{x}\left[v_{\parallel,\alpha}^{p}\right]_{i+\frac{1}{2}}\overline{\left(\tilde{f}_{\alpha}^{p+1}\right)}_{i}^{v_{\parallel,\mathrm{eff}}\delta_{x}\left(\boldsymbol{v}\right)}\right]_{j+\frac{1}{2},k}=\\
-\gamma_{x,\alpha,i+\frac{1}{2},j+\frac{1}{2},k}^{p+1}v_{\alpha,i+\frac{1}{2}}^{*,p}\hat{v}_{\parallel,\alpha,\mathrm{eff},i+\frac{1}{2},j+\frac{1}{2}}^{p+1}\delta_{x}\left[v_{\parallel,\alpha,j+\frac{1}{2}}^{p}\right]_{i+\frac{1}{2}}\times\\
\mathrm{\mathrm{Interp}}\left(-\hat{v}_{\parallel,\alpha,\mathrm{eff},i+\frac{1}{2},j+\frac{1}{2}}^{p+1}\delta_{x}\left[v_{\parallel,\alpha,j+\frac{1}{2}}^{p}\right]_{i+\frac{1}{2}},\tilde{f}_{\alpha}^{p+1}\right)_{i,j+\frac{1}{2},k}.\label{eq:discrete-vsamr-x-parallel}
\end{multline}
Here, we note the pseudo-flux involving the nonlinear constraint function
$\xi_{\alpha,i+\frac{1}{2}}$ appears through $v_{\parallel,\alpha,\mathrm{eff},i+\frac{1}{2},j+\frac{1}{2}}$.
Similarly, the perpendicular-velocity flux is defined by
\begin{multline}
-\left[\gamma_{x,\alpha,i+\frac{1}{2}}^{p+1}v_{\alpha,i+\frac{1}{2}}^{*,p}\hat{v}_{\parallel,\alpha,\mathrm{eff},i+\frac{1}{2}}^{p+1}\delta_{x}\left[v_{\bot,\alpha}^{p}\right]_{i+\frac{1}{2}}\overline{\left(\tilde{f}_{\alpha}^{p+1}\right)}_{i}^{v_{\parallel,\mathrm{eff}}\delta_{x}\left(\boldsymbol{v}\right)}\right]_{j,k+\frac{1}{2}}=\\
-\gamma_{x,\alpha,i+\frac{1}{2},j,k+\frac{1}{2}}^{p+1}v_{\alpha,i+\frac{1}{2}}^{*,p}\hat{v}_{\parallel,\alpha,\mathrm{eff},i+\frac{1}{2},j}^{p+1}\delta_{x}\left[v_{\bot,\alpha,k+\frac{1}{2}}^{p}\right]_{i+\frac{1}{2}}\times\\
\mathrm{\mathrm{Interp}}\left(-\hat{v}_{\parallel,\alpha,\mathrm{eff},i+\frac{1}{2},j}^{p+1}\delta_{x}\left[v_{\bot,\alpha,k+\frac{1}{2}}^{p}\right]_{i+\frac{1}{2}},\tilde{f}_{\alpha}^{p+1}\right)_{i,j,k+\frac{1}{2}}.\label{eq:discrete-vsamr-x-perpendicular}
\end{multline}
We note here that to evolve the normalizing speed $v_{\alpha}^{*}$
and offset velocity $u_{\parallel,\alpha}^{*}$ in space and time
we use the same strategies as in Ref\emph{.} \citep{Taitano2018a}.

\section{Derivation of constraint definitions for $\gamma_{t}$ and $\gamma_{x}$\label{app:Details-on-the}}

\subsection{Discrete momentum conservation}

First, we observe that Eqs. (\ref{eq:vlasov-1st-moment-temporal})
and (\ref{eq:vlasov-1st-moment-spatial}) may be discretely represented
as
\begin{multline}
\left\langle v_{\parallel,\alpha,i,j}^{p},\delta_{t}\tilde{f}_{\alpha,i,j,k}\right\rangle _{\delta\bm{c}}-\left\langle v_{\parallel,\alpha,i,j}^{p},\frac{1}{v_{\alpha,i}^{*,p}}\delta_{\bm{c}}\cdot\left[\gamma_{t,\alpha,i}^{p+1}\delta_{t}\left(\boldsymbol{v}_{\alpha,i}\right)\overline{\left(\tilde{f}_{\alpha}^{p+1}\right)}_{i}^{\delta_{t}\left(\boldsymbol{v}\right)}\right]_{j,k}\right\rangle _{\delta\bm{c}}\\
=\left\langle 1,\frac{c^{p+1}v_{\parallel,\alpha,i,j}^{p}\tilde{f}_{\alpha,i,j,k}^{p+1}+c^{p}v_{\parallel,\alpha,i,j}^{p-1}\tilde{f}_{\alpha,i,j,k}^{p}+c^{p-1}v_{\parallel,\alpha,i,j}^{p-2}\tilde{f}_{\alpha,i,j,k}^{p-1}}{\Delta t^{p}}\right\rangle _{\delta\bm{c}},\label{eq:discrete-1st-moment-temporal-1}
\end{multline}
and
\begin{multline}
\left\langle v_{\parallel,\alpha,i,j}^{p},\delta_{x}\left[v_{\alpha}^{*,p}\hat{v}_{\parallel,\mathrm{eff},\alpha,j}^{p+1}\overline{\left(\tilde{f}_{\alpha}^{p+1}\right)}_{j,k}^{v_{\parallel,\mathrm{eff}}}\right]_{i}\right\rangle _{\delta\bm{c}}\\
-\left\langle v_{\parallel,\alpha,i,j}^{p},\frac{1}{2v_{\alpha,i}^{*,p}}\delta_{\bm{c}}\cdot\left[\gamma_{x,\alpha,i+\frac{1}{2}}^{p+1}v_{\alpha,i+\frac{1}{2}}^{*,p}\hat{v}_{\parallel,\alpha,\mathrm{eff},i+\frac{1}{2}}^{p+1}\delta_{x}\left[\boldsymbol{v}_{\alpha}^{p}\right]_{i+\frac{1}{2}}\overline{\left(\tilde{f}_{\alpha}^{p+1}\right)}_{i}^{v_{\parallel,\mathrm{eff}}\delta_{x}\left(\boldsymbol{v}\right)}\right]_{j,k}\right\rangle _{\delta\bm{c}}\\
-\left\langle v_{\parallel,\alpha,i,j}^{p},\frac{1}{2v_{\alpha,i}^{*,p}}\delta_{\bm{c}}\cdot\left[\gamma_{x,\alpha,i-\frac{1}{2}}^{p+1}v_{\alpha,i-\frac{1}{2}}^{*,p}\hat{v}_{\parallel,\alpha,\mathrm{eff},i-\frac{1}{2}}^{p+1}\delta_{x}\left[\boldsymbol{v}_{\alpha}^{p}\right]_{i-\frac{1}{2}}\overline{\left(\tilde{f}_{\alpha}^{p+1}\right)}_{i}^{v_{\parallel,\mathrm{eff}}\delta_{x}\left(\boldsymbol{v}\right)}\right]_{j,k}\right\rangle _{\delta\bm{c}}\\
=\left\langle 1,\delta_{x}\left[\left(v_{\parallel,\alpha,j}^{p}\right)^{2}\overline{\left(\tilde{f}_{\alpha}^{p+1}\right)}_{j,k}^{v_{\parallel}}\right]_{i}\right\rangle _{\delta\bm{c}}+\left\langle 1,\delta_{x}\left[v_{\parallel,\alpha,j}^{p}\xi_{\alpha}^{p+1}\left|v_{\parallel,\alpha,j}^{p}\right|\overline{\left(\tilde{f}_{\alpha}^{p+1}\right)}_{j,k}^{\xi}\right]_{i}\right\rangle _{\delta\bm{c}},\label{eq:discrete-1st-moment-spatial-1}
\end{multline}
respectively.

If we expand the discretized form of Eq. (\ref{eq:discrete-1st-moment-temporal-1}),
we find
\begin{multline}
\left\langle v_{\parallel,\alpha,i,j}^{p},\frac{c^{p+1}\tilde{f}_{\alpha,i,j,k}^{p+1}+c^{p}\tilde{f}_{\alpha,i,j,k}^{p}+c^{p-1}\tilde{f}_{\alpha,i,j,k}^{p-1}}{\Delta t^{p}}\right\rangle _{\delta\bm{c}}\\
-\left\langle 1,\frac{c^{p+1}v_{\parallel,\alpha,i,j}^{p}\tilde{f}_{\alpha,i,j,k}^{p+1}+c^{p}v_{\parallel,\alpha,i,j}^{p-1}\tilde{f}_{\alpha,i,j,k}^{p}+c^{p-1}v_{\parallel,\alpha,i,j}^{p-2}\tilde{f}_{\alpha,i,j,k}^{p-1}}{\Delta t^{p}}\right\rangle _{\delta\bm{c}}\\
-\left\langle v_{\parallel,\alpha,i,j}^{p},\frac{1}{v_{\alpha,i}^{*,p}}\delta_{\bm{c}}\cdot\left[\gamma_{t,\alpha,i}^{p+1}\delta_{t}\left(\boldsymbol{v}_{\alpha,i}\right)\overline{\left(\tilde{f}_{\alpha}^{p+1}\right)}_{i}^{\delta_{t}\left(\boldsymbol{v}\right)}\right]_{j,k}\right\rangle _{\delta\bm{c}}=0,\label{eq:gamma_t-discrete-mom-app}
\end{multline}
which is a concise representation of the first discrete constraint
on the definition of the nonlinear constraint function $\gamma_{t}$.

To enforce Eq. (\ref{eq:discrete-1st-moment-spatial-1}), we observe
that by integrating through configuration-space (i.e., sum over $\sum\limits _{i}^{N_{x}}\Delta x$),
the right-hand side of Eq. (\ref{eq:discrete-1st-moment-spatial-1})
vanishes with periodic boundaries. Thus, expanding the individual
flux terms, we find
\begin{multline}
\sum\limits _{i}^{N_{x}}\Delta x\Bigg\{\left\langle v_{\parallel,\alpha,i,j}^{p},\frac{v_{\alpha,i+\frac{1}{2}}^{*,p}\hat{v}_{\parallel,\mathrm{eff},\alpha,i+\frac{1}{2},j}^{p+1}\overline{\left(\tilde{f}_{\alpha}^{p+1}\right)}_{i+\frac{1}{2},j,k}^{\hat{v}_{\parallel,\mathrm{eff}}}-v_{\alpha,i-\frac{1}{2}}^{*,p}\hat{v}_{\parallel,\mathrm{eff},\alpha,i-\frac{1}{2},j}^{p+1}\overline{\left(\tilde{f}_{\alpha}^{p+1}\right)}_{i-\frac{1}{2},j,k}^{\hat{v}_{\parallel,\mathrm{eff}}}}{\Delta x}\right\rangle _{\delta\bm{c}}\\
-\left\langle v_{\parallel,\alpha,i,j}^{p},\frac{1}{2v_{\alpha,i}^{*,p}}\delta_{\bm{c}}\cdot\left[\gamma_{x,\alpha,i+\frac{1}{2}}^{p+1}v_{\alpha,i+\frac{1}{2}}^{*,p}\hat{v}_{\parallel,\mathrm{eff},\alpha,i+\frac{1}{2}}^{p+1}\frac{\boldsymbol{v}_{\alpha,i+1}^{p}-\boldsymbol{v}_{\alpha,i}^{p}}{\Delta x}\overline{\left(\tilde{f}_{\alpha}^{p+1}\right)}_{i}^{v_{\parallel,\mathrm{eff}}\delta_{x}\left(\boldsymbol{v}\right)}\right]_{j,k}\right\rangle _{\delta\bm{c}}\\
-\left\langle v_{\parallel,\alpha,i,j}^{p},\frac{1}{2v_{\alpha,i}^{*,p}}\delta_{\bm{c}}\cdot\left[\gamma_{x,\alpha,i-\frac{1}{2}}^{p+1}v_{\alpha,i-\frac{1}{2}}^{*,p}\hat{v}_{\parallel,\mathrm{eff},\alpha,i-\frac{1}{2}}^{p+1}\frac{\boldsymbol{v}_{\alpha,i}^{p}-\boldsymbol{v}_{\alpha,i-1}^{p}}{\Delta x}\overline{\left(\tilde{f}_{\alpha}^{p+1}\right)}_{i}^{v_{\parallel,\mathrm{eff}}\delta_{x}\left(\boldsymbol{v}\right)}\right]_{j,k}\right\rangle _{\delta\bm{c}}\Bigg\}=0.\label{eq:discrete-1st-moment-spatial-3}
\end{multline}
Recall we have defined the effective velocity
\[
\hat{v}_{\parallel,\mathrm{eff},i+\frac{1}{2},j}^{p+1}\equiv\left(c_{\parallel,j}+\hat{u}_{\parallel,\alpha,i+\frac{1}{2}}^{*,p}+\xi_{\alpha,i+\frac{1}{2}}^{p+1}\left|c_{\parallel,j}+\hat{u}_{\parallel,\alpha,i+\frac{1}{2}}^{*,p}\right|\right).
\]
Note that in the case of $\hat{v}_{\parallel,\mathrm{eff},\alpha,i+\frac{1}{2},j}^{p+1}\overline{\left(\tilde{f}_{\alpha}^{p+1}\right)}_{i+\frac{1}{2},j,k}^{\hat{v}_{\parallel,\mathrm{eff}}}$
in the first term of Eq. (\ref{eq:discrete-1st-moment-spatial-3}),
this is simply a shorthand for the summation of the \emph{individually
interpolated }fluxes $\left[v_{\parallel,\alpha,j}^{p}\overline{\left(\tilde{f}_{\alpha}^{p+1}\right)}_{j,k}^{v_{\parallel}}\right]_{i+\frac{1}{2}}$
and $\left[\xi_{\alpha}^{p+1}\left|v_{\parallel,\alpha,j}^{p}\right|\overline{\left(\tilde{f}_{\alpha}^{p+1}\right)}_{j,k}^{\xi}\right]_{i+\frac{1}{2}}$.
If we then telescope the summation in Eq. (\ref{eq:discrete-1st-moment-spatial-3})
in configuration-space, we find
\begin{multline}
\sum\limits _{i}^{N_{x}}\Bigg\{\left\langle v_{\parallel,\alpha,i,j}^{p}-v_{\parallel,\alpha,i+1,j}^{p},\frac{1}{\Delta x}v_{\alpha,i+\frac{1}{2}}^{*,p}\hat{v}_{\parallel,\mathrm{eff},\alpha,i+\frac{1}{2},j}^{p+1}\overline{\left(\tilde{f}_{\alpha}^{p+1}\right)}_{i+\frac{1}{2},j,k}^{\hat{v}_{\parallel,\mathrm{eff}}}\right\rangle _{\delta\bm{c}}\\
-\left\langle v_{\parallel,\alpha,i,j}^{p},\frac{1}{2v_{\alpha,i}^{*,p}}\delta_{\bm{c}}\cdot\left[\gamma_{x,\alpha,i+\frac{1}{2}}^{p+1}v_{\alpha,i+\frac{1}{2}}^{*,p}\hat{v}_{\parallel,\mathrm{eff},i+\frac{1}{2}}\delta_{x}\left(\boldsymbol{v}^{p}\right)_{i+\frac{1}{2}}\overline{\left(\tilde{f}_{\alpha}^{p+1}\right)}_{i}^{v_{\parallel,\mathrm{eff}}\delta_{x}\left(\boldsymbol{v}\right)}\right]_{j,k}\right\rangle _{\delta\bm{c}}\\
-\left\langle v_{\parallel,\alpha,i+1,j}^{p},\frac{1}{2v_{\alpha,i+1}^{*,p}}\delta_{\bm{c}}\cdot\left[\gamma_{x,\alpha,i+\frac{1}{2}}^{p+1}v_{\alpha,i+\frac{1}{2}}^{*,p}\hat{v}_{\parallel,\mathrm{eff},i+\frac{1}{2}}\delta_{x}\left(\boldsymbol{v}^{p}\right)_{i+\frac{1}{2}}\overline{\left(\tilde{f}_{\alpha}^{p+1}\right)}_{i+1}^{v_{\parallel,\mathrm{eff}}\delta_{x}\left(\boldsymbol{v}\right)}\right]_{j,k}\right\rangle _{\delta\bm{c}}\Bigg\}=0,\label{eq:discrete-1st-moment-spatial-4}
\end{multline}
where the discrete constraint on $\gamma_{x,\alpha}$ is found by
enforcing that 
\begin{multline}
\left\langle v_{\parallel,\alpha,i,j}^{p}-v_{\parallel,\alpha,i+1,j}^{p},\frac{1}{\Delta x}v_{\alpha,i+\frac{1}{2}}^{*,p}\hat{v}_{\parallel,\mathrm{eff},\alpha,i+\frac{1}{2},j}^{p+1}\overline{\left(\tilde{f}_{\alpha}^{p+1}\right)}_{i+\frac{1}{2},j,k}^{\hat{v}_{\parallel,\mathrm{eff}}}\right\rangle _{\delta\bm{c}}\\
-\left\langle v_{\parallel,\alpha,i,j}^{p},\frac{1}{2v_{\alpha,i}^{*,p}}\delta_{\bm{c}}\cdot\left[\gamma_{x,\alpha,i+\frac{1}{2}}^{p+1}v_{\alpha,i+\frac{1}{2}}^{*,p}\hat{v}_{\parallel,\mathrm{eff},i+\frac{1}{2}}\delta_{x}\left(\boldsymbol{v}^{p}\right)_{i+\frac{1}{2}}\overline{\left(\tilde{f}_{\alpha}^{p+1}\right)}_{i}^{v_{\parallel,\mathrm{eff}}\delta_{x}\left(\boldsymbol{v}\right)}\right]_{j,k}\right\rangle _{\delta\bm{c}}\\
-\left\langle v_{\parallel,\alpha,i+1,j}^{p},\frac{1}{2v_{\alpha,i+1}^{*,p}}\delta_{\bm{c}}\cdot\left[\gamma_{x,\alpha,i+\frac{1}{2}}^{p+1}v_{\alpha,i+\frac{1}{2}}^{*,p}\hat{v}_{\parallel,\mathrm{eff},i+\frac{1}{2}}\delta_{x}\left(\boldsymbol{v}^{p}\right)_{i+\frac{1}{2}}\overline{\left(\tilde{f}_{\alpha}^{p+1}\right)}_{i+1}^{v_{\parallel,\mathrm{eff}}\delta_{x}\left(\boldsymbol{v}\right)}\right]_{j,k}\right\rangle _{\delta\bm{c}}=0\label{eq:discrete-1st-moment-spatial-5}
\end{multline}
is zero for each cell-face $i+\frac{1}{2}$.

\subsection{Discrete energy conservation}

We first observe that Eqs. (\ref{eq:vlasov-2nd-moment-temporal})
and (\ref{eq:vlasov-2nd-moment-spatial}) may be discretely represented
as
\begin{multline}
\left\langle m_{\alpha}\frac{1}{2}\left(\boldsymbol{v}_{\alpha,i,j,k}^{p}\right)^{2},\delta_{t}\tilde{f}_{\alpha,i,j,k}\right\rangle _{\delta\bm{c}}-\left\langle m_{\alpha}\frac{1}{2}\left(\boldsymbol{v}_{\alpha,i,j,k}^{p}\right)^{2},\frac{1}{v_{\alpha,i}^{*,p}}\delta_{\bm{c}}\cdot\left[\gamma_{t,\alpha,i}^{p+1}\delta_{t}\left(\boldsymbol{v}_{\alpha,i}\right)\overline{\left(\tilde{f}_{\alpha}^{p+1}\right)}_{i}^{\delta_{t}\left(\boldsymbol{v}\right)}\right]_{j,k}\right\rangle _{\delta\bm{c}}\\
=\left\langle 1,m_{\alpha}\frac{1}{2}\frac{c^{p+1}\left(\boldsymbol{v}_{\alpha,i,j,k}^{p}\right)^{2}\tilde{f}_{\alpha,i,j,k}^{p+1}+c^{p}\left(\boldsymbol{v}_{\alpha,i,j,k}^{p-1}\right)^{2}\tilde{f}_{\alpha,i,j,k}^{p}+c^{p-1}\left(\boldsymbol{v}_{\alpha,i,j,k}^{p-2}\right)^{2}\tilde{f}_{\alpha,i,j,k}^{p-1}}{\Delta t^{p}}\right\rangle _{\delta\bm{c}},\label{eq:discrete-2nd-moment-temporal-1}
\end{multline}
and
\begin{multline}
\left\langle m_{\alpha}\frac{1}{2}\left(\boldsymbol{v}_{\alpha,i,j,k}^{p}\right)^{2},\delta_{x}\left[v_{\parallel,\alpha,j}^{p}\overline{\left(\tilde{f}_{\alpha}^{p+1}\right)}_{j,k}^{v_{\parallel}}\right]_{i}\right\rangle _{\delta\bm{c}}+\left\langle m_{\alpha}\frac{1}{2}\left(\boldsymbol{v}_{\alpha,i,j,k}^{p}\right)^{2},\delta_{x}\left[\xi_{\alpha}^{p+1}\left|v_{\parallel,\alpha,j}^{p}\right|\overline{\left(\tilde{f}_{\alpha}^{p+1}\right)}_{j,k}^{\xi}\right]_{i}\right\rangle _{\delta\bm{c}}\\
+\left\langle m_{\alpha}\frac{1}{2}\left(\boldsymbol{v}_{\alpha,i,j,k}^{p}\right)^{2},\frac{J_{x,\parallel,\alpha,i,j+\frac{1}{2},k}^{p+1}-J_{x,\parallel,\alpha,i,j-\frac{1}{2},k}^{p+1}}{\Delta c_{\parallel}}\right\rangle _{\delta\bm{c}}\\
-\left\langle m_{\alpha}\frac{1}{2}\left(\boldsymbol{v}_{\alpha,i,j,k}^{p}\right)^{2},\frac{1}{2v_{\alpha,i}^{*,p}}\delta_{\bm{c}}\cdot\left[\gamma_{x,\alpha,i+\frac{1}{2}}^{p+1}v_{\alpha,i+\frac{1}{2}}^{*,p}\hat{v}_{\parallel,\alpha,\mathrm{eff},i+\frac{1}{2}}^{p+1}\delta_{x}\left[\boldsymbol{v}_{\alpha}^{p}\right]_{i+\frac{1}{2}}\overline{\left(\tilde{f}_{\alpha}^{p+1}\right)}_{i}^{v_{\parallel,\mathrm{eff}}\delta_{x}\left(\boldsymbol{v}\right)}\right]_{j,k}\right\rangle _{\delta\bm{c}}\\
-\left\langle m_{\alpha}\frac{1}{2}\left(\boldsymbol{v}_{\alpha,i,j,k}^{p}\right)^{2},\frac{1}{2v_{\alpha,i}^{*,p}}\delta_{\bm{c}}\cdot\left[\gamma_{x,\alpha,i-\frac{1}{2}}^{p+1}v_{\alpha,i-\frac{1}{2}}^{*,p}\hat{v}_{\parallel,\alpha,\mathrm{eff},i-\frac{1}{2}}^{p+1}\delta_{x}\left[\boldsymbol{v}_{\alpha}^{p}\right]_{i-\frac{1}{2}}\overline{\left(\tilde{f}_{\alpha}^{p+1}\right)}_{i}^{v_{\parallel,\mathrm{eff}}\delta_{x}\left(\boldsymbol{v}\right)}\right]_{j,k}\right\rangle _{\delta\bm{c}}\\
=\left\langle 1,\delta_{x}\left[m_{\alpha}\frac{1}{2}\left(\boldsymbol{v}_{\alpha,j,k}^{p}\right)^{2}v_{\parallel,\alpha,j}^{p}\overline{\left(\tilde{f}_{\alpha}^{p+1}\right)}_{j,k}^{v_{\parallel}}\right]_{i}\right\rangle _{\delta\bm{c}}\\
+\left\langle 1,\delta_{x}\left[m_{\alpha}\frac{1}{2}\left(\boldsymbol{v}_{\alpha,j,k}^{p}\right)^{2}\xi_{\alpha}^{p+1}\left|v_{\parallel,\alpha,j}^{p}\right|\overline{\left(\tilde{f}_{\alpha}^{p+1}\right)}_{j,k}^{\xi}\right]_{i}\right\rangle _{\delta\bm{c}}.\label{eq:discrete-2nd-moment-spatial-1}
\end{multline}
If we expand Eq. (\ref{eq:discrete-2nd-moment-temporal-1}), we find
\begin{multline}
\left\langle m_{\alpha}\frac{1}{2}\left(\boldsymbol{v}_{\alpha,i,j,k}^{p}\right)^{2},\frac{c^{p+1}\tilde{f}_{\alpha,i,j,k}^{p+1}+c^{p}\tilde{f}_{\alpha,i,j,k}^{p}+c^{p-1}\tilde{f}_{\alpha,i,j,k}^{p-1}}{\Delta t^{p}}\right\rangle _{\delta\bm{c}}\\
-\left\langle 1,\frac{c^{p+1}m_{\alpha}\frac{1}{2}\left(\boldsymbol{v}_{\alpha,i,j,k}^{p}\right)^{2}\tilde{f}_{\alpha,i,j,k}^{p+1}+c^{p}m_{\alpha}\frac{1}{2}\left(\boldsymbol{v}_{\alpha,i,j,k}^{p-1}\right)^{2}\tilde{f}_{\alpha,i,j,k}^{p}+c^{p-1}m_{\alpha}\frac{1}{2}\left(\boldsymbol{v}_{\alpha,i,j,k}^{p-2}\right)^{2}\tilde{f}_{\alpha,i,j,k}^{p-1}}{\Delta t^{p}}\right\rangle _{\delta\bm{c}}\\
-\left\langle m_{\alpha}\frac{1}{2}\left(\boldsymbol{v}_{\alpha,i,j,k}^{p}\right)^{2},\frac{1}{v_{\alpha,i}^{*,p}}\delta_{\bm{c}}\cdot\left[\gamma_{t,\alpha,i}^{p+1}\delta_{t}\left(\boldsymbol{v}_{\alpha,i}\right)\overline{\left(\tilde{f}_{\alpha}^{p+1}\right)}_{i}^{\delta_{t}\left(\boldsymbol{v}\right)}\right]_{j,k}\right\rangle _{\delta\bm{c}}=0,\label{eq:discrete-2nd-moment-spatial-2}
\end{multline}
which is a concise representation of the final discrete constraint
on the definition of the nonlinear constraint function $\gamma_{t}$.

As in the case of momentum conservation, Eq. (\ref{eq:discrete-2nd-moment-spatial-1})
must be enforced more carefully. Once again we will integrate through
configuration-space and assume periodic boundaries, whereupon we arrive
at
\begin{multline}
\sum\limits _{i}^{N_{x}}\Delta x\Bigg\{\left\langle \frac{1}{2}\left(\boldsymbol{v}_{\alpha,i,j,k}^{p}\right)^{2},\frac{v_{\alpha,i+\frac{1}{2}}^{*,p}\hat{v}_{\parallel,\mathrm{eff},\alpha,i+\frac{1}{2},j}^{p+1}\overline{\left(\tilde{f}_{\alpha}^{p+1}\right)}_{i+\frac{1}{2},j,k}^{\hat{v}_{\parallel,\mathrm{eff}}}-v_{\alpha,i-\frac{1}{2}}^{*,p}\hat{v}_{\parallel,\mathrm{eff},\alpha,i-\frac{1}{2},j}^{p+1}\overline{\left(\tilde{f}_{\alpha}^{p+1}\right)}_{i-\frac{1}{2},j,k}^{\hat{v}_{\parallel,\mathrm{eff}}}}{\Delta x}\right\rangle _{\delta\bm{c}}\\
-\left\langle \frac{1}{2}\left(\boldsymbol{v}_{\alpha,i,j,k}^{p}\right)^{2},\frac{1}{2v_{\alpha,i}^{*,p}}\delta_{\bm{c}}\cdot\left[\gamma_{x,\alpha,i+\frac{1}{2}}^{p+1}v_{\alpha,i+\frac{1}{2}}^{*,p}\hat{v}_{\parallel,\mathrm{eff},\alpha,i+\frac{1}{2}}^{p+1}\frac{\boldsymbol{v}_{\alpha,i+1}^{p}-\boldsymbol{v}_{\alpha,i}^{p}}{\Delta x}\overline{\left(\tilde{f}_{\alpha}^{p+1}\right)}_{i}^{v_{\parallel,\mathrm{eff}}\delta_{x}\left(\boldsymbol{v}\right)}\right]_{j,k}\right\rangle _{\delta\bm{c}}\\
-\left\langle \frac{1}{2}\left(\boldsymbol{v}_{\alpha,i,j,k}^{p}\right)^{2},\frac{1}{2v_{\alpha,i}^{*,p}}\delta_{\bm{c}}\cdot\left[\gamma_{x,\alpha,i-\frac{1}{2}}^{p+1}v_{\alpha,i-\frac{1}{2}}^{*,p}\hat{v}_{\parallel,\mathrm{eff},\alpha,i-\frac{1}{2}}^{p+1}\frac{\boldsymbol{v}_{\alpha,i}^{p}-\boldsymbol{v}_{\alpha,i-1}^{p}}{\Delta x}\overline{\left(\tilde{f}_{\alpha}^{p+1}\right)}_{i}^{v_{\parallel,\mathrm{eff}}\delta_{x}\left(\boldsymbol{v}\right)}\right]_{j,k}\right\rangle _{\delta\bm{c}}\Bigg\}=0.\label{eq:discrete-2nd-moment-spatial-3}
\end{multline}
Again telescoping the summation in configuration-space and equating
the quantity inside braces to zero, we find 
\begin{multline}
\left\langle \frac{\left(\boldsymbol{v}_{\alpha,i,j,k}^{p}\right)^{2}}{2}-\frac{\left(\boldsymbol{v}_{\alpha,i+1,j,k}^{p}\right)^{2}}{2},\frac{1}{\Delta x}v_{\alpha,i+\frac{1}{2}}^{*,p}\hat{v}_{\parallel,\mathrm{eff},\alpha,i+\frac{1}{2},j}^{p+1}\overline{\left(\tilde{f}_{\alpha}^{p+1}\right)}_{i+\frac{1}{2},j,k}^{\hat{v}_{\parallel,\mathrm{eff}}}\right\rangle \\
-\left\langle \frac{\left(\boldsymbol{v}_{\alpha,i,j,k}^{p}\right)^{2}}{2},\frac{1}{2v_{\alpha,i}^{*,p}}\delta_{\bm{c}}\cdot\left[\gamma_{x,\alpha,i+\frac{1}{2}}^{p+1}v_{\alpha,i+\frac{1}{2}}^{*,p}\hat{v}_{\parallel,\mathrm{eff},i+\frac{1}{2}}\delta_{x}\left(\boldsymbol{v}^{p}\right)_{i+\frac{1}{2}}\overline{\left(\tilde{f}_{\alpha}^{p+1}\right)}_{i}^{v_{\parallel,\mathrm{eff}}\delta_{x}\left(\boldsymbol{v}\right)}\right]_{j,k}\right\rangle _{\delta\bm{c}}\\
-\left\langle \frac{\left(\boldsymbol{v}_{\alpha,i+1,j,k}^{p}\right)^{2}}{2},\frac{1}{2v_{\alpha,i+1}^{*,p}}\delta_{\bm{c}}\cdot\left[\gamma_{x,\alpha,i+\frac{1}{2}}^{p+1}v_{\alpha,i+\frac{1}{2}}^{*,p}\hat{v}_{\parallel,\mathrm{eff},i+\frac{1}{2}}\delta_{x}\left(\boldsymbol{v}^{p}\right)_{i+\frac{1}{2}}\overline{\left(\tilde{f}_{\alpha}^{p+1}\right)}_{i+1}^{v_{\parallel,\mathrm{eff}}\delta_{x}\left(\boldsymbol{v}\right)}\right]_{j,k}\right\rangle _{\delta\bm{c}}=0,\label{eq:discrete-2nd-moment-spatial-4}
\end{multline}
which is the final discrete constraint on $\gamma_{x}$.

\section{Derivation of constraint definitions for $\xi$, $\phi$, and $\gamma_{q}$\label{app:Derivation-of-constraint}}

\subsection{Discrete charge \& mass conservation \label{subsec:Discrete-charge-=000026}}

To demonstrate a discrete mass conservation, we apply to Eq. (\ref{eq:Vlasov-disc-simp})
the discrete moment $\left\langle m_{\alpha},\cdots\right\rangle _{\delta\bm{c}}$
. First, we observe that with appropriate discrete boundary conditions
(i.e., zero mass flux in the velocity space and periodic boundaries
in the configuration space), all velocity-space divergence terms vanish
under the discrete moment (as in the continuum case). Thus, we are
left with
\begin{equation}
\left\langle m_{\alpha},\delta_{t}\tilde{f}_{\alpha,i,j,k}\right\rangle _{\delta\bm{c}}+\left\langle m_{\alpha},\delta_{x}\left[v_{\parallel,\alpha,j}^{p}\overline{\left(\tilde{f}_{\alpha}^{p+1}\right)}_{j,k}^{v_{\parallel}}\right]_{i}\right\rangle _{\delta\bm{c}}+\left\langle m_{\alpha},\delta_{x}\left[\xi_{\alpha}^{p+1}\left|v_{\parallel,\alpha,j}^{p}\right|\overline{\left(\tilde{f}_{\alpha}^{p+1}\right)}_{j,k}^{\xi}\right]_{i}\right\rangle _{\delta\bm{c}}=0.\label{eq:discrete-mass-cons-2}
\end{equation}
Clearly, if we sum over all species $\alpha$ and integrate over the
configuration space,$\sum\limits _{\alpha}^{N_{sp}}\sum\limits _{i=1}^{N_{x}}$,
we will obtain the proper discrete mass conservation (assuming a periodic
domain in the configuration space):
\[
\delta_{t}M_{tot}=\frac{c^{p+1}M_{tot}^{p+1}+c^{p}M_{tot}^{p}+c^{p-1}M_{tot}^{p-1}}{\Delta t^{p}}=0,
\]
where $M_{tot}^{p}\equiv\sum\limits _{\alpha}^{N_{sp}}\sum\limits _{i=1}^{N_{x}}m_{\alpha}n_{\alpha,i}^{p}=\sum\limits _{\alpha}^{N_{sp}}\sum\limits _{i=1}^{N_{x}}m_{\alpha}\left\langle \tilde{f}_{\alpha,i,j,k}^{p}\right\rangle _{\boldsymbol{v}_{j,k}}$.
Defining the discrete moments
\begin{align}
n_{\alpha,i}^{p+1} & \equiv\left\langle 1,\tilde{f}_{\alpha,i,j,k}^{p+1}\right\rangle _{\delta\bm{c}},\label{eq:n-discrete}\\
\widetilde{\Gamma}_{\parallel,\alpha,i+\frac{1}{2}}^{p+1} & \equiv\left\langle 1,v_{\alpha,i+\frac{1}{2}}^{*,p}\left(c_{\parallel,j}+\hat{u}_{\parallel,\alpha,i+\frac{1}{2}}^{*,p}\right)\overline{\left(\tilde{f}_{\alpha}^{p+1}\right)}_{i+\frac{1}{2},j,k}^{v_{\parallel,i+\frac{1}{2},j}^{p}}\right\rangle _{\delta\bm{c}},\label{eq:nu-cf-discrete}\\
\Pi_{\xi,\parallel,\alpha,i+\frac{1}{2}}^{p+1} & \equiv\left\langle 1,v_{\alpha,i+\frac{1}{2}}^{*,p}\left|c_{\parallel,j}+\hat{u}_{\parallel,\alpha,i+\frac{1}{2}}^{*,p}\right|\overline{\left(\tilde{f}_{\alpha}^{p+1}\right)}_{i+\frac{1}{2},j,k}^{\xi_{\alpha,i+\frac{1}{2}}^{p+1}}\right\rangle _{\delta\bm{c}},\label{eq:nu-gamma-xi}
\end{align}
we may further express Eq. (\ref{eq:discrete-mass-cons-2}) in terms
of discrete moment quantities: 
\begin{multline}
m_{\alpha}\frac{c^{p+1}n_{\alpha,i}^{p+1}+c^{p}n_{\alpha,i}^{p}+c^{p-1}n_{\alpha,i}^{p-1}}{\Delta t^{p}}\\
+m_{\alpha}\left(\frac{\widetilde{\Gamma}_{\parallel,\alpha,i+\frac{1}{2}}^{p+1}-\widetilde{\Gamma}_{\parallel,\alpha,i-\frac{1}{2}}^{p+1}}{\Delta x}+\frac{\xi_{\alpha,i+\frac{1}{2}}^{p+1}\Pi_{\xi,\parallel,\alpha,i+\frac{1}{2}}^{p+1}-\xi_{\alpha,i-\frac{1}{2}}^{p+1}\Pi_{\xi,\parallel,\alpha,i-\frac{1}{2}}^{p+1}}{\Delta x}\right)=0.\label{eq:discrete-mass-cons-3}
\end{multline}
If we recall Sec. \ref{subsec:Charge-=000026-mass-conservation},
we know that the species particle flux density density that forms
the current density in Ampère's equation must be identical to the
momentum density that appears in the continuity equation. We observe
that the discrete particle flux density forming the current in Ampère's
equation, $\widehat{\Gamma}_{\parallel,\alpha,i+\frac{1}{2}}^{p+1}$,
must therefore be
\begin{equation}
\widehat{\Gamma}_{\parallel,\alpha,i+\frac{1}{2}}^{p+1}=\widetilde{\Gamma}_{\parallel,\alpha,i+\frac{1}{2}}^{p+1}+\xi_{\alpha,i+\frac{1}{2}}^{p+1}\Pi_{\xi,\parallel,\alpha,i+\frac{1}{2}}^{p+1}.\label{eq:xi-discrete-1}
\end{equation}
Thus, the purpose of the nonlinear constraint function $\xi$ is to
enforce that the truncation error between the discrete representations
of particle flux density, $\widetilde{\Gamma}_{\parallel,\alpha,i+\frac{1}{2}}$
and $\widehat{\Gamma}_{\parallel,\alpha,i+\frac{1}{2}}$, vanishes.
The precise discrete definition of $\widehat{\Gamma}_{\parallel,\alpha,i+\frac{1}{2}}$
is given in \ref{subsec:Discrete-energy-conservation}. The constraint
function $\xi$ and its `pseudo-advection' operator are critical to
enforcing charge conservation, as we will see in Sec. \ref{subsec:Ion-acoustic-shockwave}.

\subsection{Discrete momentum conservation\label{subsec:Discrete-momentum-conservation}}

To demonstrate a discrete momentum conservation, we apply to Eq. (\ref{eq:Vlasov-disc-simp})
the discrete moment $\left\langle m_{\alpha}v_{\parallel,\alpha,i,j}^{p},\cdots\right\rangle _{\delta\bm{c}}$,
and note that $\boldsymbol{v}_{\alpha,i,j,k}^{p}=v_{\alpha,i}^{*,p}\left(\bm{c}_{j,k}+\boldsymbol{e}_{\parallel}\hat{u}_{\parallel,\alpha,i}^{*,p}\right)$,
i.e., $v_{\parallel,\alpha,i,j}^{p}=v_{\alpha,i}^{*,p}\left(c_{\parallel,j}+\hat{u}_{\parallel,\alpha,i}^{*,p}\right)$
and $v_{\bot,\alpha,i,k}^{p}=v_{\alpha,i}^{*,p}\tilde{v}_{\bot,k}$.
Employing the discrete moment in the configuration space, we find
\begin{multline}
m_{\alpha}\sum\limits _{i}^{N_{x}}\Delta x\Bigg\{\frac{c^{p+1}\Gamma_{\parallel,\alpha,i}^{p+1}+c^{p}\Gamma_{\parallel,\alpha,i}^{p}+c^{p-1}\Gamma_{\parallel,\alpha,i}^{p-1}}{\Delta t^{p}}+\left\langle v_{\parallel,\alpha,i,j}^{p},\frac{q_{\alpha}}{m_{\alpha}}\frac{E_{\parallel,i}^{p+1}}{v_{\alpha,i}^{*,p}}\delta_{c_{\parallel}}\left[\overline{\left(\tilde{f}_{\alpha}^{p+1}\right)}_{i,k}^{q_{\alpha}E_{\parallel}}\right]_{j}\right\rangle _{\delta\bm{c}}\\
+\left\langle v_{\parallel,\alpha,i,j}^{p},\delta_{c_{\parallel}}\left[\phi_{\alpha,i}^{p+1}\overline{\left(\tilde{f}_{\alpha}^{p+1}\right)}_{i,k}^{\phi}\right]_{j}\right\rangle _{\delta\bm{c}}+\left\langle v_{\parallel,\alpha,i,j}^{p},\delta_{c_{\parallel}}\left[\gamma_{q,\alpha,i}^{p+1}\overline{\left(\tilde{f}_{\alpha}^{p+1}\right)}_{i,k}^{\gamma_{q}}\right]_{j}\right\rangle _{\delta\bm{c}}\Bigg\}=0,\label{eq:discrete-mom-cons-2}
\end{multline}
where we have defined 
\[
\Gamma_{\parallel,\alpha,i}^{p+1}\equiv\left\langle 1,v_{\parallel,\alpha,i,j}^{p}\tilde{f}_{\alpha,i,j,k}^{p+1}\right\rangle _{\delta\bm{c}}.
\]
Expanding the individual flux operators and defining the discrete
number density based on the moment of the acceleration operator,
\begin{equation}
\overline{n}_{\alpha,i}^{p+1}\equiv-\left\langle \frac{v_{\parallel,\alpha,i,j}^{p}}{v_{\alpha,i}^{*,p}},\frac{\overline{\left(\tilde{f}_{\alpha}^{p+1}\right)}_{i,j+\frac{1}{2},k}^{q_{\alpha}E_{\parallel}}-\overline{\left(\tilde{f}_{\alpha}^{p+1}\right)}_{i,j-\frac{1}{2},k}^{q_{\alpha}E_{\parallel}}}{\Delta c_{\parallel}}\right\rangle _{\delta\bm{c}},\label{eq:n-bar-discrete}
\end{equation}
we find 
\begin{multline}
m_{\alpha}\sum\limits _{i}^{N_{x}}\Delta x\Bigg\{\frac{c^{p+1}\Gamma_{\parallel,\alpha,i}^{p+1}+c^{p}\Gamma_{\parallel,\alpha,i}^{p}+c^{p-1}\Gamma_{\parallel,\alpha,i}^{p-1}}{\Delta t^{p}}-\overline{n}_{\alpha,i}^{p+1}E_{\parallel,i}^{p+1}\frac{q_{\alpha}}{m_{\alpha}}\\
-\left(\phi_{\alpha,i}^{+,p+1}+1\right)n_{\alpha,i}^{+,p+1}-\left(\gamma_{q,\alpha,i}^{-,p+1}+1\right)n_{\alpha,i}^{-,p+1}\Bigg\}=0,\label{eq:discrete-mom-cons-4}
\end{multline}
Here, we defined $\phi$ and $\gamma_{q}$ to be split in $v_{\parallel}$-space
as
\begin{align}
\phi_{\alpha,i,j+\frac{1}{2}}^{p+1} & =\begin{cases}
\phi_{\alpha,i}^{+,p+1} & \text{if }v_{\parallel,\alpha,i,j+\frac{1}{2}}\geq u_{\parallel,\alpha,i}^{p}\\
1 & \text{otherwise }
\end{cases},\nonumber \\
\gamma_{q,\alpha,i,j+\frac{1}{2}}^{p+1} & =\begin{cases}
1 & \text{if }v_{\parallel,\alpha,i,j+\frac{1}{2}}\geq u_{\parallel,\alpha,i}^{p}\\
\gamma_{q,\alpha,i}^{-,p+1} & \text{otherwise}
\end{cases},\label{eq:phi-gammaq-split}
\end{align}
where the definitions for $\phi_{\alpha,i}^{+,p+1}$ and $\gamma_{q,\alpha,i}^{-,p+1}$
will be determined shortly. The rationale for splitting $\phi$ and
$\gamma_{q}$ in this way is, as shall be seen in Sec. \ref{subsec:Discrete-energy-conservation},
for solvability of the resulting $2\times2$ linear system from which
$\phi$ and $\gamma_{q}$ are calculated. From this splitting, we
define the ``upper'' and ``lower'' densities as
\begin{align}
n_{\alpha,i}^{+,p+1} & \equiv\left\langle v_{\parallel,\alpha,i,j}^{p},\frac{1}{v_{\alpha,i}^{*,p}}\frac{\overline{\left(\tilde{f}_{\alpha}^{p+1}\right)}_{i,j+\frac{1}{2},k}^{\mathrm{central}}-\overline{\left(\tilde{f}_{\alpha}^{p+1}\right)}_{i,j-\frac{1}{2},k}^{\mathrm{central}}}{\Delta c_{\parallel}}\right\rangle _{\delta\bm{c}\text{ where }v_{\parallel,\alpha,i,j+\frac{1}{2}}\geq u_{\parallel,\alpha,i}^{p}},\label{eq:n-tilde-plus}\\
n_{\alpha,i}^{-,p+1} & \equiv\left\langle v_{\parallel,\alpha,i,j}^{p},\frac{1}{v_{\alpha,i}^{*,p}}\frac{\overline{\left(\tilde{f}_{\alpha}^{p+1}\right)}_{i,j+\frac{1}{2},k}^{\mathrm{central}}-\overline{\left(\tilde{f}_{\alpha}^{p+1}\right)}_{i,j-\frac{1}{2},k}^{\mathrm{central}}}{\Delta c_{\parallel}}\right\rangle _{\delta\bm{c}\text{ where }v_{\parallel,\alpha,i,j+\frac{1}{2}}<u_{\parallel,\alpha,i}^{p}}.\label{eq:n-tilde-minus}
\end{align}

If we then sum Eq. (\ref{eq:discrete-mom-cons-4}) over all species
$\alpha$, we find 
\begin{multline}
\delta_{t}P_{\mathrm{total}}-\sum\limits _{\alpha}^{N_{sp}}m_{\alpha}\sum\limits _{i}^{N_{x}}\Delta x\Bigg\{\overline{n}_{\alpha,i}^{p+1}E_{\parallel,i}^{p+1}\frac{q_{\alpha}}{m_{\alpha}}+\left(\phi_{\alpha,i}^{+,p+1}+1\right)n_{\alpha,i}^{+,p+1}+\left(\gamma_{q,\alpha,i}^{-,p+1}+1\right)n_{\alpha,i}^{-,p+1}\Bigg\}=0,\label{eq:discrete-mom-cons-5}
\end{multline}
where
\[
\delta_{t}P_{\mathrm{total}}\equiv\sum\limits _{\alpha}^{N_{sp}}m_{\alpha}\sum\limits _{i}^{N_{x}}\Delta x\frac{c^{p+1}\Gamma_{\parallel,\alpha,i}^{p+1}+c^{p}\Gamma_{\parallel,\alpha,i}^{p}+c^{p-1}\Gamma_{\parallel,\alpha,i}^{p-1}}{\Delta t^{p}}
\]
is the discrete time derivative of the total momentum of the system.
Recalling the symmetry with Gauss' law in Eq. (\ref{eq:vlasov-1st-moment-2}),
we realize that Eq. (\ref{eq:discrete-mom-cons-5}) must become
\begin{multline}
\delta_{t}P_{\mathrm{total}}-\sum\limits _{\alpha}^{N_{sp}}m_{\alpha}\sum\limits _{i}^{N_{x}}\Delta x\Bigg\{ n_{\alpha,i}^{p+1}E_{\parallel,i}^{p+1}\frac{q_{\alpha}}{m_{\alpha}}\Bigg\}=\delta_{t}P_{\mathrm{total}}-\sum\limits _{i}^{N_{x}}\Delta x\Bigg\{\epsilon_{0}E_{\parallel,i}^{p+1}\frac{E_{\parallel,i+\frac{1}{2}}^{p+1}-E_{\parallel,i-\frac{1}{2}}^{p+1}}{\Delta x}\Bigg\}=0,\label{eq:discrete-mom-cons-6}
\end{multline}
where the final summation vanishes \textendash{} assuming periodic
boundaries \textendash{} when we recall that $E_{\parallel,i}^{p+1}$
is the average of adjacent cell-face values. Thus we achieve $\delta_{t}P_{\mathrm{total}}=0$,
which is a discrete statement of momentum conservation. The equivalence
between Eqs. (\ref{eq:discrete-mom-cons-5}) and (\ref{eq:discrete-mom-cons-6})
provides us the first discrete constraint for the nonlinear constraint
functions $\phi_{\alpha}$ and $\gamma_{q,\alpha}$:
\begin{equation}
\left(\phi_{\alpha,i}^{+,p+1}+1\right)n_{\alpha,i}^{+,p+1}+\left(\gamma_{q,\alpha,i}^{-,p+1}+1\right)n_{\alpha,i}^{-,p+1}=\left(n_{\alpha,i}^{p+1}-\overline{n}_{\alpha,i}^{p+1}\right)E_{\parallel,i}^{p+1}\frac{q_{\alpha}}{m_{\alpha}}.\label{eq:phi-gammaq-discrete-1}
\end{equation}
Thus, $\phi$ and $\gamma_{q}$ act in concert to enforce that the
truncation error between the two discrete representations of density,
$n_{\alpha,i}$ and $\overline{n}_{\alpha,i}$, vanishes.

\subsection{Discrete energy conservation\label{subsec:Discrete-energy-conservation}}

To demonstrate discrete energy conservation, we apply the discrete
moment  $\left\langle m_{\alpha}\frac{1}{2}\left(\boldsymbol{v}_{\alpha,i,j,k}^{p}\right)^{2},\cdots\right\rangle _{\delta\bm{c}}$to
Eq. (\ref{eq:Vlasov-disc-simp}), where $\frac{1}{2}\left(\boldsymbol{v}_{\alpha,i,j,k}^{p}\right)^{2}=\frac{1}{2}\left[\left(v_{\parallel,\alpha,i,j}^{p}\right)^{2}+\left(v_{\bot,\alpha,i,k}^{p}\right)^{2}\right]$.
Utilizing the preceding developments and evaluating the discrete moment
in the configuration space while summing over all species, we find
\begin{multline}
\delta_{t}U_{\mathrm{fluid}}+\sum\limits _{\alpha}^{N_{sp}}\sum\limits _{i}^{N_{x}}\Delta x\Bigg\{\left\langle m_{\alpha}\frac{1}{2}\left(\boldsymbol{v}_{\alpha,i,j,k}^{p}\right)^{2},\frac{q_{\alpha}}{m_{\alpha}}\frac{E_{\parallel,i}^{p+1}}{v_{\alpha,i}^{*,p}}\delta_{c_{\parallel}}\left[\overline{\left(\tilde{f}_{\alpha}^{p+1}\right)}_{i,k}^{q_{\alpha}E_{\parallel}}\right]_{j}\right\rangle _{\delta\bm{c}}\\
+\left\langle m_{\alpha}\frac{1}{2}\left(\boldsymbol{v}_{\alpha,i,j,k}^{p}\right)^{2},\delta_{c_{\parallel}}\left[\phi_{\alpha,i}^{p+1}\overline{\left(\tilde{f}_{\alpha}^{p+1}\right)}_{i,k}^{\phi}\right]_{j}\right\rangle _{\delta\bm{c}}\\
+\left\langle m_{\alpha}\frac{1}{2}\left(\boldsymbol{v}_{\alpha,i,j,k}^{p}\right)^{2},\delta_{c_{\parallel}}\left[\gamma_{q,\alpha,i}^{p+1}\overline{\left(\tilde{f}_{\alpha}^{p+1}\right)}_{i,k}^{\gamma_{q}}\right]_{j}\right\rangle _{\delta\bm{c}}\Bigg\}=0,\label{eq:discrete-ener-cons-2}
\end{multline}
where we defined the quantities 
\[
\delta_{t}U_{\mathrm{fluid}}\equiv\sum\limits _{\alpha}^{N_{sp}}\sum\limits _{i}^{N_{x}}\Delta x\frac{c^{p+1}\varepsilon_{\alpha,i}^{p+1}+c^{p}\varepsilon_{\alpha,i}^{p}+c^{p-1}\varepsilon_{\alpha,i}^{p-1}}{\Delta t^{p}},
\]
and
\[
\varepsilon_{\alpha,i}^{p+1}\equiv\left\langle 1,m_{\alpha}\frac{1}{2}\left(\boldsymbol{v}_{\alpha,i,j,k}^{p}\right)^{2}\tilde{f}_{\alpha,i,j,k}^{p+1}\right\rangle _{\boldsymbol{v}_{j,k}}.
\]
Again, expanding the individual flux operators and defining the discrete
particle flux density obtained from the moment of the acceleration
operator,
\[
\overline{\Gamma}_{\parallel,\alpha,i}^{p+1}\equiv-\left\langle \frac{\left(\boldsymbol{v}_{\alpha,i,j,k}^{p}\right)^{2}}{2v_{\alpha,i}^{*,p}},\frac{\overline{\left(\tilde{f}_{\alpha}^{p+1}\right)}_{i,j+\frac{1}{2},k}^{q_{\alpha}E_{\parallel}}-\overline{\left(\tilde{f}_{\alpha}^{p+1}\right)}_{i,j-\frac{1}{2},k}^{q_{\alpha}E_{\parallel}}}{\Delta c_{\parallel}}\right\rangle _{\delta\bm{c}},
\]
we obtain
\begin{multline}
\delta_{t}U_{\mathrm{fluid}}-\sum\limits _{\alpha}^{N_{sp}}\sum\limits _{i}^{N_{x}}\Delta x\Bigg\{ q_{\alpha}E_{\parallel,i}^{p+1}\overline{\Gamma}_{\parallel,\alpha,i}^{p+1}\\
+m_{\alpha}\left[\left(\phi_{\alpha,i}^{+,p+1}+1\right)\Gamma_{\parallel,\alpha,i}^{+,p+1}+\left(\gamma_{q,\alpha,i}^{-,p+1}+1\right)\Gamma_{\parallel,\alpha,i}^{-,p+1}\right]\Bigg\}=0.\label{eq:discrete-ener-cons-4}
\end{multline}
Here, as before, we defined the ``upper'' and ``lower'' momenta as
\begin{align}
\Gamma_{\parallel,\alpha,i}^{+,p+1} & \equiv\left\langle \frac{1}{2}\left(\boldsymbol{v}_{\alpha,i,j,k}^{p}\right)^{2},\frac{1}{v_{\alpha,i}^{*,p}}\frac{\overline{\left(\tilde{f}_{\alpha}^{p+1}\right)}_{i,j+\frac{1}{2},k}^{\mathrm{central}}-\overline{\left(\tilde{f}_{\alpha}^{p+1}\right)}_{i,j-\frac{1}{2},k}^{\mathrm{central}}}{\Delta c_{\parallel}}\right\rangle _{\delta\bm{c}\text{ where }v_{\parallel,\alpha,i,j+\frac{1}{2}}\geq u_{\parallel,\alpha,i}^{p}},\label{eq:nu-tilde-plus}\\
\Gamma_{\parallel,\alpha,i}^{-,p+1} & \equiv\left\langle \frac{1}{2}\left(\boldsymbol{v}_{\alpha,i,j,k}^{p}\right)^{2},\frac{1}{v_{\alpha,i}^{*,p}}\frac{\overline{\left(\tilde{f}_{\alpha}^{p+1}\right)}_{i,j+\frac{1}{2},k}^{\mathrm{central}}-\overline{\left(\tilde{f}_{\alpha}^{p+1}\right)}_{i,j-\frac{1}{2},k}^{\mathrm{central}}}{\Delta c_{\parallel}}\right\rangle _{\delta\bm{c}\text{ where }v_{\parallel,\alpha,i,j+\frac{1}{2}}<u_{\parallel,\alpha,i}^{p}}.\label{eq:nu-tilde-minus}
\end{align}
Thus we see Eq. (\ref{eq:discrete-ener-cons-4}) is the discrete time
derivative of the total \emph{fluid} energy of the system, including
the total thermal and kinetic energy of the plasma. Recalling the
definition of $E_{\parallel,i}^{p+1}$ we may rearrange the summation
in Eq. (\ref{eq:discrete-ener-cons-4}), which gives us 
\begin{multline}
\delta_{t}U_{\mathrm{fluid}}-\sum\limits _{\alpha}^{N_{sp}}\sum\limits _{i}^{N_{x}}\Delta x\Bigg\{\overline{\Gamma}_{\parallel,\alpha,i+\frac{1}{2}}^{p+1}E_{\parallel,i+\frac{1}{2}}^{p+1}q_{\alpha}\Bigg\}\\
-\sum\limits _{\alpha}^{N_{sp}}m_{\alpha}\sum\limits _{i}^{N_{x}}\Delta x\Bigg\{\left(\phi_{\alpha,i}^{+,p+1}+1\right)\Gamma_{\parallel,\alpha,i}^{+,p+1}+\left(\gamma_{q,\alpha,i}^{-,p+1}+1\right)\Gamma_{\parallel,\alpha,i}^{-,p+1}\Bigg\}=0,\label{eq:discrete-ener-cons-5}
\end{multline}
where we defined
\begin{equation}
\overline{\Gamma}_{\parallel,\alpha,i+\frac{1}{2}}^{p+1}\equiv\frac{\overline{\Gamma}_{\parallel,\alpha,i}^{p+1}+\overline{\Gamma}_{\parallel,\alpha,i+1}^{p+1}}{2}.\label{eq:nu-cf-accel-APP}
\end{equation}
Next, we recall the symmetry with Ampère's equation in Eq. (\ref{eq:vlasov-2nd-moment-2})
and define
\begin{equation}
\overline{\Gamma}_{\parallel,\alpha,i+\frac{1}{2}}^{p+1}=\widehat{\Gamma}_{\parallel,\alpha,i+\frac{1}{2}}^{p+1}\label{eq:nu-hat-discrete}
\end{equation}
to enforce particle fluxes in the current in Ampère's equation that
come from the energy moment of the acceleration operator. Thus, we
find that the nonlinear constraint function $\xi_{\alpha}$ is completely
determined by requiring
\begin{equation}
\xi_{\alpha,i+\frac{1}{2}}^{p+1}=\frac{\overline{\Gamma}_{\parallel,\alpha,i+\frac{1}{2}}^{p+1}-\widetilde{\Gamma}_{\parallel,\alpha,i+\frac{1}{2}}^{p+1}}{\Pi_{\xi,\parallel,\alpha,i+\frac{1}{2}}^{p+1}}.\label{eq:xi-discrete-2}
\end{equation}
We now observe that Eq. (\ref{eq:discrete-ener-cons-5}) must become
\begin{multline}
\delta_{t}U_{\mathrm{fluid}}+\sum\limits _{i}^{N_{x}}\Delta x\Bigg\{\epsilon_{0}E_{\parallel,i+\frac{1}{2}}^{p+1}\delta_{t}E_{\parallel,i+\frac{1}{2}}-E_{\parallel,i+\frac{1}{2}}^{p+1}\overline{j}_{\parallel}^{p+1}\Bigg\}\\
-\sum\limits _{i}^{N_{x}}\Delta x\Bigg\{\sum\limits _{\alpha}^{N_{sp}}m_{\alpha}\left[\left(\phi_{\alpha,i}^{+,p+1}+1\right)\Gamma_{\parallel,\alpha,i}^{+,p+1}+\left(\gamma_{q,\alpha,i}^{-,p+1}+1\right)\Gamma_{\parallel,\alpha,i}^{-,p+1}\right]\Bigg\}=0.\label{eq:discrete-ener-cons-6}
\end{multline}
In a previous implementation of this method, as was done in Refs.
\citep{Taitano2015a,Taitano2015b}, the use of a Crank-Nicolson integration
scheme ensured the equivalence 
\[
E_{\parallel,i+\frac{1}{2}}^{p+1}\delta_{t}E_{\parallel,i+\frac{1}{2}}=\frac{1}{2}\delta_{t}\left(E_{\parallel,i+\frac{1}{2}}^{2}\right)
\]
in the discrete. However, BDF2 in the current development does not
ensure this relation. Thus, the second purpose of the quantities $\phi$
and $\gamma_{q}$ is to enforce the equivalence of Eq. (\ref{eq:discrete-ener-cons-6})
to the equation
\begin{multline}
\delta_{t}U_{\mathrm{fluid}}+\sum\limits _{i}^{N_{x}}\Delta x\Bigg\{\delta_{t}\left(\epsilon_{0}\frac{1}{2}E_{\parallel,i+\frac{1}{2}}^{2}\right)-E_{\parallel,i+\frac{1}{2}}^{p+1}\overline{j}_{\parallel}^{p+1}\Bigg\}\\
=\delta_{t}U_{\mathrm{fluid}}+\delta_{t}U_{E_{\parallel}}-\overline{E}_{\parallel}^{p+1}\overline{j}_{\parallel}^{p+1}=\delta_{t}U_{\mathrm{total}}-\overline{E}_{\parallel}^{p+1}\overline{j}_{\parallel}^{p+1}=0.\label{eq:discrete-ener-cons-7}
\end{multline}
Note, that in the absence of an external electric field, $\overline{E}_{\parallel}^{p+1}\overline{j}_{\parallel}^{p+1}$
vanishes discretely, preserving the discrete conservation principle.
The final constraint on $\phi$ and $\gamma_{q}$ is thus 
\begin{equation}
\left(\phi_{\alpha,i}^{+,p+1}+1\right)\Gamma_{\parallel,\alpha,i}^{+,p+1}+\left(\gamma_{q,\alpha,i}^{-,p+1}+1\right)\Gamma_{\parallel,\alpha,i}^{-,p+1}=\frac{1}{m_{\alpha}N_{sp}}\left\{ \epsilon_{0}\left[E_{\parallel,i+\frac{1}{2}}^{p+1}\delta_{t}E_{\parallel,i+\frac{1}{2}}-\delta_{t}\left(\frac{1}{2}E_{\parallel,i+\frac{1}{2}}^{2}\right)\right]\right\} ,\label{eq:phi-gammaq-discrete-2}
\end{equation}
where we have pulled the temporal derivatives into the species summation
by dividing by the number of species, $N_{sp}$. 

We see from Eqs. (\ref{eq:phi-gammaq-discrete-1}) and (\ref{eq:phi-gammaq-discrete-2})
that $\phi$ and $\gamma_{q}$ are determined for each species and
locally at each point in the configuration space by a simple $2\times2$
linear system of equations:
\begin{equation}
\left[\begin{array}{cc}
n_{\alpha,i}^{+,p+1} & n_{\alpha,i}^{-,p+1}\\
\Gamma_{\parallel,\alpha,i}^{+,p+1} & \Gamma_{\parallel,\alpha,i}^{-,p+1}
\end{array}\right]\left[\begin{array}{c}
\left(\phi_{\alpha,i}^{+,p+1}+1\right)\\
\left(\gamma_{q,\alpha,i}^{-,p+1}+1\right)
\end{array}\right]=\left[\begin{array}{c}
\left(n_{\alpha,i}^{p+1}-\overline{n}_{\alpha,i}^{p+1}\right)E_{\parallel,i}^{p+1}\frac{q_{\alpha}}{m_{\alpha}}\\
\frac{1}{m_{\alpha}N_{sp}}\left\{ \epsilon_{0}\left[E_{\parallel,i+\frac{1}{2}}^{p+1}\delta_{t}E_{\parallel,i+\frac{1}{2}}-\delta_{t}\left(\frac{1}{2}E_{\parallel,i+\frac{1}{2}}^{2}\right)\right]\right\} 
\end{array}\right].\label{eq:phi-gamma-discrete-final}
\end{equation}
To ensure solvability of Eq. (\ref{eq:phi-gamma-discrete-final}),
the determinant of the system 
\[
\mathrm{Det}\left[\begin{array}{cc}
n_{\alpha,i}^{+,p+1} & n_{\alpha,i}^{-,p+1}\\
\Gamma_{\parallel,\alpha,i}^{+,p+1} & \Gamma_{\parallel,\alpha,i}^{-,p+1}
\end{array}\right]=n_{\alpha,i}^{+,p+1}\Gamma_{\parallel,\alpha,i}^{-,p+1}-n_{\alpha,i}^{-,p+1}\Gamma_{\parallel,\alpha,i}^{+,p+1},
\]
must be strictly finite. It can be shown that if the splitting velocity
for Eq. (\ref{eq:phi-gammaq-split}) is within the discrete bounds
of $v_{\parallel,\alpha,i,j}^{p+1}$ then the system in Eq. (\ref{eq:phi-gamma-discrete-final})
is well-posed (see \ref{app:Well-posedness-of-}). To reduce nonlinearity
of the algorithm we use $u_{\parallel,\alpha,i}^{p}$ as the splitting
velocity, which is sufficiently close to $u_{\parallel,\alpha,i}^{p+1}$
for the constraint functions $\phi$ and $\gamma_{q}$ to remain well-behaved.

\section{Well-posedness of $2\times2$ system for $\phi^{+}$ and $\gamma_{q}^{-}$\label{app:Well-posedness-of-}}

In Sec. \ref{sec:Discrete-conservation-strategy} we presented a $2\times2$
linear system to be solved (locally, for each species) for $\phi_{\alpha,i}^{-,p+1}$
and $\gamma_{q,\alpha,i}^{-,p+1}$. This may be expressed in simpler
notation as
\begin{equation}
\left[\begin{array}{cc}
n^{+} & n^{-}\\
\Gamma^{+} & \Gamma^{-}
\end{array}\right]\left[\begin{array}{c}
\left(\phi^{+}+1\right)\\
\left(\gamma_{q}^{-}+1\right)
\end{array}\right]=\left[\begin{array}{c}
R_{1}\\
R_{2}
\end{array}\right].\label{eq:phi-gamma-2x2}
\end{equation}
To be well-posed, the determinant
\begin{equation}
\mathrm{Det}\left[\begin{array}{cc}
n^{+} & n^{-}\\
\Gamma^{+} & \Gamma^{-}
\end{array}\right]=n^{+}\Gamma^{-}-n^{-}\Gamma^{+}\label{eq:phigamma-1d1v-det}
\end{equation}
must be strictly non-zero. To show when this is the case, we will
for simplicity consider a 1D-1V system, with moments $n$ and $nu$
defined as
\begin{align}
n & =\int_{-\infty}^{\infty}f(v)dv,\label{eq:n-1d1v}\\
\Gamma & =\int_{-\infty}^{\infty}vf(v)dv.\label{eq:nu-1d1v}
\end{align}
We define the bulk velocity $u\equiv\frac{\Gamma}{n}$, and the split
quantities $n^{+/-}$ and $\Gamma^{+/-}$ by
\begin{align}
n^{-} & =\int_{-\infty}^{u^{*}}f(v)dv,\label{eq:nminus-1d1v}\\
n^{+} & =\int_{u^{*}}^{\infty}f(v)dv,\label{eq:nplus-1d1v}\\
\Gamma^{-} & =\int_{-\infty}^{u^{*}}vf(v)dv,\label{eq:numinus-1d1v}\\
\Gamma^{+} & =\int_{u^{*}}^{\infty}vf(v)dv.\label{eq:nuplus-1d1v}
\end{align}
We may then perform a coordinate transformation $v'=v-u^{*}$ of the
integration for Eqs. (\ref{eq:numinus-1d1v}) and (\ref{eq:nuplus-1d1v}):
\begin{align}
\Gamma^{-} & =\int_{-\infty}^{0}\left(v'+u^{*}\right)f(v')dv'=\int_{-\infty}^{0}v'f(v')dv'+n^{-}u^{*},\label{eq:numinus-trans-1d1v}\\
\Gamma^{+} & =\int_{0}^{\infty}\left(v'+u^{*}\right)f(v')dv'=\int_{0}^{\infty}v'f(v')dv'+n^{+}u^{*}.\label{eq:nuplus-trans-1d1v}
\end{align}
If we define 
\begin{align}
\left(nw\right)^{-} & =\int_{-\infty}^{0}v'f(v')dv',\label{eq:nwminus}\\
\left(nw\right)^{+} & =\int_{0}^{\infty}v'f(v')dv',\label{eq:nwplus}
\end{align}
we see that $\left(nw\right)^{-}$ is negative definite and $\left(nw\right)^{+}$
is positive definite. Equation (\ref{eq:phigamma-1d1v-det}) now becomes
\begin{equation}
n^{+}\left[\left(nw\right)^{-}+n^{-}u^{*}\right]-n^{-}\left[\left(nw\right)^{+}+n^{+}u^{*}\right]=n^{+}\left(nw\right)^{-}-n^{-}\left(nw\right)^{+}+n^{+}n^{-}\left(u^{*}-u^{*}\right),\label{eq:phi-gamma-det-final}
\end{equation}
which is negative definite. Thus, we find that Eq. (\ref{eq:phi-gamma-2x2})
is well-posed for arbitrary (\emph{finite}) $u^{*}$. We note that
in the discrete system $u_{\alpha,i}^{*}$ must lie within the discrete
bounds of $v_{\parallel}$ for the given species $\alpha$ at configuration-space
index $i$. In practice any choice near $u_{\parallel,\alpha,i}^{p+1}$
should be suitable.

\section{Two-stream instability for cold Maxwellian beams\label{app:Two-stream-instability-for}}

Recall that the general dispersion relation for the electron-electron
two stream instability \citep{Chen1989} is
\begin{equation}
1+\frac{\omega_{p,b}^{2}}{k^{2}v_{th,b}^{2}}\left[2+\zeta_{+}Z\left(\zeta_{+}\right)+\zeta_{-}Z\left(\zeta_{-}\right)\right]=0,\label{eq:dispersion-two-stream-appendix}
\end{equation}
where
\[
\zeta_{\pm}\equiv\frac{\omega\mp kv_{b}}{kv_{th,b}},
\]
and $\omega_{p,b}$ is the beam plasma frequency. Recall also that
if the electron beams are delta functions (i.e. in the limit $v_{th,b}\rightarrow0$)
Eq. (\ref{eq:dispersion-two-stream-appendix}) becomes

\begin{equation}
1-\frac{1}{\left(\omega+v_{b}k\right)^{2}}-\frac{1}{\left(\omega-v_{b}k\right)^{2}}=0.\label{eq:dispersion-two-stream-delta-appendix}
\end{equation}

Based on the delta-function dispersion relation, Eq. (\ref{eq:dispersion-two-stream-delta-appendix}),
the growth rate of electric field energy is $\gamma=0.353\omega_{p,b}$.
However, as we mentioned in Sec. (\ref{subsec:Two-stream-instability}),
for thermalized beams there will be some deviation, and we expect
that as the ratio $v_{th,b}/v_{b}$ increases that the system will
become more stable (i.e. $\gamma$ will decrease). Figure \ref{fig:Two-Stream-Instability-alphasweep}
indeed shows that if we increase $v_{th,b}/v_{b}$ towards some critical
ratio near $1$, the growth rate decreases precipitously. Here we
present a semi-analytic analysis of the generalized two-stream dispersion
relation in Eq. (\ref{eq:dispersion-two-stream-appendix}).

First, we will rearrange Eq. (\ref{eq:dispersion-two-stream-appendix}):
\begin{equation}
k^{2}\lambda_{D}^{2}+2+\zeta_{+}Z\left(\zeta_{+}\right)+\zeta_{-}Z\left(\zeta_{-}\right)=0,\label{eq:dispersion-general-rearranged}
\end{equation}
where $\lambda_{D}^{2}\equiv\frac{v_{th,b}^{2}}{\omega_{p,b}^{2}}$.
For convenience we then recast $\zeta_{\pm}$ in terms of dimensionless
quantities:
\begin{equation}
\zeta_{\pm}=\left(\frac{\beta_{r}}{\delta}\mp\alpha\right)+i\left(-\frac{\beta}{\delta}\right),\label{eq:zeta-redefined}
\end{equation}
where 
\begin{align*}
\omega\equiv\omega_{r}-i\gamma, & \alpha\equiv\frac{v_{b}}{v_{th,b}},\\
\beta_{r}\equiv\frac{\omega_{r}}{\omega_{p,b}}, & \beta\equiv\frac{\gamma}{\omega_{p,b}},\\
\delta\equiv k\lambda_{D}.
\end{align*}
We next observe that the plasma dispersion function $Z(\zeta)$ is
given as
\begin{equation}
Z\left(\zeta\right)=\frac{1}{\sqrt{\pi}}\int_{-\infty}^{\infty}\frac{e^{-z^{2}}}{z-\zeta}dz,\label{eq:dispersion-function}
\end{equation}
 which may be expressed in terms of the complex error function $\mathrm{erf}\left(z\right)$
as \citep{Fried1961}
\begin{equation}
Z\left(\zeta\right)=i\sqrt{\pi}e^{-\zeta^{2}}\left[1+\mathrm{\mathrm{erf}}\left(i\zeta\right)\right].\label{eq:dispersion-function-erf}
\end{equation}
 Thus, combining Eqs. (\ref{eq:zeta-redefined}) and (\ref{eq:dispersion-function-erf})
with Eq. (\ref{eq:dispersion-general-rearranged}), we obtain
\begin{multline}
\delta^{2}+2+i\sqrt{\pi}\zeta_{+}e^{-\left[\left(\frac{\beta_{r}}{\delta}-\alpha\right)+i\left(-\frac{\beta}{\delta}\right)\right]^{2}}\left[1+\mathrm{\mathrm{erf}}\left(i\left(\frac{\beta_{r}}{\delta}-\alpha\right)-\left(-\frac{\beta}{\delta}\right)\right)\right]\\
+i\sqrt{\pi}\zeta_{-}e^{-\left[\left(\frac{\beta_{r}}{\delta}+\alpha\right)+i\left(-\frac{\beta}{\delta}\right)\right]^{2}}\left[1+\mathrm{\mathrm{erf}}\left(i\left(\frac{\beta_{r}}{\delta}+\alpha\right)-\left(-\frac{\beta}{\delta}\right)\right)\right]=0.\label{eq:dispersion-normalized}
\end{multline}

Equation (\ref{eq:dispersion-normalized}) may be separated into its
real and imaginary components, and for a given $\alpha$ and $\delta$
(i.e., given the beam velocity $v_{b}$, beam thermal speed $v_{th,b}$,
wavenumber $k$, and Debye length $\lambda_{D}$), we may solve for
the instability growth rate $\beta$ (as well as the oscillatory component
$\beta_{r}$).

For the case in Sec. \ref{subsec:Two-stream-instability}, we have
$1/\alpha=v_{th,b}/v_{b}=[0.15,0.3,0.5,0.65,0.8]$ and $k=2\pi$.
For this case, we have $\omega_{p,b}=\omega_{p,e}=1$, defined using
the total electron density, which with a fixed $\left|v_{b}\right|=0.1$,
will give us $\delta=2\pi\times\left[0.015,0.03,0.05,0.065,0.08\right]$.
As we saw in Table \ref{tab:two-stream}, this will produce 
\[
\beta=\left[0.3488,0.3318,0.2734,0.1953,0.08911\right].
\]